\documentclass[a4paper,11pt]{article}
\usepackage{jinstpub} 
\usepackage{lineno}
\usepackage{subfigure}
\usepackage{multirow}
\usepackage{tabularx}

\title{\boldmath The CMS Phase-2 Fast Beam Condition Monitor prototype test with beam}

\author[a]{G. Auzinger}
\author[h]{H. Bakhshiansohi}
\author[a]{A. E. Dabrowski}
\author[t]{A. G. Delannoy}
\author[h]{V. Dalavi}
\author[a, p]{N. Dienemann} 
\author[f]{M. Dragicevic} 
\author[g]{M. F. Garcia} 
\author[c]{M. Guthoff}
\author[b]{B. Gyöngyösi} 
\author[m]{M. Jenihhin}
\author[d]{Á. Kadlecsik}
\author[a]{J. Kaplon}
\author[a]{O. Karacheban}
\author[b]{B. Korcsm\'aros} 
\author[d]{A. Lokhovitskiy}
\author[a, o]{W. H. Liu}
\author[a]{R. Loos}
\author[i]{S. Mallows}
\author[m]{D. Mihhailov}
\author[a, q]{M. Obradovic}
\author[a]{S. Orfanelli} 
\author[a]{M. Pari}
\author[d]{G. P\'asztor}
\author[a,s]{F. L. Pereira Carneiro} 
\author[a, r]{M. Princova}
\author[l]{B. Ristic}
\author[k]{C. Romero}
\author[e]{J. Schwandt} 
\author[h]{M. Sedghi}
\author[u]{A. Shevelev}
\author[m]{K. Shibin}
\author[t]{S. Spanier}
\author[e]{G. Steinbrueck} 
\author[j]{D. P. Stickland}
\author[a]{A. Tsirou}
\author[b]{B. Ujvári} 
\author[k]{M. Velasco}
\author[n]{P. G. Verdini}
\author[a]{G. J. Wegrzyn}
\author[a]{M. Wiehe} 
\author[k]{D. Wilbern}
\author[a,p]{P. D. Winney}

\affiliation[a]{CERN, European Organization for Nuclear Research, Geneva, Switzerland}
\affiliation[b]{University of Debrecen, Debrecen, Hungary} 
\affiliation[c]{Deutsches Elektronen-Synchrotron, Hamburg, Germany} 
\affiliation[d]{ELTE E\"otv\"os Loránd University, Budapest, Hungary} 
\affiliation[e]{University of Hamburg, Hamburg, Germany} 
\affiliation[f]{Institute for High Energy Physics, Vienna, Austria} 
\affiliation[g]{University of Cantabria and Spanish National Research Council (CSIC), Santander, Spain} 
\affiliation[h]{Isfahan University of Technology, Isfahan, Iran} 
\affiliation[i]{Karlsruher Institut für Technologie, Karlsruhe, Germany} 
\affiliation[j]{Princeton University, Princeton, NJ, USA} 
\affiliation[k]{Northwestern University, Evanston, IL, USA} 
\affiliation[l]{ETH, Z\"urich, Switzerland} 
\affiliation[m]{Tallinn University of Technology, Tallinn, Estonia} 
\affiliation[n]{University of Pisa and INFN, Pisa, Italy} 
\affiliation[o]{Oxford University, Oxford, United Kingdom} 
\affiliation[p]{Karlsruhe University of Applied Sciences, Karlsruhe, Germany}
\affiliation[q]{University of Belgrade, Belgrade, Serbia} 
\affiliation[r]{Palack\'y University Olomouc, Olomouc, Czech Republic} 
\affiliation[s]{University of Minho, Braga, Portugal} 
\affiliation[t]{University of Tennessee, Knoxville, TN, U.S.A.} 
\affiliation[u]{University of Maryland, College Park, MD, U.S.A.}

\emailAdd{Olena.Karacheban@cern.ch}

\enlargethispage{12pt}

\abstract{
The Fast Beam Condition Monitor (FBCM) is a standalone luminometer for the High Luminosity LHC (HL-LHC) program of the CMS Experiment at CERN. The detector is under development and features a new, radiation-hard, front-end application-specific integrated circuit (ASIC) designed for beam monitoring applications. The achieved timing resolution of a few nanoseconds enables the measurement of both  the luminosity and the beam-induced background. 
The ASIC, called FBCM23, features six channels with adjustable shaping times, enabling in-field fine-tuning.
 
Each ASIC channel outputs a single binary asynchronous signal encoding time-of-arrival and time-over-threshold information. The FBCM is based on silicon-pad sensors, with two sensor designs presently being considered.
This paper presents the results of tests of the FBCM detector prototype using both types of silicon sensors with hadron, muon, and electron beams. Irradiated FBCM23 ASICs and silicon-pad sensors were also tested to simulate the expected conditions near the end of the detector's lifetime in the HL-LHC radiation environment. Based on test results, direct bonding between the sensor and ASIC was chosen, and an optimal bias voltage and ASIC threshold for FBCM operation were proposed. The current design of the front-end test board was validated following the beam test and is now being used for the first front-end module, which is expected to be produced in summer 2025.
These results represent a major step forward in validating the FBCM concept, first version of the firmware and establishing a reliable design path for the final detector.
}

\keywords{beam-line instrumentation, radiation-hard detectors, si microstrip and pad detectors, detector design and construction technologies and materials}

\begin{document}
\maketitle
\flushbottom
\section{Introduction}
\label{sec:intro}

\subsection{Motivation and requirements for the dedicated CMS luminosity detector}

The Fast Beam Condition Monitor (FBCM) is a standalone luminometer designed for precision luminosity measurements and beam-induced background (BIB)~\cite{BRIL-TDR} monitoring that exploits silicon (Si) pad sensors with a fast, untriggered readout. It is part of the upgrade of the CMS experiment for the High-Luminosity LHC (HL-LHC) era, developed within the Beam Radiation, Instrumentation, and Luminosity (BRIL) project. A precise luminosity measurement is critical for numerous CMS measurements, and the target precision on the luminosity is set to about 2\% in real time (online) and below 1\% after data post-processing (offline)~\cite{BRIL-TDR}. 

During the HL-LHC phase, a large range of average rate of interactions per bunch crossing (also referred to as pileup, $\langle {PU}\rangle$) must be covered: from $\langle {PU}\rangle = 0.5$ up to $\langle {PU}\rangle = 200$. 
The linearity and the stability of the luminometer response to the full $\langle {PU}\rangle$ range are two key aspects which define its design. The linearity across a wide pileup range is essential as the specialized calibration runs are conducted at a low pileup of $\langle \text{PU} \rangle\approx 0.5$ a few times per year, while standard physics data taking in proton-proton collisions during HL-LHC operations will occur at $\langle \text{PU} \rangle\approx 140-200$. The linearity of FBCM was studied and optimized in simulations~\cite{BRIL-TDR}. The detector is shown to be linear up to $\langle {PU}\rangle = 200$ within the uncertainty band. The stability of the detector is defined as the ability to maintain its metrological properties, such as accuracy and performance, over an extended period of time, on the scale of several years for luminometer. The stability of the luminometer is influenced by operational aspects, as well as the aging of the sensors and electronics under radiation exposure. 
 
The simple and modular design of FBCM uses radiation-hard components for efficient operation until its scheduled replacement in the middle of the HL-LHC period. The expected total ionizing dose (TID) at that time is estimated to be approximately 80 MRad at the location of the Si-pad sensors at a radius of $R=14.5$~cm, corresponding to a 1 MeV neutron equivalent fluence of about $ 1.5 \times 10^{15} cm^{-2}$ and an integrated luminosity of about 2000 $\text{fb}^{-1}$ (see Fig 3.1 and Fig.18.12 in ~\cite{BRIL-TDR}).  The mixed radiation field at this location is composed of various particle types with a charged lepton to charged hadron to neutron fluence ratio of approximately 0.5: 0.75: 1.

To facilitate real-time validation and monitoring of the luminometer performance, and to provide precise estimates of systematic biases, multiple independent luminometers using different technologies are necessary at a hadron collider experiment. Multiple CMS subsystems, designed and used primarily for physics measurements, will also be utilized as luminometers: the outermost layer of the outer tracker -- consisting of double-sided Si-strip modules; the endcaps of the inner Si-pixel tracker -- including the innermost ring of the last (fourth) disk that will be fully dedicated to luminosity measurements; the forward hadron calorimeter; the barrel muon system; as well as the level-1 trigger data scouting system -- providing access to various trigger objects at 40~MHz~\cite{BRIL-TDR}. This approach allows robust and continuous luminosity reporting to the LHC during collisions and monitoring of the beam-induced background at all times when beams are present in the accelerator. Since the BRIL project does not control the operation of the CMS subsystems, a critical need arises for at least one dedicated detector optimized for luminosity measurement with sub-bunch-crossing time resolution, such as the Fast Beam Condition Monitor discussed in this paper.

\subsection{FBCM detector overview}

A first design of the FBCM was presented in 2021 in the BRIL Phase-2 Technical Design Report~\cite{BRIL-TDR}. Since then, extensive development work was performed to improve, validate, and finalize the design. It led to the (pre-)production and testing of a dedicated front-end ASIC~\cite{FBCM-ASIC} and various printed circuit boards (PCB) for prototyping the individual components~\cite{ shibin2024_1ASIC_test_board}, the finalization of the readout, control and powering systems, as well as the cabling plan, and the production of a real size mock-up~\cite{JINST_Andr_2024}. 

The FBCM is designed to operate independently of the CMS central trigger and data acquisition systems, functioning continuously with triggerless readout. It employs a dedicated front-end application-specific integrated circuit (ASIC) to amplify signals from CO$_2$-cooled silicon-pad sensors, offering nanosecond-level timing resolution to measure luminosity and beam-induced background. The modular system consists of two half-discs, each containing twelve modules, installed at both ends of the CMS detector, approximately 280 cm from the interaction point.
Service electronics are positioned near the outer edge of the half-discs to mitigate radiation-related aging. The electronics design reuses components from the CMS Tracker for power, control, and readout. The FBCM23 ASIC features six channels with adjustable shaping times, enabling fine tuning to match the sensor leakage current.
The FBCM mechanics are described in more detail in Subsec.~\ref{subsec:mechanics}, electronics in Subsec.~\ref{subsec:ASIC_carrier_board}, and sensors in Subsec.~\ref{subsec:sensors}. 

\subsubsection{FBCM detector mechanics}
\label{subsec:mechanics}

\begin{figure}[htbp]
\centering
\includegraphics[width=1.0\textwidth]{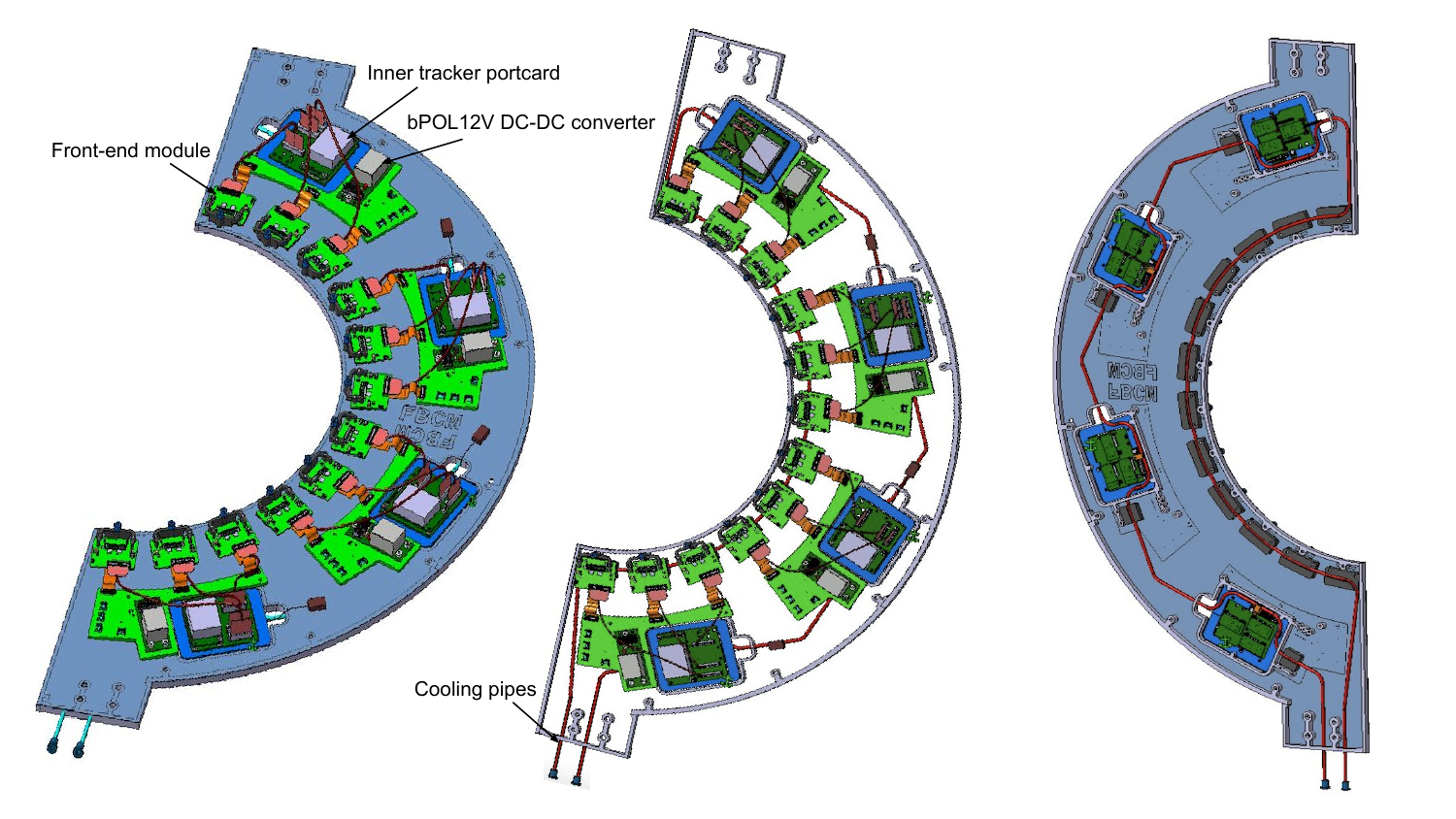}
\caption{The updated design of the FBCM half disks. Left: front view. Middle: front view without the upper carbon support sheet to reveal the cooling loop. Right: back view without the lower carbon support sheet to reveal the cooling loop. 
\label{fig:FBCM_dee}}
\end{figure}

\begin{figure}[htbp]
\centering
\includegraphics[width=1.0\textwidth]{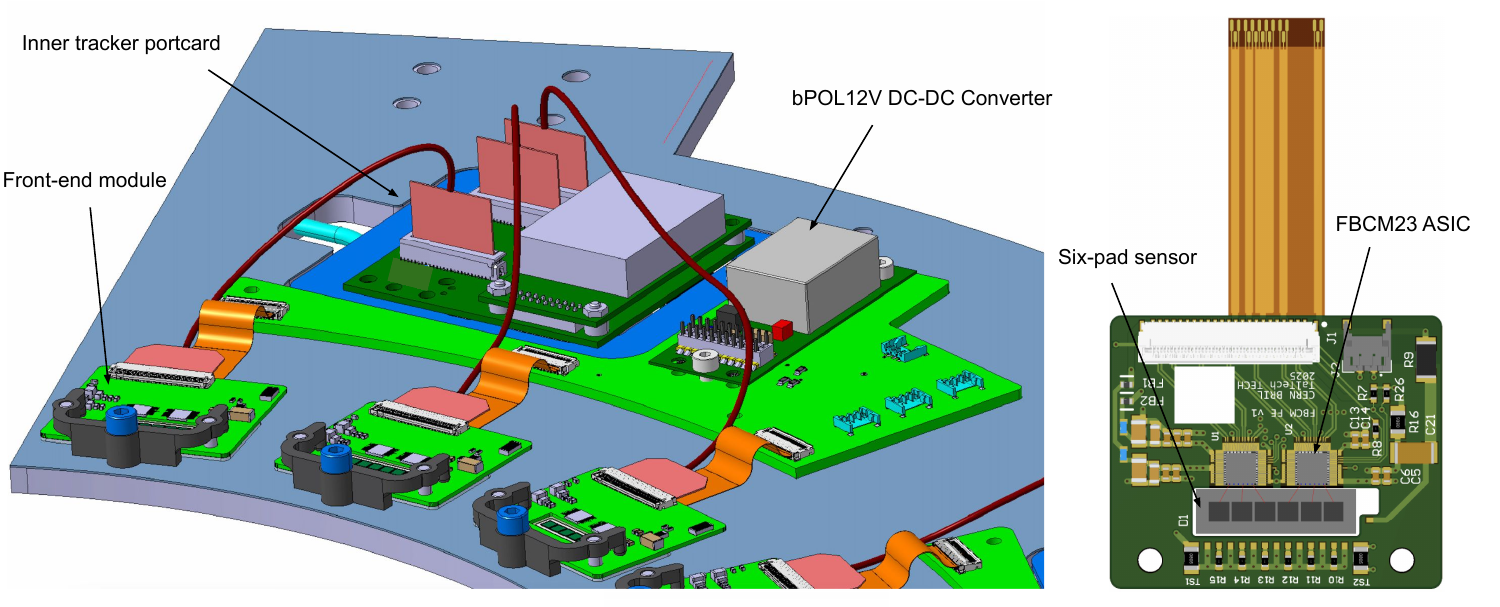}
\caption{ Left: an FBCM segment with three front-end modules, an inner tracker portcard with its cooling frame and a bPOL12V DC-DC converter housed on a service board. Right: the updated design of the front-end module with a six-pad silicon sensor, two FBCM23 ASICs, a flexible tail for powering, and a connector to provide a data link to the portcard.
\label{fig:FBCM_FE_module}}
\end{figure}

The latest design of the FBCM half disks is shown in Fig.~\ref{fig:FBCM_dee} both from front and back views, the former with and without the carbon support sheets to reveal the position of the cooling loop. The cooling loop passes behind the front-end modules, close to the inner edges, and is bent to cover the inner half of the portcard cooling frames, maximizing the cooling efficiency~\cite{JINST_Andr_2024}. FBCM is modular, with four repeating segments on each half disc. One of them is shown up close in Fig.~\ref{fig:FBCM_FE_module} on the left, together with the updated design of the front-end module on the right. Each segment consists of three front-end modules, each equipped with a six-pad silicon sensor and two FBCM23 ASICs~\cite{FBCM-ASIC}, one inner tracker (IT) portcard~\cite{TRACKER-TDR, lpGBT_manual} for ASIC signal sampling and electrical-to-optical conversion, and one bPOL12V DC-DC converter~\cite{Faccio:2020rae} (12~V $\rightarrow$ 1.25~V) for powering the front-end and one service board to distribute high-voltage and low-voltage lines. 
The step-by-step construction process of the half disk is described in Ref.~\cite{JINST_Andr_2024}.

\subsubsection{FBCM front-end electronics}
\label{subsec:ASIC_carrier_board}

\begin{figure}[bhtp]
\centering
\includegraphics[width=1.0\textwidth]{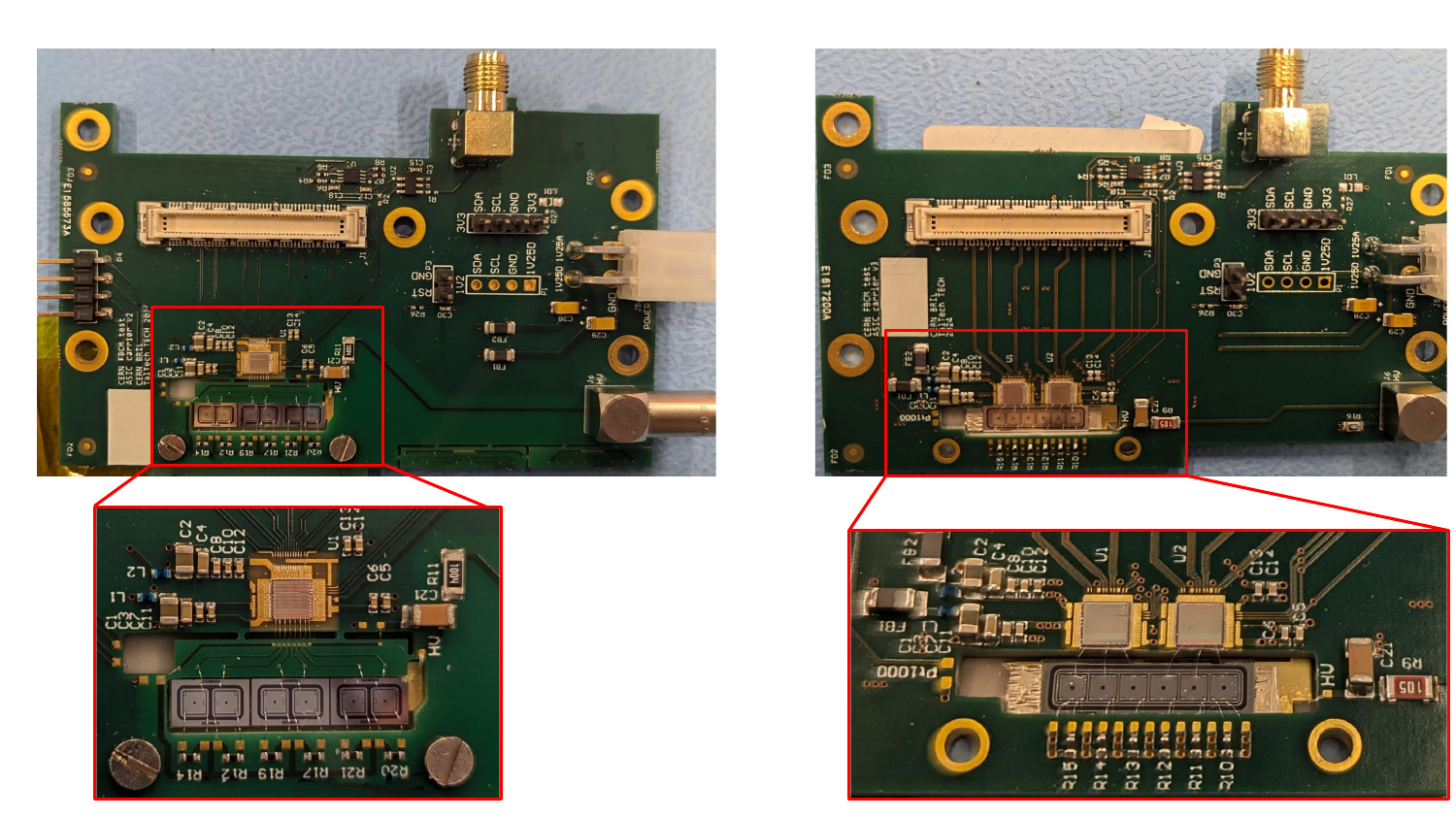}
\caption{ ASIC carrier board versions. Left: first test board design with an ASIC and three two-pad silicon sensors arranged in a row, wire-bonded to the ASIC inputs via a pitch adapter. Right: new test board with two ASICs and a six-pad silicon sensor with direct wire-bonding. The lower part of each figure zooms on the area with the ASIC(s) and the sensor(s).  
\label{fig:test_boards}}
\end{figure}

The configuration of the FBCM ASIC registers~\cite{FBCM-ASIC} requires $I^2C$ communication with the chip, which will be provided via the portcard in the final version of the detector. In 2024, a pre-production version of the portcard was used, which did not have $I^2C$ lines routed to the connector. As a result, external connectors were added to the FBCM FE module prototype (hereafter referred to as the ASIC carrier board or test board). The first version of the ASIC carrier board is shown in Fig.~\ref{fig:test_boards} (left) and described in Ref.~\cite{shibin2024_1ASIC_test_board}: it was designed to have a single ASIC to read out all six Si-pads, thus utilizes all six ASIC channels. In addition to the external $I^2C$ lines, it also features multiple test points, an analogue multiplexer (AMUX) connector for the direct readout of the ASIC voltage levels, and a connector for exchangeable mezzanine boards. Several mezzanines were designed, including one to connect the ASIC outputs to the portcard inputs via a flexible flat cable, and another with two sub-miniature version A (SMA) connectors per channel to allow direct readout of the raw  differential signals from the ASIC outputs for testing and characterization purposes.

The six sensor pads of the front-end module can be assembled in two different ways using the two types of sensors available for FBCM: either a single six-pad sensor or three two-pad sensors arranged in a row. The latter configuration results in a sensor panel that is 4 mm longer due to the additional edges at the borders between the sensors. As a result, the sensor choice impacts the design of the ASIC carrier board. A detailed description of the sensors is provided in Subsec.~\ref{subsec:sensors}.

The first version of the ASIC carrier board had a pitch adapter integrated into the PCB, allowing the wire-bonding of the sensor electrodes to the small ASIC input pads located very close to each other. The pitch adapter was designed to be interchangeable to accommodate different spacings between the silicon pads for the two sensor options.  
The pitch adapter is clearly visible on the lower subpicture in Fig.~\ref{fig:test_boards} (left) which zooms on the relevant area. On the upper side of the pitch adapter, eight wire bonds to the ASIC are visible. There are two wire-bonds to provide grounding connections (leftmost and rightmost), while the remaining six wire-bonds connect the AC-coupled electrodes of the six sensor pads to the six ASIC inputs. The wire-bonds to ground the guard ring and the DC-connections of the pads through resistors are shown on the lower side of the sensors. For the six-pad sensor, a different layout of the pitch adapter was used. For both sensor designs, pitch adapters with and without an extra grounding layer integrated among the PCB layers were produced.

After the first measurements with beam in April 2024, it was concluded that the pickup noise is too high and depends on the length of the signal path on the pitch adapter (more details in Sec.~\ref{subsec:ToT}). The minimum threshold to cut off the noise was between 2.5~fC and 3~fC, with a slightly lower threshold when using a pitch adapter with the extra grounding layer. Such a threshold is too close to the most probable value (MPV) of about 2.5-3~fC of the signal created by a minimum ionizing particle (MIP) in 290~$\mu$m of silicon. To mitigate the noise, it was proposed to switch to direct bonding from the sensor to the ASIC and remove the pitch adapter. To accommodate this solution and minimize the length of the bonds, a new design of the ASIC carrier board was developed featuring two ASICs, with only three input channels of each ASIC used. An example of such a test board with a six-pad sensor is shown in Fig.~\ref{fig:test_boards} (right). 

The second test with beam was performed in July 2024 for multiple ASIC carrier board designs (described in detail in Sec.~\ref{sec:test_beam}). 
The test boards differed in sensor types (two-pad or six-pad), sensor irradiation doses (between 0 and 100 MRad), number of ASICs (one or two), ASIC irradiation doses (0 or 100 MRad), and pitch adapter versions for single-ASIC designs (with grounding layer matching the two-pad or six-pad pitch, or no pitch adapter).

A list of all the test boards which are described in the paper with tracking number and important characteristics is presented in Table~\ref{table:testboards}. For the boards with irradiated sensors and ASICs, the approximate total ionizing dose is given. The TID ranges presented for the irradiated sensors are broad, as the 24~GeV proton beam used at CERN IRRAD facilities~\cite{IRRAD} did not cover the sensors uniformly, causing variations across the pads of a single sensor. In contrast, the ASICs were irradiated under a wide X-ray beam~\cite{X-ray}, resulting in a more precise TID.

\begin{table}[h!]
\centering
\begin{tabular}{| c | l | l | c |} 
 \hline
 Tracking number & Sensor type /  & Number of ASICs /   & Pitch adapter   \\ 
(high voltage)      &   irradiation   &  irradiation      & type             \\ 
 \hline\hline
          & six-pad   /          &                       & \\ 
 board 21 &  Ch0-3: 100 MRad,   &    2 ASICs /  & not used  \\ 
 (300 V)  &  Ch4: 60-80 MRad,    &   100 MRad          &         \\
          &  Ch5: >50 MRad     &                       &         \\          
  \hline
 board 23 & six-pad /             &   2 ASICs /     & not used \\ 
 (100 V)  &  Ch0-5: unirradiated     &    unirradiated         &         \\        
  \hline
          & two-pad /                &                          & \\ 
 board 24 & Ch0-1: unirradiated,        &  2 ASICs /        &  not used  \\ 
 (800 V)  & Ch3-4: 50-70 MRad,       &    100 MRad  &         \\           
          & Ch4-5: 80-100 MRad      &        &         \\  
  \hline
 board 25 & two-pad /                & 1 ASIC / & with grounding layer, \\ 
 (300 V)  &  Ch0-5: unirradiated     &  unirradiated     & two-pad pitch   \\        
  \hline
 board 26 & six-pad                    & 1 ASIC /     & with grounding layer, \\ 
 (300 V)  &  Ch0-5: unirradiated          &           unirradiated  & six-pad pitch       \\        
  \hline 
\end{tabular}
\caption{Summary table of ASIC carrier board properties used at the beam test in July 2024, including the tracking number and operational high voltage of each board, as well as the type and received irradiation dose of the main components. For irradiation scale reference, the TID of 80 MRad at the location of the Si-pad sensors at a radius of $R=14.5$~cm, corresponds to a 1 MeV neutron equivalent fluence of about $ 1.5 \times 10^{15} cm^{-2}$ and an integrated luminosity of about 2000 $\text{fb}^{-1}$. }
\label{table:testboards}
\end{table}

\subsubsection{FBCM silicon sensors}
\label{subsec:sensors}

\begin{figure}[htbp]
\centering
\includegraphics[width=1.0\textwidth]{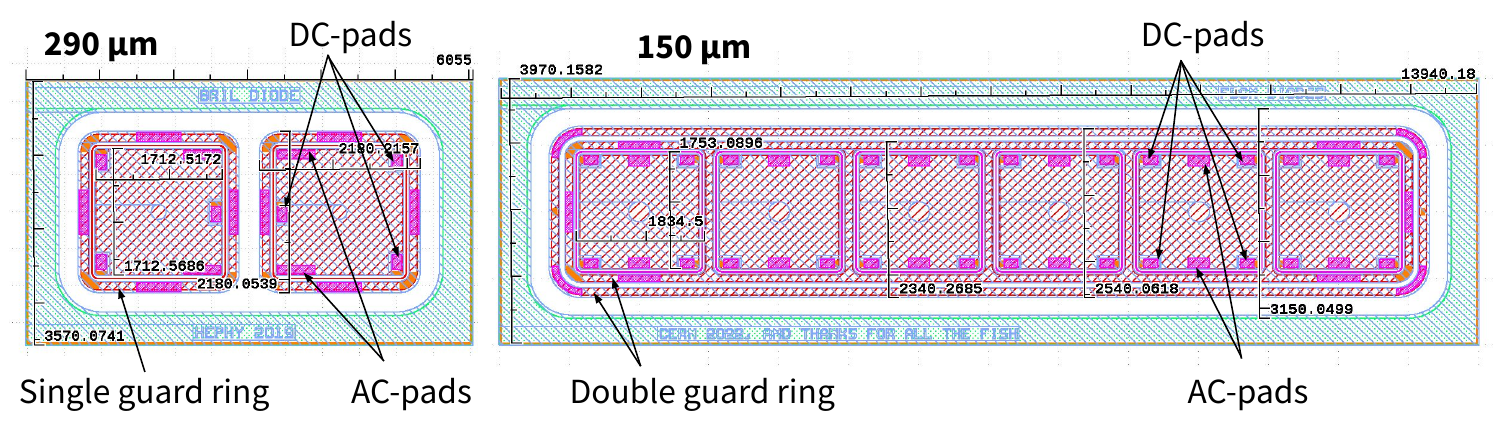}
\caption{ Left: 290 $\mu$m thick two-pad silicon sensor design. Right: 150 $\mu$m thick six-pad silicon sensor design. Areas with direct contact to the bulk of the sensor are shown in orange. Pads shown in pink are dedicated for wire-bonding. The thin lines around each pad show the location of the p-stop implant between the pad and the guard ring. Areas without metallization are indicated in white. 
\label{fig:fbcm_sensors}}
\end{figure}

There are two types of n-on-p sensors available for FBCM, as shown in Fig.~\ref{fig:fbcm_sensors}. Both sensor types use float zone thinned (FTH150) wafer material~\cite{TRACKER-TDR,6-pad-paper} and all sensors are produced by a single vendor, Hamamatsu Photonics K.K. (HPK). The main difference between them is the thickness and the guard ring structure, which is required to restrict the active area of the sensor. The active area per pad, defined by the size of the metallization, is the same for both sensor options --- each pad measures  1.7 x 1.7 mm$^2$. The FBCM ASIC is designed to be compatible with both types of sensor. However, shot noise increases with increasing leakage current under accumulated sensor exposure to radiation. The initial ASIC noise is 650 e$^-$~\cite{FBCM-ASIC}, but after exposure to  $10^{15}$ cm$^{-2}$ 1 MeV neutron equivalent ($ \textrm{n}_\textrm{eq} $) fluence, it increases to about 1500 e$^-$ for the 290 $\mu$m thick two-pad sensors and 850 e$^-$ for 150 $\mu$m thick six-pad sensors. The expected signal in electrons for a particular sensor can be estimated by multiplying the sensor thickness by the number of electron-hole pairs produced in silicon $n_{e^{-}h^{+}}$=78~$\mu$m$^{-1}$~\cite{Otarid:2884106}. The expected signal-to-noise (S/N) ratio and other characteristics of the sensors, both before and after irradiation, are listed in  Table~\ref{table:sensors_comparison}, taking into account the reduced charge collection efficiency (CCE) of the sensors after irradiation~\cite{CCE_1}. The capacitance is given based on the simple reverse-biased diode capacitance model.

\begin{table}[h!]
\centering
\begin{tabular}{|>{\raggedright}m{8cm}|>{\raggedright\arraybackslash}m{2.5cm}|>{\raggedright\arraybackslash}m{2.5cm}|}
 \hline
 Parameter          & two-pad sensor         &   six-pad sensor       \\         
  \hline \hline
 Thickness         &     290 $\mu$m        & 150 $\mu$m           \\         
  \hline
 Guard ring structure around each pad     &     single  & double  \\   
   \hline
Full depletion voltage     &   250-280 V    & 50-70 V  \\   
  \hline 
 Capacitance at full depletion         &     1.2 pF        &  2.2 pF           \\         
  \hline
 S/N$_\text{new}$ before irradiation               &   35           &   17            \\             
\hline
 Irradiation fluence (1 MeV $ \textrm{n}_\textrm{eq}) $               &   $1 \times 10^{15}$ cm$^{-2}$  &   $1.5 \times 10^{15}$ cm$^{-2}$            \\                 
\hline
 Estimated S/N$_\text{irrad}$ after irradiation               &   11          &   8           \\                 
  \hline
 Estimated CCE after irradiation at 600 V            &   $\approx$ 50\%     &  $\approx$ 90\% \\         
  \hline
Measured leakage current after irradiation   at -20~$^\circ C$ at 500~V   &   1 $\mu A$     &  ~0.1 $\mu A$  \\         
  \hline 
\end{tabular}
\caption{Comparison of two-pad and six-pad sensors (see also Fig.~\ref{fig:fbcm_sensors}). The values given for the S/N$_\text{irrad}$ and CCE after irradiation are estimations
based on the extrapolation between available data points in the literature for 120~$\mu$m, 200~$\mu$m and 300~$\mu$m thick sensors~\cite{CCE_1}. }
\label{table:sensors_comparison}
\end{table}

\begin{figure}[htbp]
\centering
    \subfigure{%
        \includegraphics[width=0.95\linewidth]{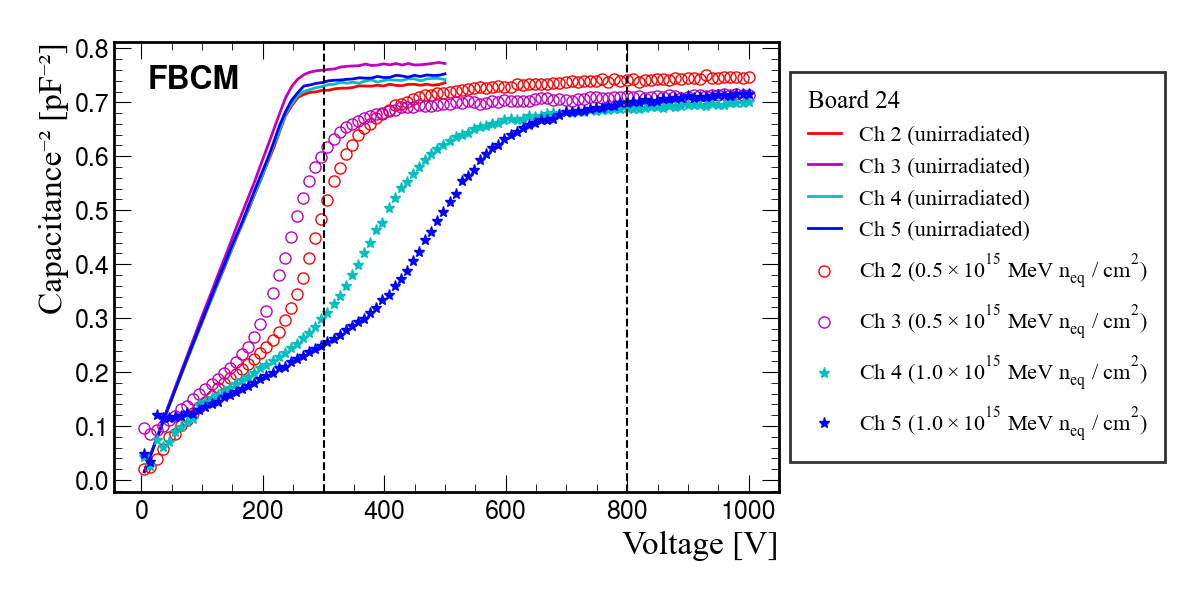}}
\caption{ Inverted capacitance squared of the two-pad sensors as a function of the applied  voltage. The vertical dashed lines correspond to the operational voltage applied during the test beam in July 2024 for unirradiated (300~V) and irradiated (800~V) sensors.}
\label{fig:2_pad_CV}
\end{figure}

\begin{figure}[htbp]
\centering
    \subfigure{%
        \includegraphics[width=0.95\linewidth]{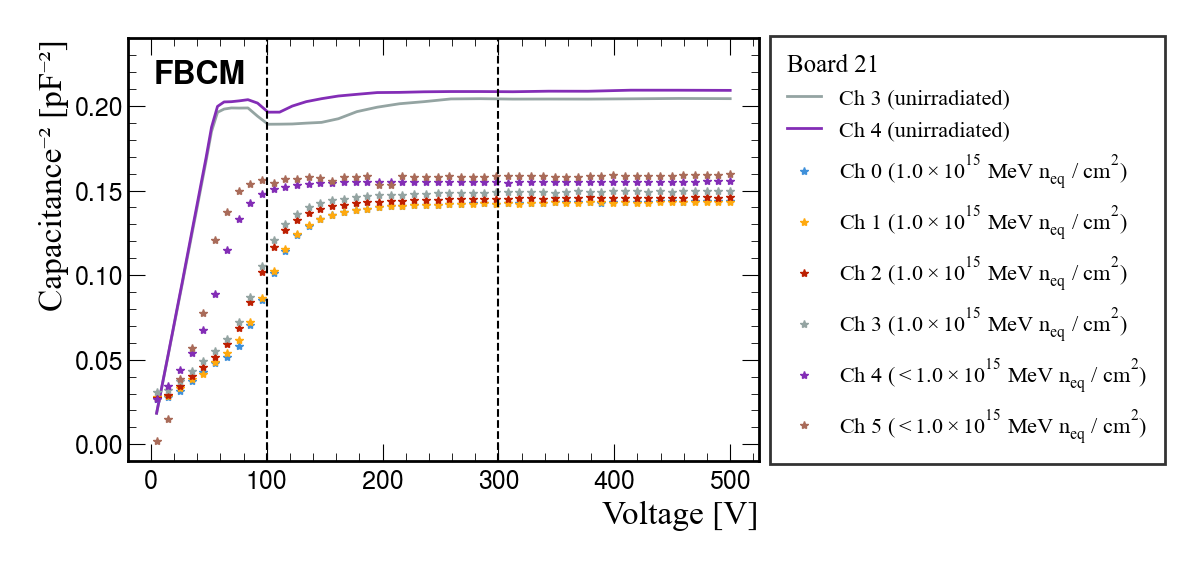}}
\caption{ Inverted capacitance squared of the six-pad sensors as a function of the applied voltage. The vertical dashed lines correspond to the operational voltage applied during the test beam in July 2024 for unirradiated (100~V) and irradiated (300~V) sensors. Capacitance change after irradiation is not expected and to be checked in the next irradiation campaign. }
\label{fig:6_pad_CV}
\end{figure}

Examples of the measured inverted capacitance squared of the sensors as a function of the applied voltage (C-V curves) are shown with solid lines for unirradiated and, with multiple markers, for irradiated sensor pads in Fig.~\ref{fig:2_pad_CV} for two-pad sensors and in Fig.~\ref{fig:6_pad_CV} for six-pad sensors. The minimum high voltage (HV) required to reach the plateau of these curves corresponds to the full depletion of the sensor. The first vertical dashed line on the plots (100 V for six-pad and 300 V for two-pad sensors) indicates the operational voltage applied at the beam test for unirradiated sensors, and the second dashed line for irradiated sensors (300 V for six-pad and 800 V for two-pad sensors). 

Two-pad sensors were previously used in CMS for the BCM1F detector~\cite{BCM1F-RUN3} during Run 3 of the LHC and initially operated at bias voltage of 400~V. As they exhibited current instability after exposure to a fluence of about $2 \times 10^{13}$ cm$^{-2}$ 1~MeV neutron equivalent, operational bias voltage was lowered to 350~V. This instability~\cite{BCM1F-RUN3-paper} manifested itself in the unpredicted rise of the leakage current and eventual trip due to having reached the set power supply current limit (50~$\mu$A). This was attributed to the build up of surface current and the formation of a conductive path, which allowed current to flow from the edges of the sensor to the guard ring. Reduction of the bias voltage minimizes the frequency of the trips, but does not mitigate them. Further bias voltage reduction was not possible due decline in sensors efficiency. To reproduce this behavior under controlled laboratory conditions, the two-pad sensors were tested under an X-ray lamp~\cite{X-ray}, which generates only surface charge. Over ten hours of X-ray exposure, a slow increase of the current was observed. Intensity of X-ray was selected to match the leakage current close to that observed in operations at the beginning of collisions (10~$\mu$A). When the X-ray lamp was switched off, a gradual current decrease  was observed, lasting about one hour.
To verify that the double guard ring structure of the six-pad sensor resolves this issue, the same test was repeated. The results showed that the current through the guard ring and DC-pads remained stable after hours of the surface irradiation, making the six-pad sensor the preferred choice for FBCM construction. However, both types of sensors were thoroughly tested in the proton beam test in July 2024.
\section{Beam test with the ASIC carrier boards}
\label{sec:test_beam}

To validate the performance of the FBCM23 ASIC and test its response in realistic conditions, a beam test was performed for a series of test boards. These included boards with either three two-pad sensors or a six-pad sensor, with either fresh or irradiated sensors and ASICs, as specified in Table~\ref{table:testboards}. The experiment took place at the T9 beam line in the East Area of the Proton Synchrotron~\cite{T9:PS} at CERN. The beam had a defined structure: particles arrived in short 400 ms wide time windows (spills). Normally, there were 2-3 spills per minute, reduced to 1 spill per minute when the LHC accelerator ring was being filled. 

Beams of various particle types and energies were available: electrons, muons, and protons. Most of the data was taken with 15 GeV protons since the available beam intensity was highest in that configuration, reaching 2 $\times$ 10$^5$ protons per spill. Several short tests were conducted with other beam types. 

\begin{figure}[htbp]
\centering
    \subfigure{%
        \includegraphics[width=0.45\linewidth]{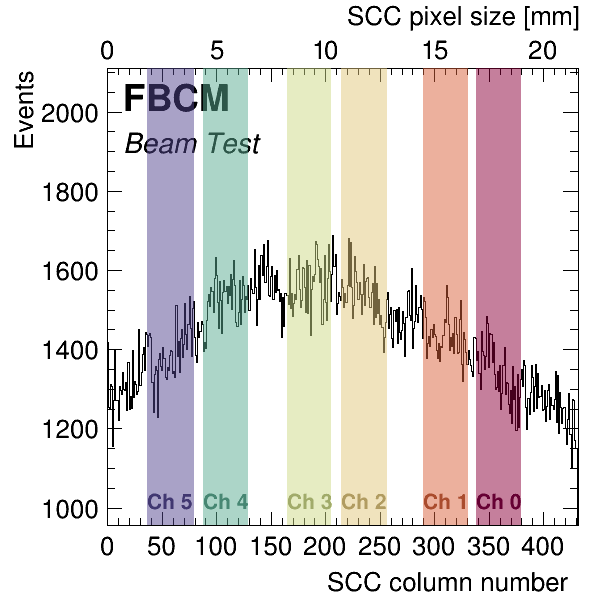}}
    \subfigure{%
        \includegraphics[width=0.45\linewidth]{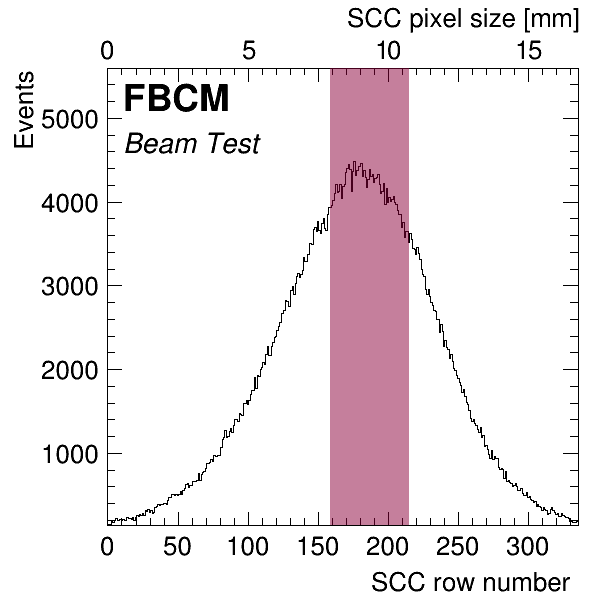}}
\caption{ The profile of the proton beam in X (left) and in Y (right) as measured using an Inner Tracker Single Chip Card (SCC) placed upstream of the cold box. The colored bands correspond to the position of the FBCM sensor pads. The channel numbering used in the paper is also indicated on the left. SCC pixel size is 50 $\times$ 50 $\mu$m (for raw and column mm scale reference).}
\label{fig:beamprof_x_y}
\end{figure}

The settings of the focusing magnets and the collimators were adjusted to produce an oval beam shape centered at the position of the FBCM sensors. It was important to have a horizontally (in X coordinate) flattened beam to achieve a maximally uniform hit rate along the 18 mm long panel of the FBCM Si-pads. In contrast, the vertical (Y coordinate) component of the beam had to be focused to maximize the hit rate in the less than 2 mm wide vertical active area of the sensor. The achieved beam profile is shown in Fig.~\ref{fig:beamprof_x_y}.  
The beam width was approximately 15 mm in X and 3 mm in Y at the location of the FBCM sensors.

\subsection{Experimental setup}
\label{subsec:test_beam_setup}

\begin{figure}[htbp]
\centering
\includegraphics[width=1.0\textwidth]{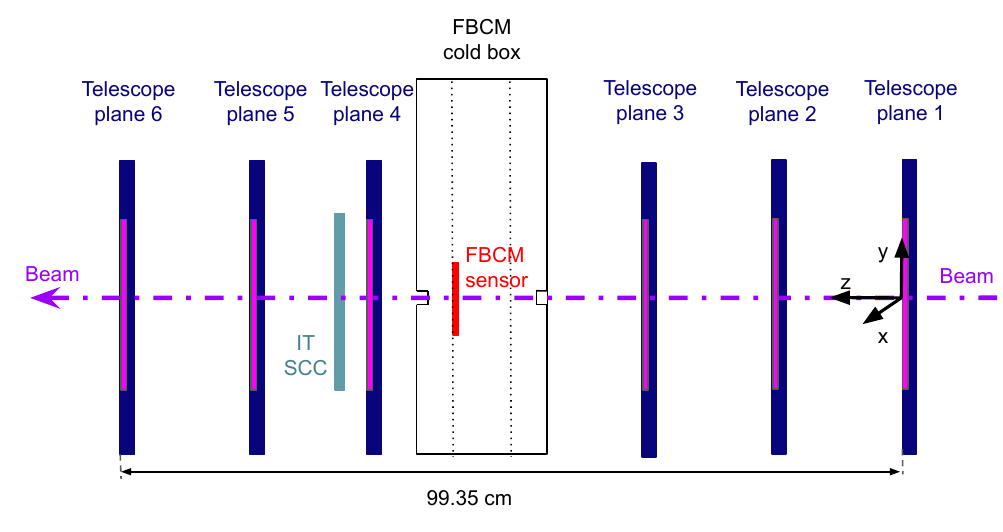}
\caption{Beam test setup: six telescope planes (dark blue with the active pixel detectors shown in magenta), the FBCM cold box with sensors attached to the ASIC carrier board (red), and an Inner Tracker Single Chip Card (SCC) (teal-turquoise) used as a timing reference. }
\label{fig:testbeam_setup_drawing}
\end{figure}

\begin{figure}[htbp]
\centering
\includegraphics[width=1.0\textwidth]{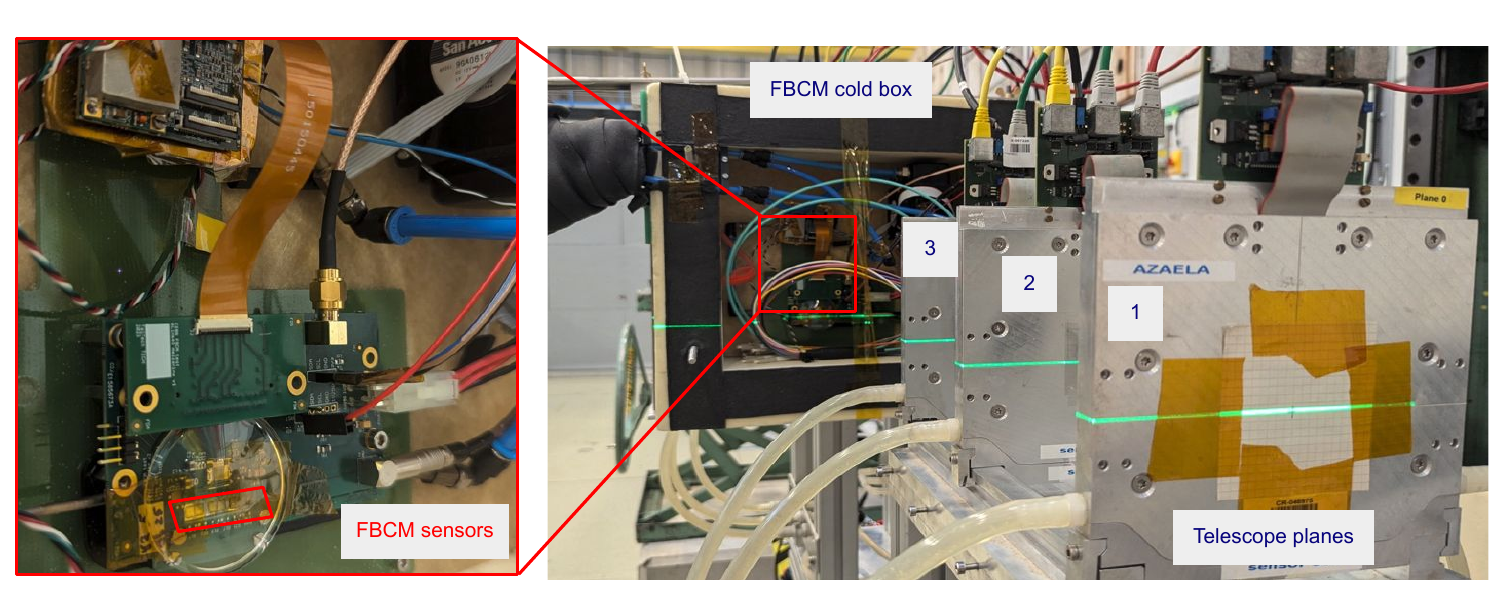}
\caption{Photograph of the beam test setup. Right: three telescope planes are visible, followed by the (open) FBCM cold box. Left: a close-up of the ASIC carrier board with three two-pad Si sensors is shown, along with the  mezzanine that connects the ASIC outputs to the portcard inputs via a flexible flat cable.}
\label{fig:testbeam_setup_photo}
\end{figure}
    
\begin{figure}[htbp]
\centering
    \subfigure{%
        \includegraphics[width=0.45\linewidth]{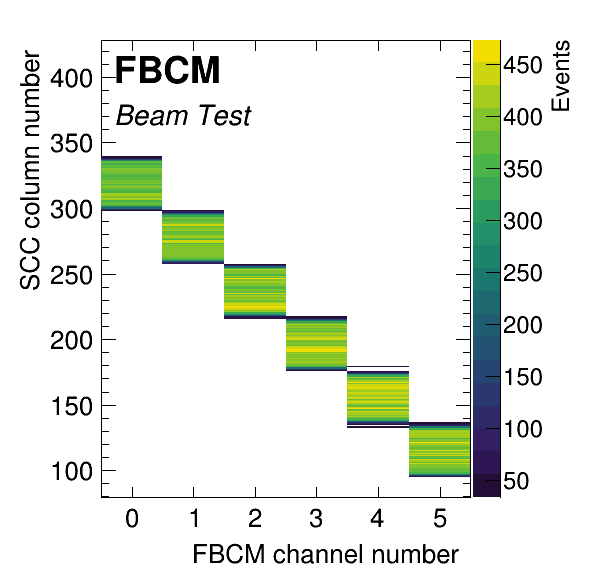}}
    \subfigure{%
        \includegraphics[width=0.45\linewidth]{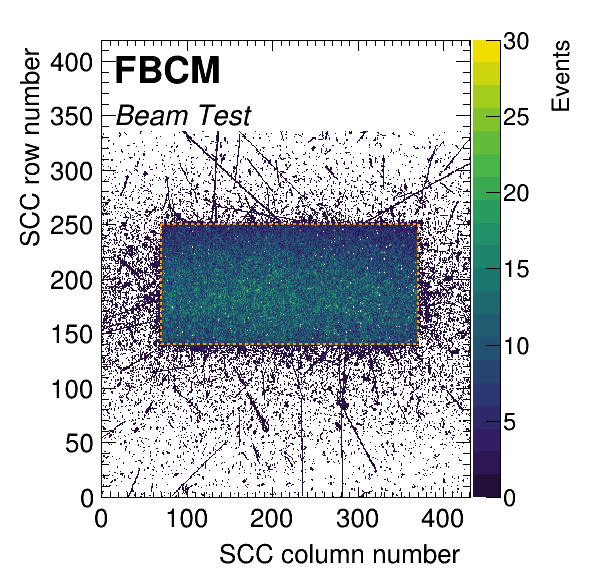}}
\caption{  Left: correlation map between hits in the SCC and signals from the FBCM Si-pads. Right: hit map in the SCC for events with a correlated signal from the FBCM sensor pads to set the trigger window. Hits outside the hot rectangular area arise when a coinciding hit is also present within that area.}
\label{fig:Hitmap_CROC}
\end{figure}

Fig.~\ref{fig:testbeam_setup_drawing} shows a sketch of the beam test setup, illustrating the FBCM sensors positioned in the cold box and two auxiliary detectors. The six-plane AIDA~\cite{Rubinskiy:Telescope} telescope, with 18.4 $\mu$m $\times$ 18.4 $\mu$m Si-pixels, is used to reconstruct the trajectory of charged particles, while a CMS Inner Tracker Single Chip Card (SCC) hosting an RD53B chip with 50 $\mu$m $\times$ 50 $\mu$m Si-pixels~\cite{TRACKER-TDR} and 25 ns time resolution, is utilized for triggering on particles that pass through the region aligned with the location of the FBCM sensors. The Inner Tracker SCC, hereafter simply SCC, was attached to the telescope plane positioned after the FBCM cold box, as shown in Fig.~\ref{fig:testbeam_setup_drawing}, and aligned with the center of the telescope using a mechanical holder. The FBCM cold box was mounted on an X-Y movable stage between the two groups of three telescope planes. A laser was used for preliminary alignment of the FBCM sensors with the center of the AIDA telescope and the beam line. It is visible as the horizontal green line passing through the center of the first telescope plane on the photo of the beam test setup in Fig.~\ref{fig:testbeam_setup_photo}. The final alignment of the FBCM cold box was done with the beam, based on correlation maps between the triggered hits in the SCC and signals from the FBCM Si-pads. 
These correlation maps were also used to define the active area of the inner tracker pixel module (pixel mask), which guaranteed that a trigger was only generated when a particle crossed the FBCM sensor area. An example of such a correlation map is shown in Fig.~\ref{fig:Hitmap_CROC} (left). A pixel mask defined from such a correlation map to cover all FBCM sensors is visible in Fig.~\ref{fig:Hitmap_CROC} (right), framed by an orange rectangle.

\begin{figure}[htbp]
\centering
    \resizebox{.9\textwidth}{!}{%
        \includegraphics[height=.9\textheight]{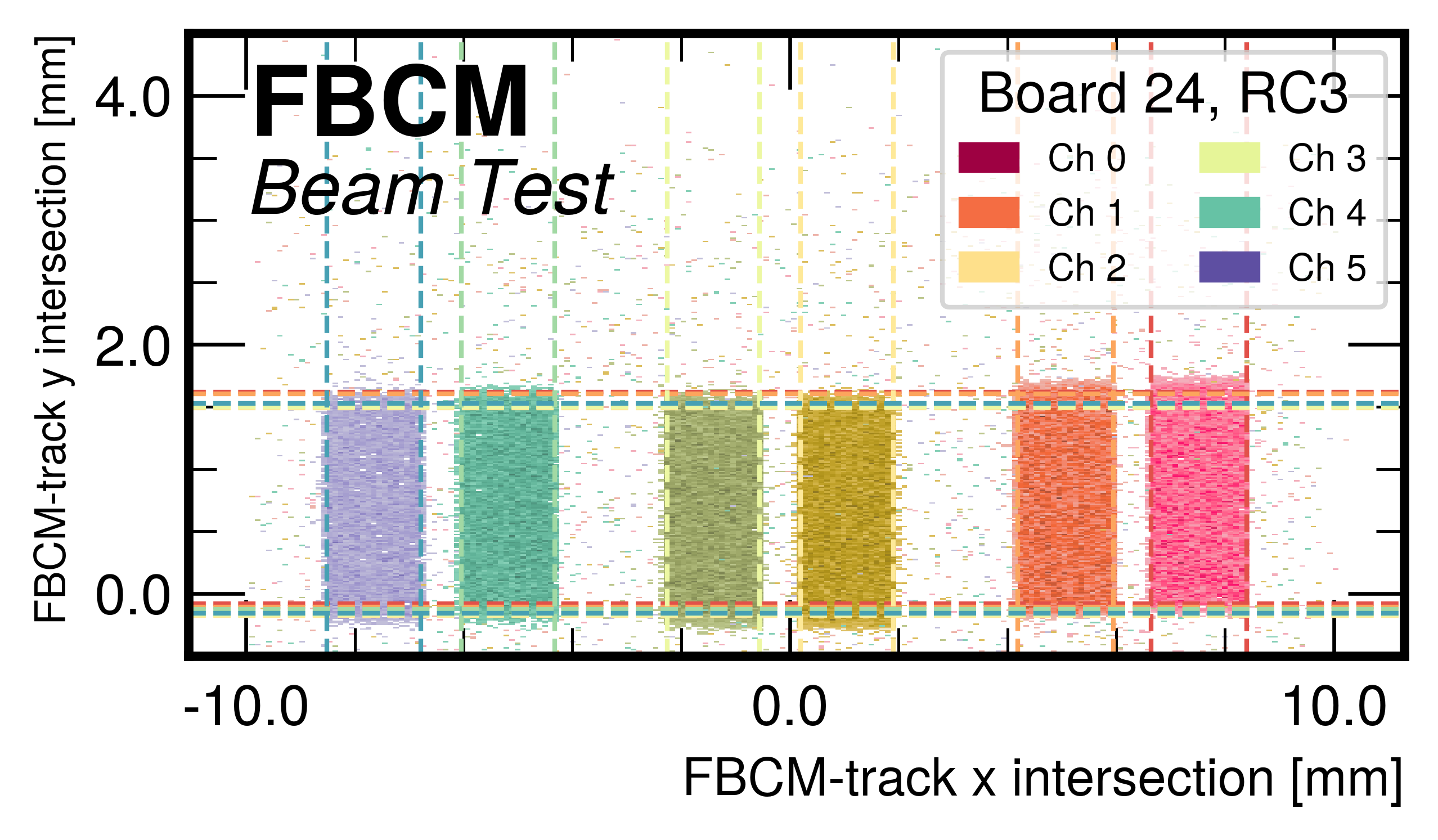}%
        \includegraphics[height=.9\textheight]{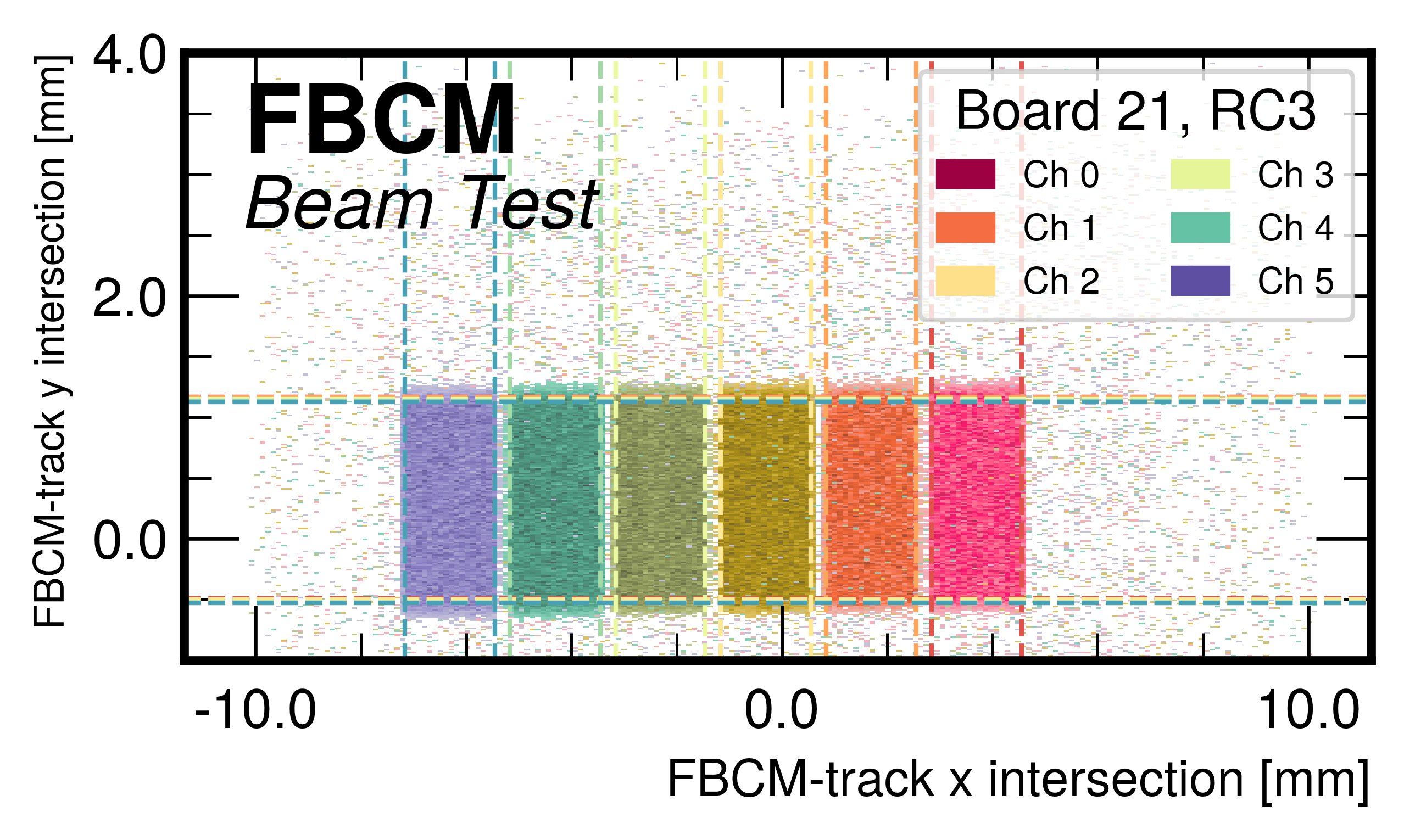}%
    }
\caption{  Hit map associated with signal in any of the FBCM channels for board 24 (with two-pad sensors) on the left and for board 21 (with a six-pad sensor) on the right. Data is aggregated for all threshold with high sensor efficiency to maximize the statistics. The tracks are reconstructed with the AIDA telescope and the expected hit positions are interpolated to the FBCM plane. Dashed lines indicate the position of the sensor edges defined by a step function fit to the hit map distribution.}
\label{fig:Hitmap_FBCM}
\end{figure}

All track positions with an associated signal in one of the FBCM channels are interpolated to the FBCM plane and their coordinates are shown in Fig.~\ref{fig:Hitmap_FBCM}: the hit map on the left corresponds to board 24 with three two-pad sensors, and the one on the right to board 21 with a six-pad sensor. Hits in each channel are shown in colors, using the same scheme as Fig.~\ref{fig:beamprof_x_y}. The geometrical difference in the position of the sensitive pads for the two types of boards is clearly visible. These maps were used to define the per-channel fiducial region, as described in detail in Sec.~\ref{subsec:sensor_efficiency}.

\subsection{Signal processing and data acquisition }
\label{subsec:data_acquisition}

\begin{figure}[htbp]
\centering
\includegraphics[width=1.0\textwidth]{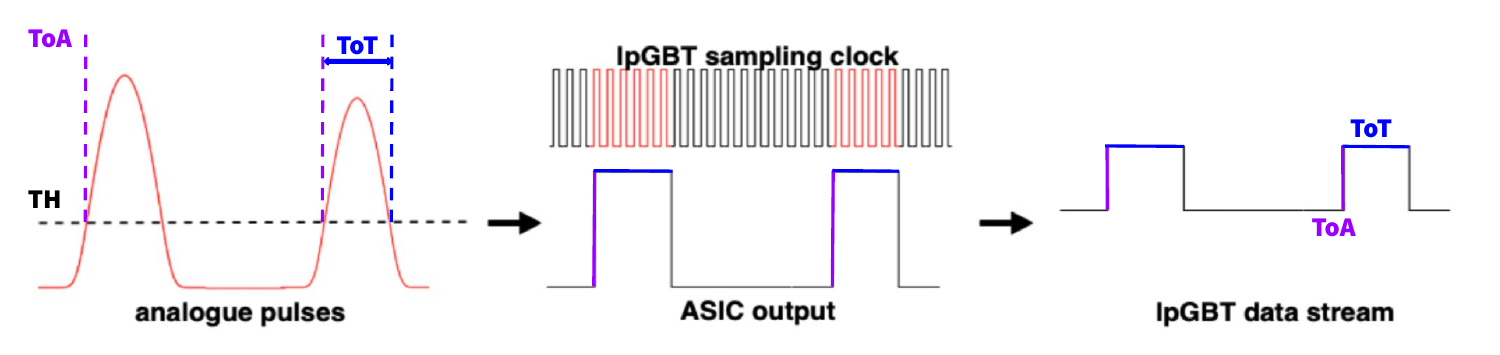}

\caption{Sketch of the ASIC data processing and sampling of the FBCM23 signal with the lpGBT on the portcard. The time of arrival (ToA) is indicated in violet for the adjustable threshold (TH) value and the time over threshold (ToT) is indicated in blue. }
\label{fig:ASIC_output}
\end{figure}

A primary electric signal is induced in the silicon sensors when a charged particle passes through them. This signal is processed by the FBCM23 ASIC~\cite{FBCM-ASIC}, which outputs rectangular pulses with the time of arrival (ToA) indicating when the signal crosses the threshold, while the time over threshold (ToT) corresponds to the amount of time during which the pulse is above the threshold. 
The definitions of ToA and ToT for two analogue pulses of different lengths are illustrated in Fig.~\ref{fig:ASIC_output}. In this example, a single value of the threshold is shown, however the threshold for FBCM23 ASIC is an adjustable parameter. It is important to emphasize that the ToT and ToA are threshold dependent, as explained in detail in Sec.~\ref{sec:results}. 

To cover a wide operational range and to reduce noise, the ASIC contains an internal low-pass RC filter, comprised of a 2.1~k$\Omega$ resistor and a capacitance that is adjustable in increments of 430~fF within a range between 0 and 6 $\times$ 430 = 2580~fF. While these capacitance and resistance values are provided for reference, the RC filter is part of a larger analog signal processing chain which includes amplification and shaping stages~\cite{FBCM-ASIC}. As a result, the effective filter characteristics -- and consequently its cutoff frequency -- cannot be directly inferred from these individual parameter values in a straightforward manner. 
Adjustment of the filter capacitance is done via an internal register and can be viewed as altering the ``strength" of the filter, with the lowest setting having no capacitance and being referred to as RC0, and the highest setting having the full capacitance of 6 $\times$ 430~fF and being referred to as RC6. Laboratory tests showed that the RC0 setting is not optimal due to high noise levels. The RC3 setting was found to be optimal and used as a baseline setting at the beam test. Several tests with the RC6 setting were also performed during the beam test to study the effect of the filter on the performance.

\begin{figure}[htbp]
\centering
\includegraphics[width=1.0\textwidth]{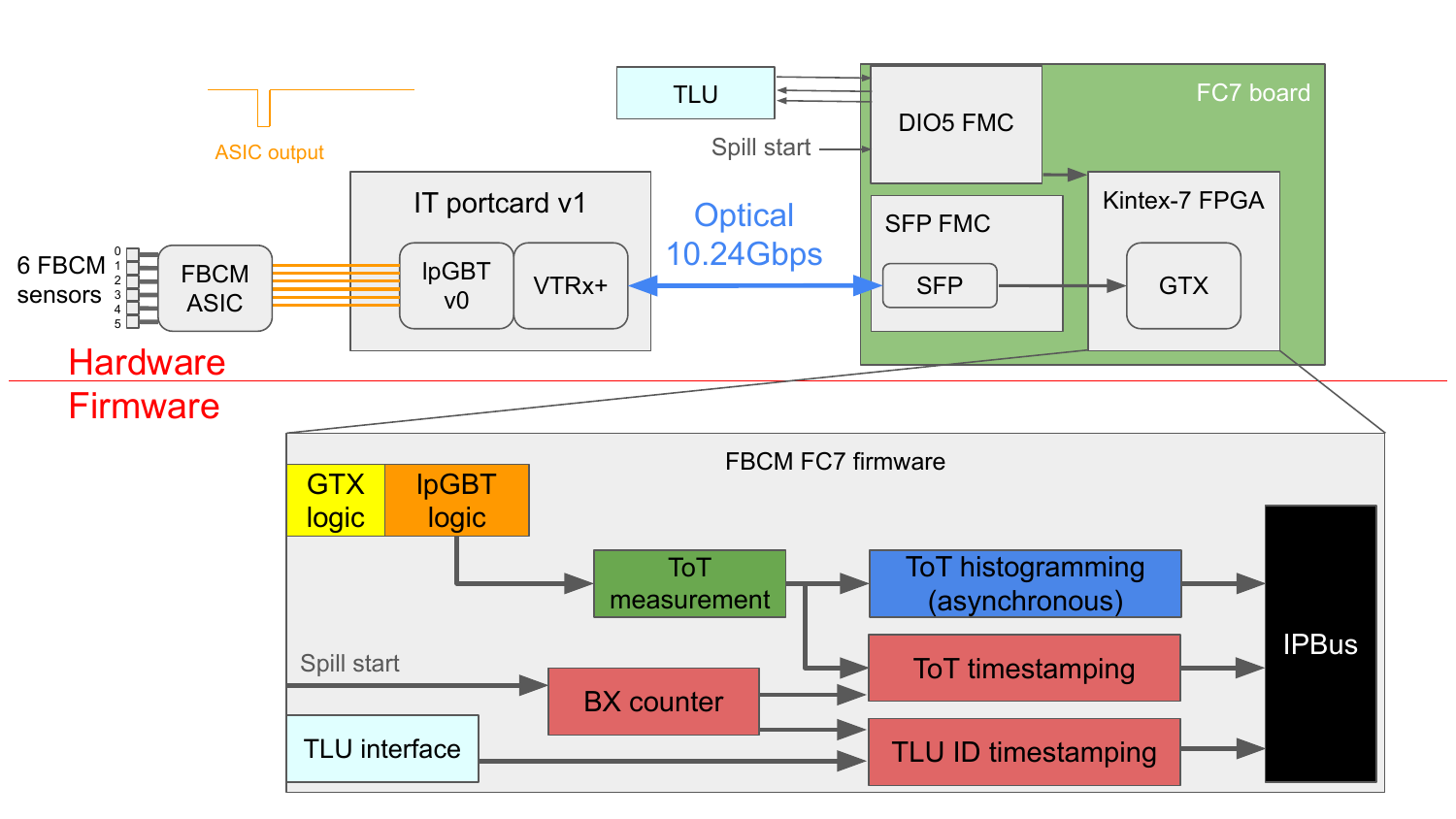}

\caption{FBCM hardware (top part of the block diagram) and firmware (bottom part of the block diagram) data path. All data is transferred to the FC7 processing board via two FPGA Mezzanine Cards: DIO5 and SFP. Data processing is carried out in the Kintex-7 FPGA firmware.}
\label{fig:FW_test_beam}
\end{figure}

The FBCM23 ASIC output is sampled at 1.28 GHz rate by the Low Power GigaBit Transceiver (lpGBT) chip~\cite{lpGBT_manual} on the IT portcard and converted into an optical signal by the Versatile Link Plus Transceiver (VTRx+)~\cite{VTRx}. The optical signal is then transmitted to the FC7 advanced mezzanine card (AMC)~\cite{FC7} processing board via small form-factor pluggable (SFP) on the FPGA mezzanine card (FMC), at a maximum bandwidth of 10.24 Gbps. The second I/O FMC (DIO5 FMC) is used to receive trigger number and ``spill start" signals from the trigger logic unit (TLU).   
Data received by the gigabit transceiver (GTX) is processed in the Xilinx Kintex-7 Field Programmable Gate Array (FPGA) firmware. The full hardware and firmware data path is shown in Fig.~\ref{fig:FW_test_beam}.

The pulses generated by the FBCM23 ASIC are asynchronous and untriggered. The firmware receives a multi-gigabit data stream from the lpGBT. From this continuous stream of binary data, the firmware logic captures a series of consecutive ``1" bits which correspond to the FBCM23 ASIC output signal in the active state (incoming signal from the sensor above the selected threshold). The first ``1" in each sequence gives the ToA. For the ToT measurement, the firmware calculates the length of each such series, which corresponds to the FBCM23 ASIC output pulse length (in 0.78 ns units). This number is then stored in memory and subsequently read out.

Since protons arrive in 400 ms spills during the beam test, the firmware must be able to accept an external signal representing the start of a spill. This signal was used to reset internal counters in the firmware and start collecting a new ToT histogram. 
Therefore, the prototype FBCM firmware was adjusted with special features tailored to the beam test. In particular, the signal indicating start of the spill was used to enable per-spill ToT histogram readout, while  the TLU input was used for the subset of events triggered by an external detector (Inner Tracker SCC) for cross-checking.
In the absence of an end of the spill signal, an internal counter was implemented as a stop signal for per-spill data collection and a ``data ready" flag for readout.
Additionally, the ToA measurement was disabled as there was no relationship to a bunch-crossing clock in the beam test setup.

The ToT measured in the firmware was stored in two FPGA memory locations: (1) as a ToT histogram entry (for all events), and (2) as  timestamped-triggered ``raw" data (for a subset of events) allowing later correlation studies with the beam telescope data (see Sec.~\ref{subsec:hit_position_study}).

A TLU trigger was initiated when a particle crossed the active pixels in the SCC, indicated by the orange box in Fig.~\ref{fig:Hitmap_CROC} (right). 
After the TLU trigger was issued, the FBCM-readout software opened a short time window (of about 400 ns) and associated all detected hits within this window to the received TLU trigger number. The delay in the arrival of the trigger, due to cable length and processing time, was measured experimentally by shifting the 400 ns time window in the estimated delay range and maximizing the number of registered hits. Each trigger was tagged by a ``bunch crossing" (BX) identification number corresponding to 25 ns time slices, while the time over threshold is sampled at 1.28 GHz, resulting in a time granularity of 0.78 ns. The TLU trigger number was timestamped using the same BX counter value as the ASIC signal pulses in the firmware, and saved alongside the timestamp in a separate buffer for triggered readout.

While timestamped events were only collected during the active spill time, the ToT histograming logic is designed to work continuously, as it would during the normal operation of FBCM. Every time the collection of a histogram was stopped, it was stored in a buffer to be read out and another histogram was immediately started. During the beam test data collection, it resulted in two histograms being collected per each spill: one for the the signal received during the spill and another for the background received between the two spills.

\begin{figure}[htbp]
\centering
\includegraphics[width=1.0\textwidth]{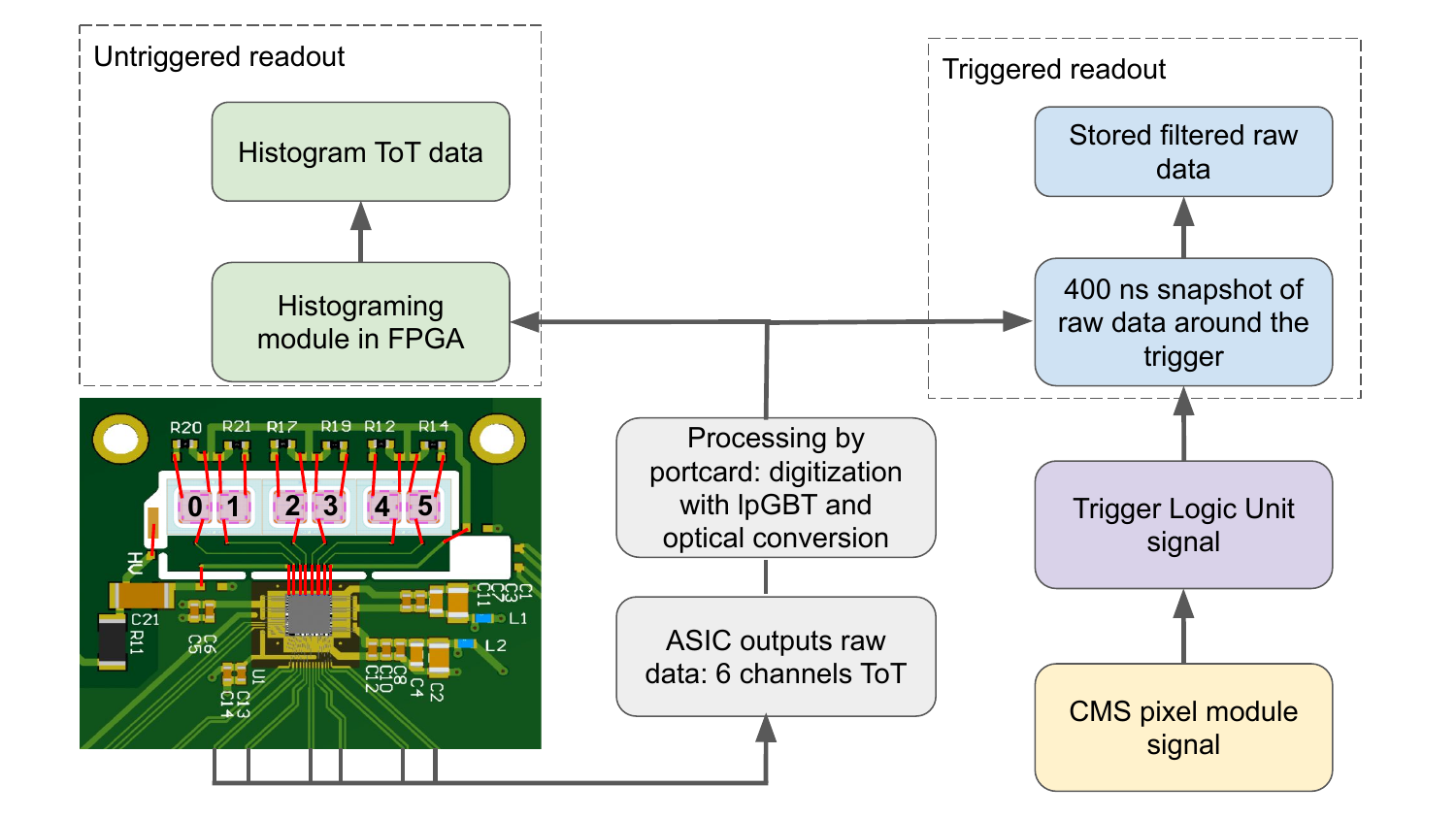}
\caption{Simplified sketch of the data readout during beam test. }
\label{fig:readout}
\end{figure}

All collected data (ToT histograms for spills and between spills, timestamped filtered ToT data, and TLU ID for triggered events) were internally transferred in the firmware to first-in-first-out (FIFO) buffers synchronized with the spill start signal. The host PC, connected to the FC7 board via a local area network, read out the data via the IPBus protocol for future offline processing in software. The untriggered histogram data readout, as well as the triggered raw data readout, are shown in Fig.~\ref{fig:readout}.

\subsection{Beam telescope alignment}
\label{subsec:telescope_alignment}

The telescope planes and the SCC were mounted in the beamline using laser sights to align their centers with the center of the beam profile. However, this mechanical alignment is not perfect and relative displacements between the detectors in $x$ and $y$ must be measured precisely. The final alignment was based on the reconstruction of the trajectories of charged particles. To reduce effects of multiple scatterings, a data-taking run dedicated to alignment was carried out with the cold box taken out of the beamline. Tracks were fit with the General Broken Lines~\cite{Kleinwort:2012np} model implemented in the Corryvreckan software package~\cite{Dannheim:2020jlk}, which is also used to perform the alignment.

\begin{figure}[htbp]
\centering
    \subfigure{%
        \includegraphics[width=0.45\linewidth]{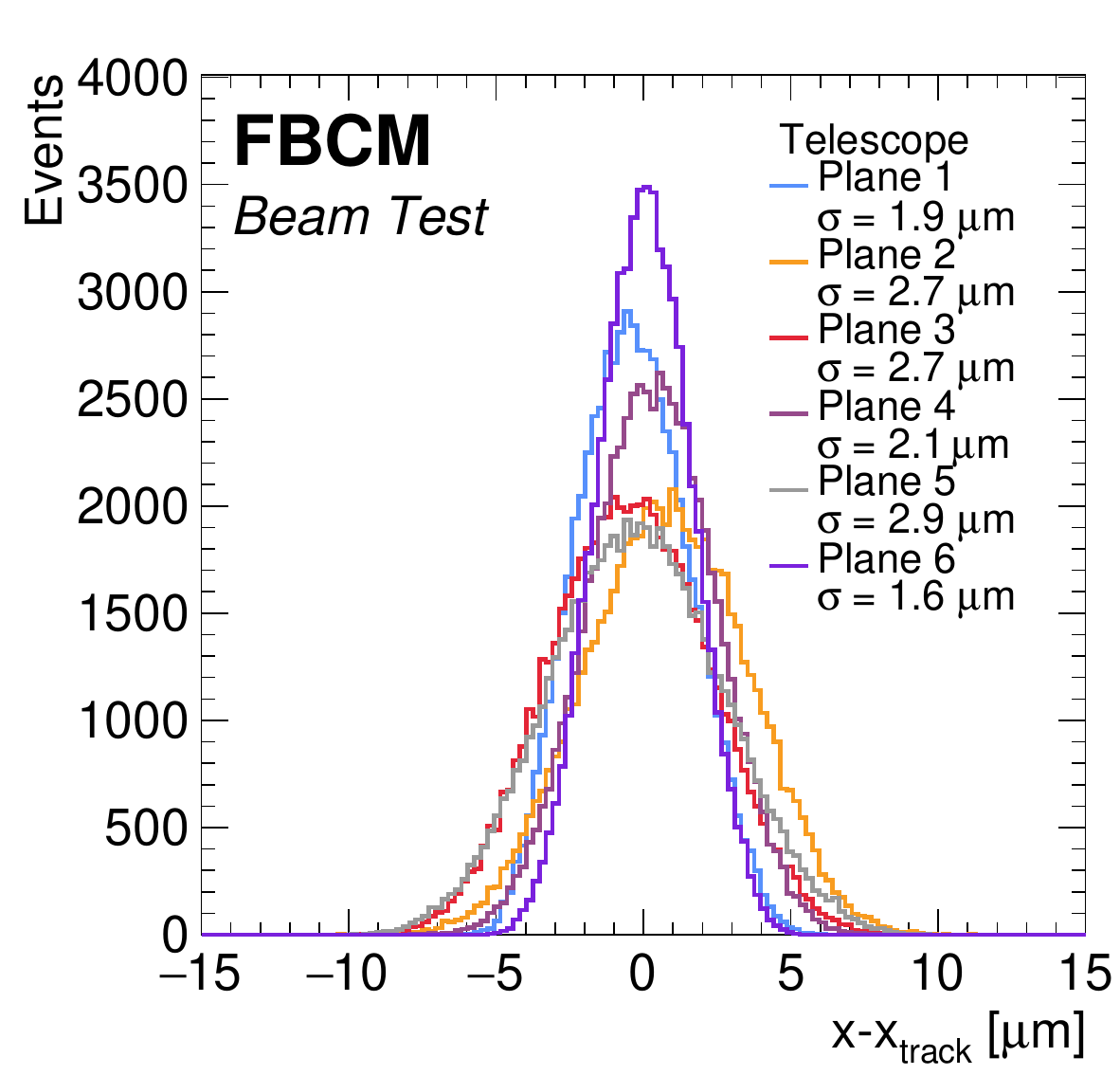}}
    \subfigure{%
        \includegraphics[width=0.45\linewidth]{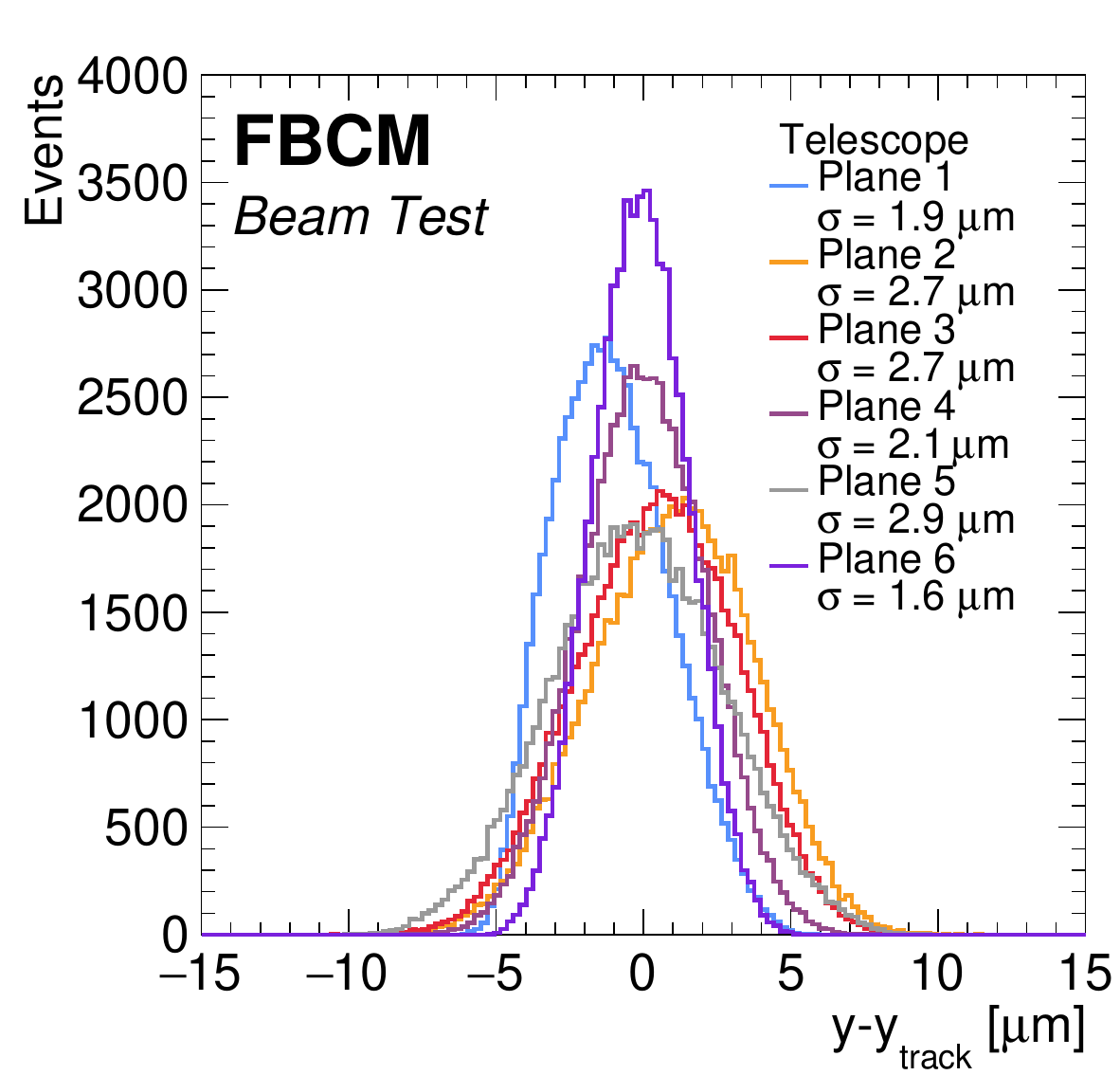}}
\caption{Residuals between the measured hit position and the fitted track in the x (left) and y (right) coordinates in each telescope plane. To quantify the resolution of the tracking, the standard deviation for each distribution is shown in the legend.
\label{fig:telescope_residuals}}
\end{figure}

First, a coarse pre-alignment step was performed, where the mean values of the pixel correlation histograms in the $x$ and $y$ directions with respect to a telescope plane designated as the reference are taken as the displacements. The full alignment procedure then consists of iteratively refitting tracks in the telescope with the aim of minimizing the $\chi^2$ of the track fits with respect to the displacements and rotations of the telescope planes. The rotation angles of the telescope planes were only allowed to vary in the last iteration of the alignment. Hits in the telescope planes are associated to track candidates based on a spatial tolerance cut, which was set at the order of the pixel pitch for the final tracking. Due to the poor performance of the track fitting algorithm with misaligned telescope planes, the spatial tolerance was relaxed in early iterations of the alignment. Starting from 500~$\mu$m, the tolerance was reduced in steps of 50~$\mu$m with 12 iterations of track refitting at each point. Rotations and shifts of the telescope planes obtained from the alignment run were applied as corrections in all data-taking runs. The shifts obtained by the final alignment range from about 10~$\mu$m up to about 500~$\mu$m, and the rotation angles were below $1^\circ$. The track residual distributions after the final alignment are presented in Fig.~\ref{fig:telescope_residuals}.

\section{Results}
\label{sec:results}

Multiple ASIC carrier boards were tested with beam to verify the different design options and quantify the change of the response for several doses of irradiation on the sensors and ASICs. The summary table of the tested ASIC carrier boards (Table~\ref{table:testboards}) has been presented in~Sec.~\ref{subsec:ASIC_carrier_board}. In this section the main results, observations, and important conclusions for operations are summarized.

\subsection{Time over threshold}
\label{subsec:ToT}

Generally, a larger amount of energy deposited in the sensor results in an analogue pulse with a higher amplitude at the ASIC input and, therefore, a longer ToT measured by the ASIC at a fixed threshold. ToT can be used during detector operation to monitor the system performance, ensuring that the signal threshold is set just above the noise and tracking radiation damage relative to the start of operation.  

\begin{figure}[htbp]
\centering
    \subfigure{%
        \includegraphics[width=0.45\linewidth]{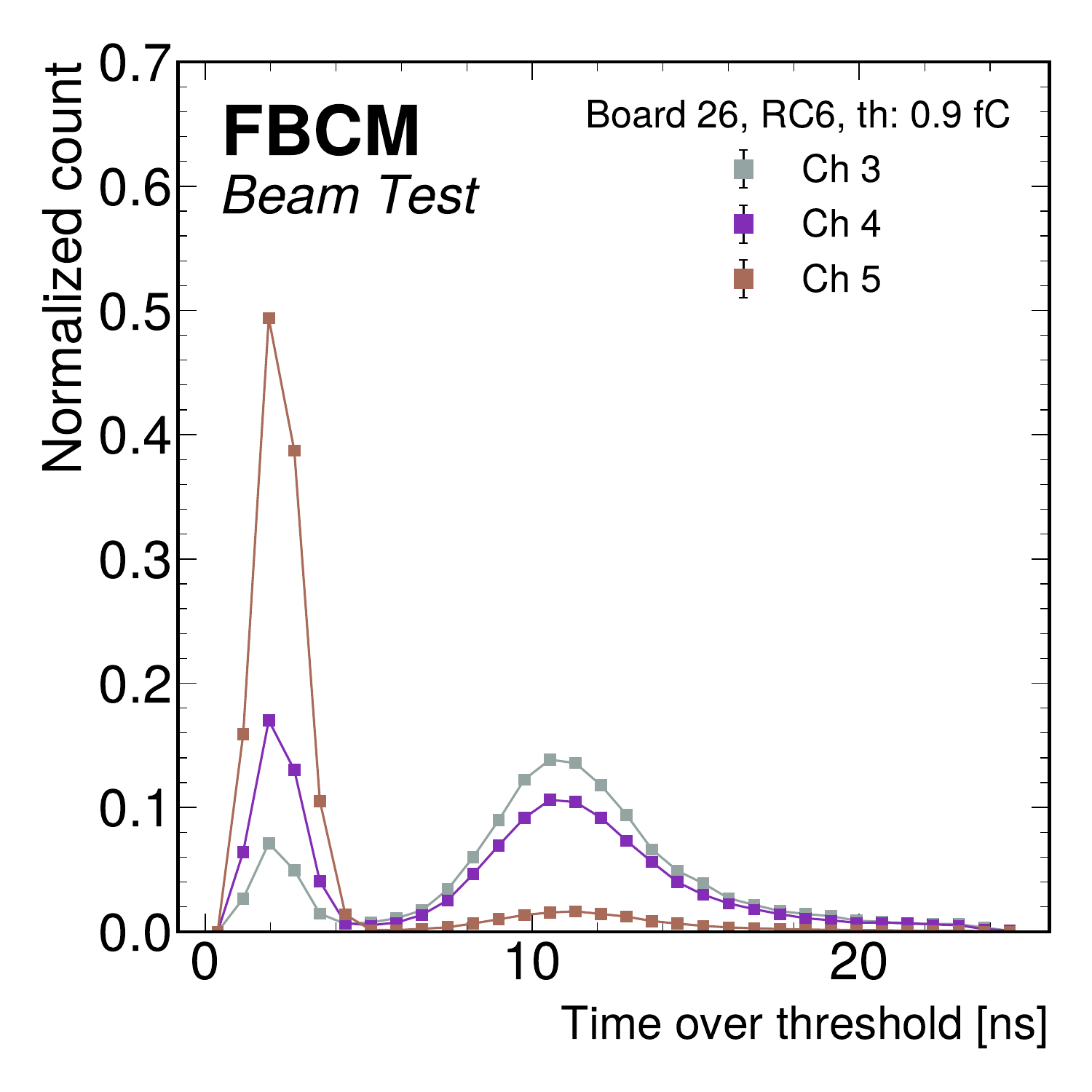}}
    \subfigure{%
        \includegraphics[width=0.45\linewidth]{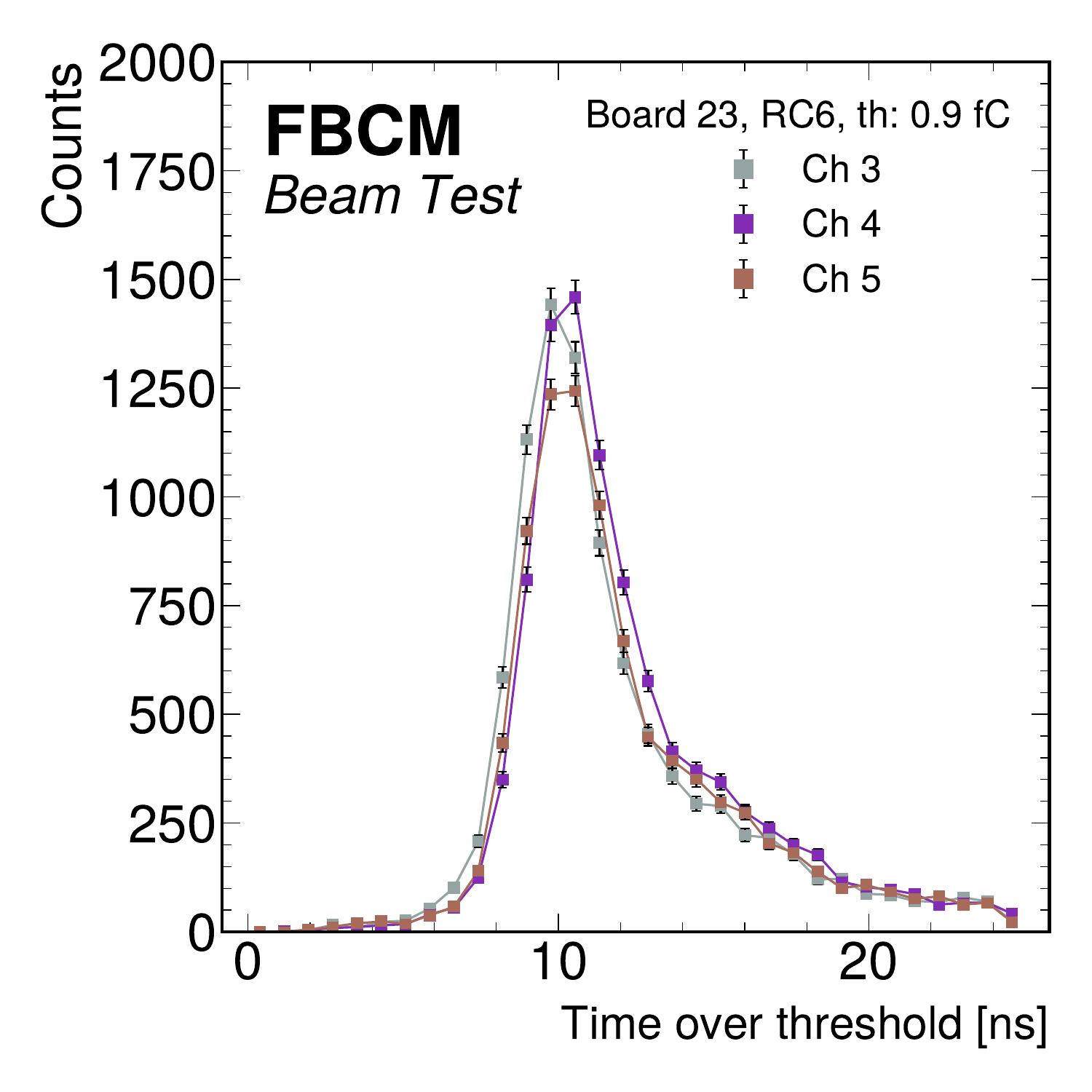}}
\caption{Examples of ToT histograms for channels with different signal path lengths due to the presence of pitch adapter (left) and for channels with direct bonding to the ASIC (right). On the left, channel 3 has the shortest and channel 5 the longest path length.}
\label{fig:ToT_intro}
\end{figure}

The ASIC signal threshold and the internal low-pass RC filter capacitance are configurable (see Sec.~\ref{subsec:data_acquisition}) and several of those settings were tested with beam to validate the response. The ASIC response to analogue pulses has been illustrated in Fig.~\ref{fig:ASIC_output}.    
Examples of the ToT histograms aggregated during the beam test for three channels from two different test boards with six-pad sensors, using a threshold of 0.9 fC and RC6 ASIC setting are shown in Fig.~\ref{fig:ToT_intro}: a board with pitch adapter (Board 26) on the left, and a board without pitch adapter (Board 23) on the right. 
From here after, all figures show the most important parameters of the measurement: the ASIC signal threshold (th) and the RC filter settings, as well as the board and channel identifiers.
Short ToT pulses below 5 ns correspond to noise. The ToT distributions of the board with a pitch adapter show a clear dependence of the noise on the channel number, which arises due to the different lengths of the signal paths between the sensor and the ASIC via the pitch adapter (connected on both ends with wire bonds). Channel 3, with the shortest signal path, shows significantly lower noise when compared to channel 5, with the longest (see Fig.~\ref{fig:readout} for channel mapping). The ToT distributions of the board without a pitch adapter confirm the significant noise reduction for the sensors directly bonded to the ASIC. No noise peak is observed with the same ASIC threshold of 0.9 fC. The lower noise at identical ASIC thresholds for all channels motivated the change in design, with direct sensor to ASIC bonding being chosen for future FBCM prototypes. Note that most of the measurements during the beam test were done for this type of test boards.

\begin{figure}[htbp]
\centering
    \subfigure{%
        \includegraphics[width=0.45\linewidth]{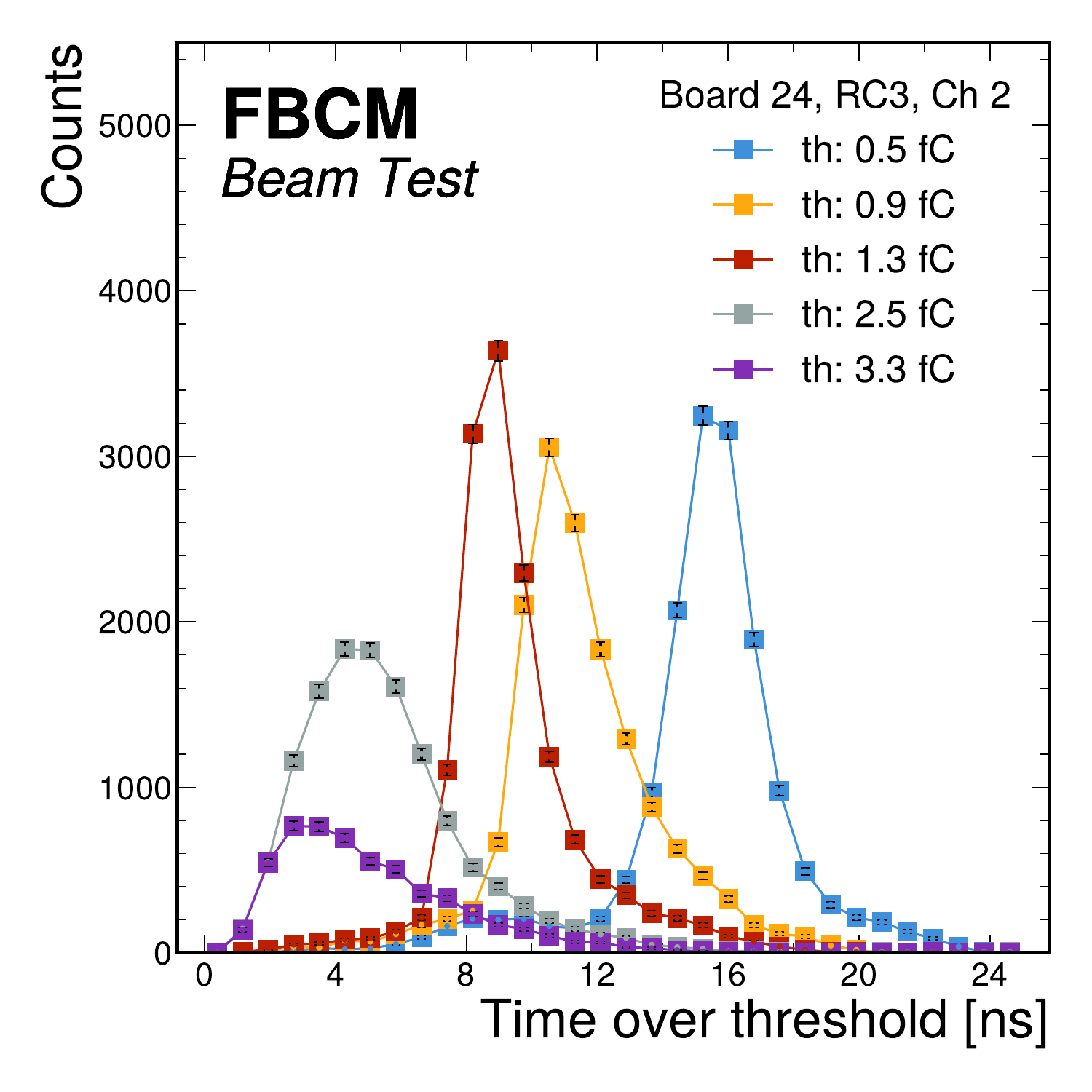}}
    \subfigure{%
        \includegraphics[width=0.45\linewidth]{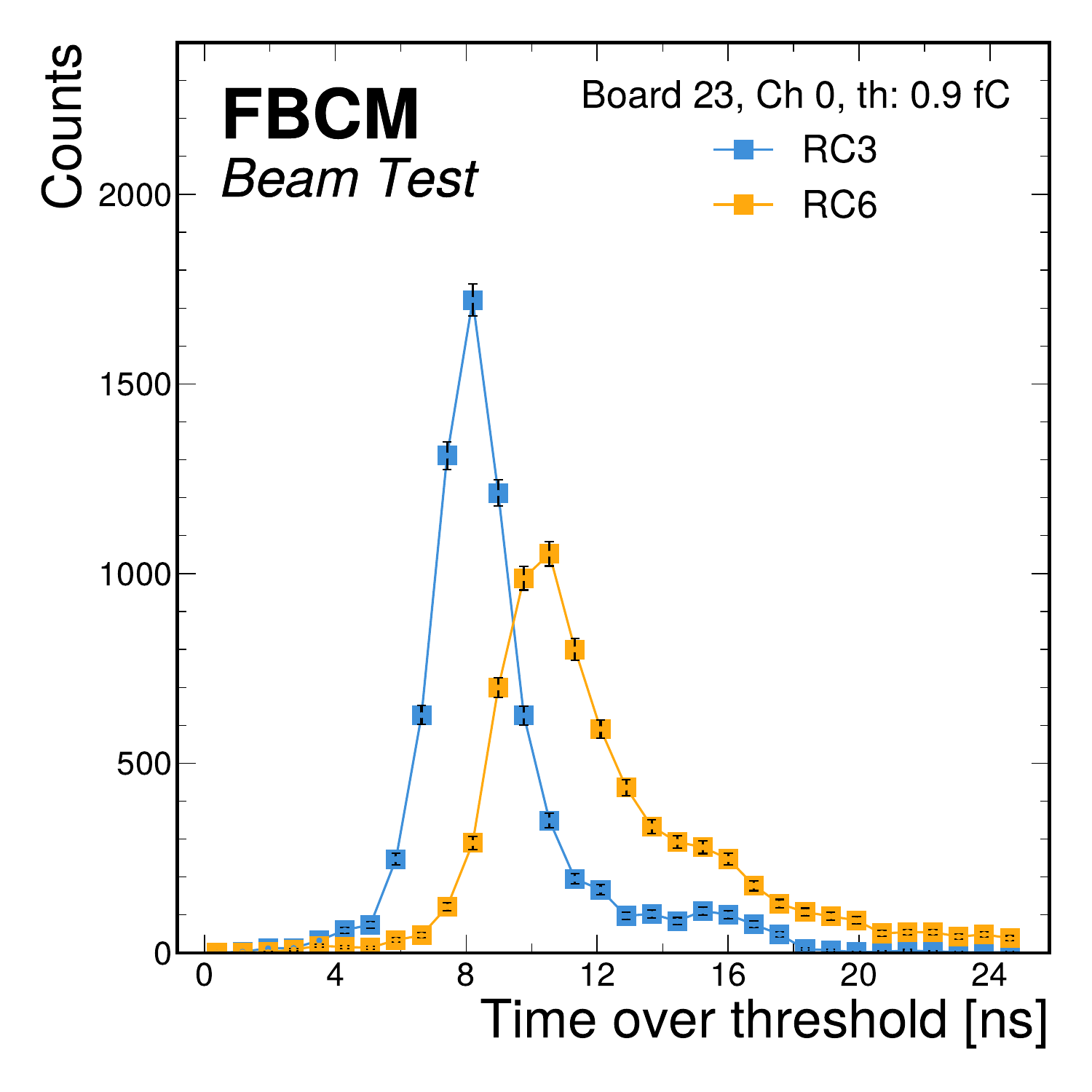}} 
\caption{ Examples of the ToT histograms for the same channel obtained at different ASIC thresholds (left) and different ASIC RC filter settings (right). }
\label{fig:ToT_vs_TH_and_RC}
\end{figure}

As the measured ToT depends on the threshold, the ToT for the same analogue pulse shape is shorter for higher thresholds, and the pulse can be even completely missed if the threshold is set too high.
Examples of the ToT histograms for the same input channel, measured at five different thresholds, are shown in Fig.~\ref{fig:ToT_vs_TH_and_RC} (left). As expected, the ToT distribution shifts to lower values with increasing threshold and, for the largest thresholds of 2.5 fC  and 3.3 fC, the loss of efficiency leads to a reduced number of counts. 

Fig.~\ref{fig:ToT_vs_TH_and_RC} (right) shows the effect of RC3 versus RC6 ASIC settings on the ToT histogram for the same input channel and at a fixed 0.9 fC threshold value. As the shaping time of the signal is different for these two settings, the expected shift of the ToT to higher value for RC6 is observed. This is beneficial for irradiated sensors, as it compensates for the lower charge collection efficiency, but has to be operated in combination with higher thresholds to cut the noise. In this particular example utilizing a not irradiated sensor, the RC6 integral is about 15\% lower than that of RC3, indicating a drop in efficiency. It was checked that for the irradiated sensors the integrals of the histograms aggregated for the same number of spills with RC3 and RC6 settings agree within 5\%.      

\subsection{Time walk }

The main deliverable of the FBCM for HL-LHC will be ToA information, which will be used for bunch-by-bunch luminosity and beam-induced background measurements.
The ToA with respect to the bunch crossing will be aggregated in sub-bunch crossing (foreseen to be 2.5 ns) granularity histograms during 1 second time periods. During the beam test ToA is not meaningful in the absence of a time reference. During LHC operation, the beginning of the LHC orbit and the LHC bunch clock will serve as such references.  

\begin{figure}[htbp]
\centering
    \subfigure{%
        \includegraphics[width=0.45\linewidth]{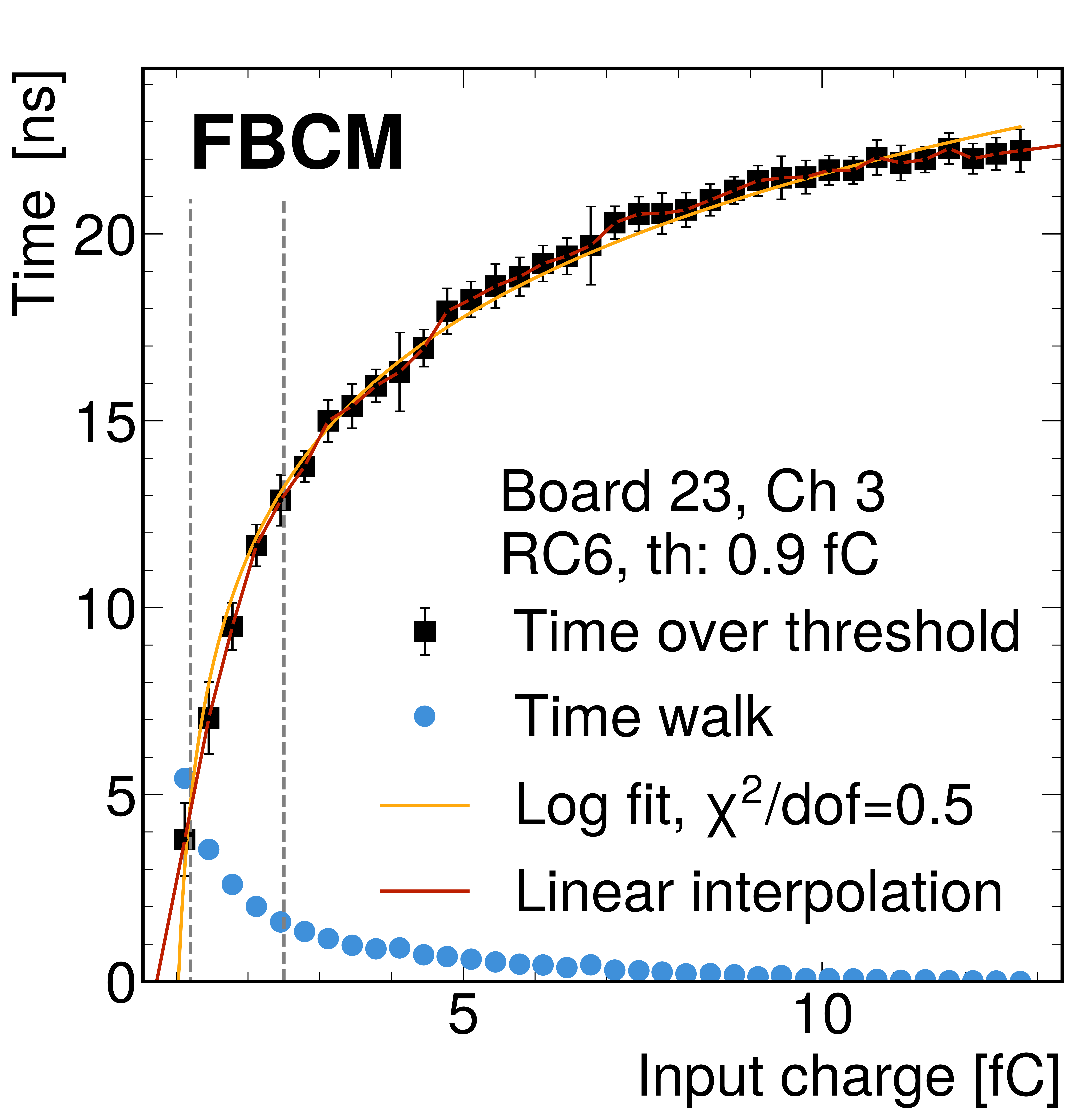}}
    \subfigure{%
        \includegraphics[width=0.45\linewidth]{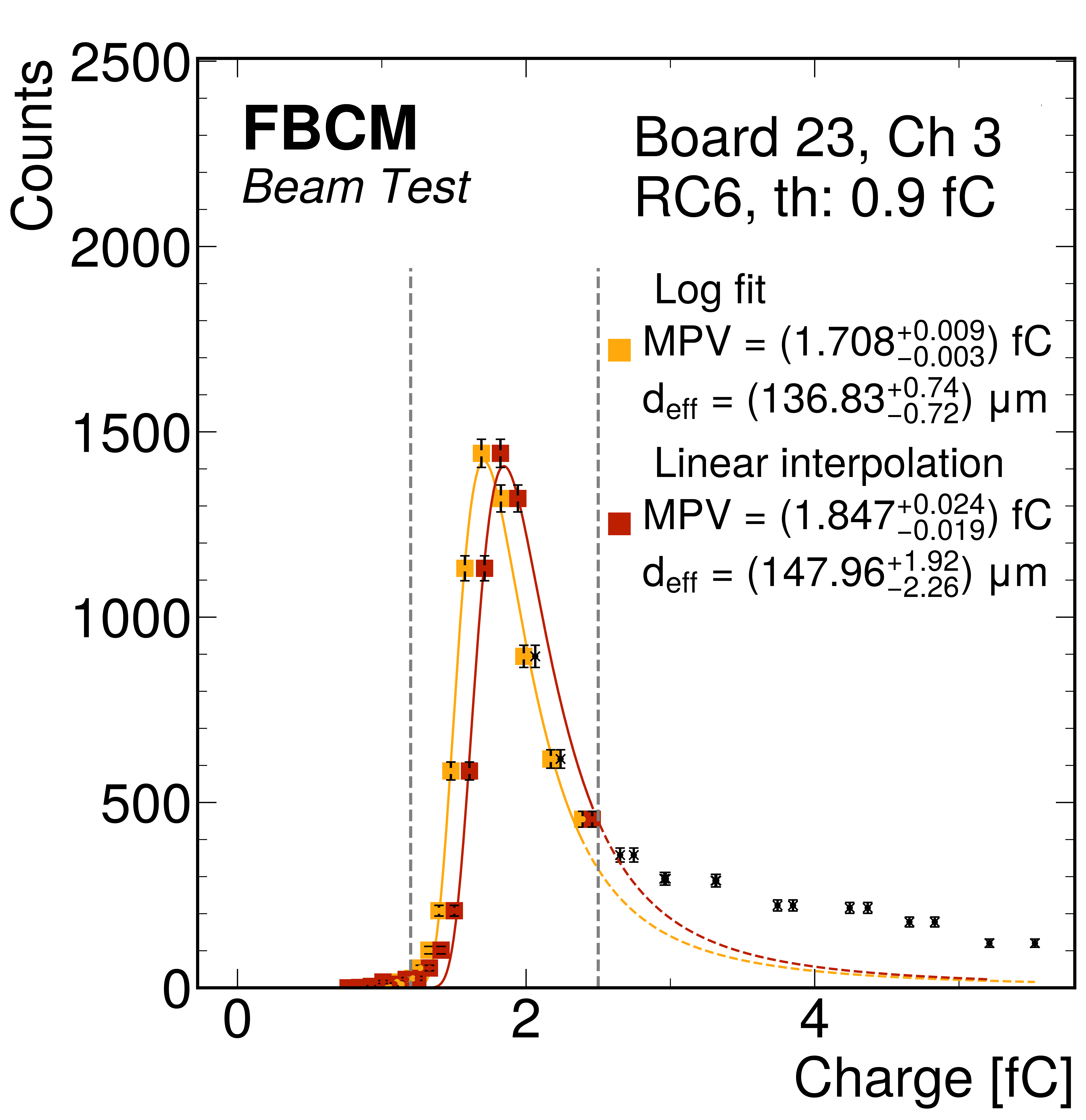}} 
\caption{ Left: time over threshold and time walk as a function of the input charge measured at the laboratory using calibration pulses at a threshold of 0.9 fC with RC6 setting for board 23 with 150~$\mu$m thick Si-sensor. The uncertainty is defined as the standard deviation of 15 measurements. Right: the charge deposition calculated using the two different approximations of the ToT calibration curve shown on the left (see text). The MPVs and their uncertainties from the fit are listed in the legend. Vertical dashed lines indicate the position of the bulk of the signal. }
\label{fig:ToT_TW_lab_measurements}
\end{figure}

The ToA is affected by the input charge due to ``time walk": the lower the input charge, the later the threshold is reached. 
Time walk was measured in the laboratory and defined as a delay between the trigger initiating the injection of a fixed-charge calibration pulse and the detected arrival time of the pulse, shifted such that the time walk is zero at maximum charge injection. Calibration pulses from 1 to 12.5 fC charge were tested. For an injected charge of 1.9 fC, as expected in a sensor with a width of 150 $\mu$m, the measured time walk is about 2 ns, as shown with blue dots in Fig.~\ref{fig:ToT_TW_lab_measurements} (left). This matches the FBCM23 ASIC specifications~\cite{FBCM-ASIC}.   
At an input charge of 1 fC, the delay reaches about 5 ns. On this scale the time walk does not affect the beam induced background measurements or per bunch crossing luminosity measurements, as bunch spacing is 25 ns.

\subsection{Charge deposition calibration }

To convert ToT measured in units of ns into charge deposition in fC, a calibration curve is required. An example of such a calibration curve is shown with black dots in Fig.~\ref{fig:ToT_TW_lab_measurements} (left) and the corresponding charge deposition distributions based on fits to the calibration curve are shown in Fig.~\ref{fig:ToT_TW_lab_measurements} (right). Both a logarithmic fit to all data points and a linear interpolation between neighboring measurements were used to approximate the calibration curve. They deviate significantly at the turn-on between the gray dashed lines at 1.2 and 2.5~fC. This difference influences the two reconstructed charge deposition distributions, as the bulk of the six-pad sensor signal lies exactly in this range. 
To evaluate the accuracy of the calibration curve fit, the effective width $\textrm{d}_\textrm{eff}$ of the sensor is calculated and used as a reference.  
A convolution of a Landau and a Gaussian function is used to fit the resulting charge distribution and to extract its most probable value ($Q_\text{MPV}$). Based on the electron charge $k_{e^{-}}$ and the average number of electron-hole pairs per $\mu$m created by a MIP crossing the silicon sensor pad $n_{e^{-}h^{+}}$, the effective sensor width can be calculated as:

\begin{equation}
    \text{d}_{\text{eff}} [\mu \text{m}]= \frac{Q_\text{MPV}[\text{fC}]}{n_{e^{-}h^{+}} [\mu \text{m}^{-1}] \cdot k_{e^{-}} [\text{fC}]},
\end{equation}
where $k_{e^{-}}=1.6\cdot 10^{-4} \ \text{fC}$, and $n_{e^{-}h^{+}}$ = 78 $\mu$m$^{-1}$~\cite{Otarid:2884106}. The obtained MPV for the deposited charge and $\textrm{d}_\textrm{eff}$ are shown in the legend for both approximations of the calibration curve.
 
The difference between the expected 150~$\mu m$ thickness of a six-pad sensor and the calculated effective thickness lies within 9\% (using the logarithmic fit) and 2\% (using the linear interpolation) for this example. However, for other channels, boards and RC filter settings, the difference could be as large as 30\%, as indicated in Table~\ref{table:testboards_d_eff}. Higher deviation is observed for the board 21 with irradiated sensors. The discrepancy is caused by several factors: the uncertainty in the approximation of the calibration curve; the difference between the signal shape of the fast injected charges directly to the ASIC input during calibration and the actual shape induced by charges moving through the Si-sensor. The latter having a larger time constant, especially for irradiated sensors explains why observed difference from the expected thickness increases for irradiated sensors. 

\begin{table}[h!]

\centering
\begin{tabular}{| c | c | c | c | c | c | c | c | c |} 
 \hline
 Test board  &  ASIC   & \multicolumn{6}{|c|}{deviation from the expected thickness [\%]} & |average dev.[\%]|  \\ 
   tracking number & RC setting & Ch0 & Ch1 & Ch2 & Ch3 & Ch4 & Ch5 & $\pm$ std. dev. \\ 
 \hline\hline
  Board 23  &  RC3 &   -9 & 2  & 8  & -7 & 2 & -2 & 1 $\pm$ 6 \\ 
  Board 23  &  RC6 &   7  & 12 & 13 & 9 & 9 & 12 & 10 $\pm$ 3 \\ 
  Board 21  &  RC3 &   17 & 21  & 12  & 11 & 12 & 17 & 15 $\pm$ 4 \\ 
  Board 21  &  RC6 &   28  & 26 & 27 & 24 & 24 & 21 & 25 $\pm$ 2  \\ 
  \hline 
\end{tabular}
\caption{Summary of the observed deviations of the effective sensor thickness $\textrm{d}_\textrm{eff}$ from the expected 150~$\mu m$ of a six-pad Si-sensor at a threshold of 0.9 fC. The MPV of the charge deposit used in the $\textrm{d}_\textrm{eff}$ calculation was obtained using the linear interpolation of the ToT - charge calibration curve. The average deviations from the expected thickness and the corresponding standard deviations for the six channels of a sensor are given in the last column. }
\label{table:testboards_d_eff}
\end{table}

\begin{figure}[htbp]
\centering
    \subfigure{%
        \includegraphics[width=0.45\linewidth]{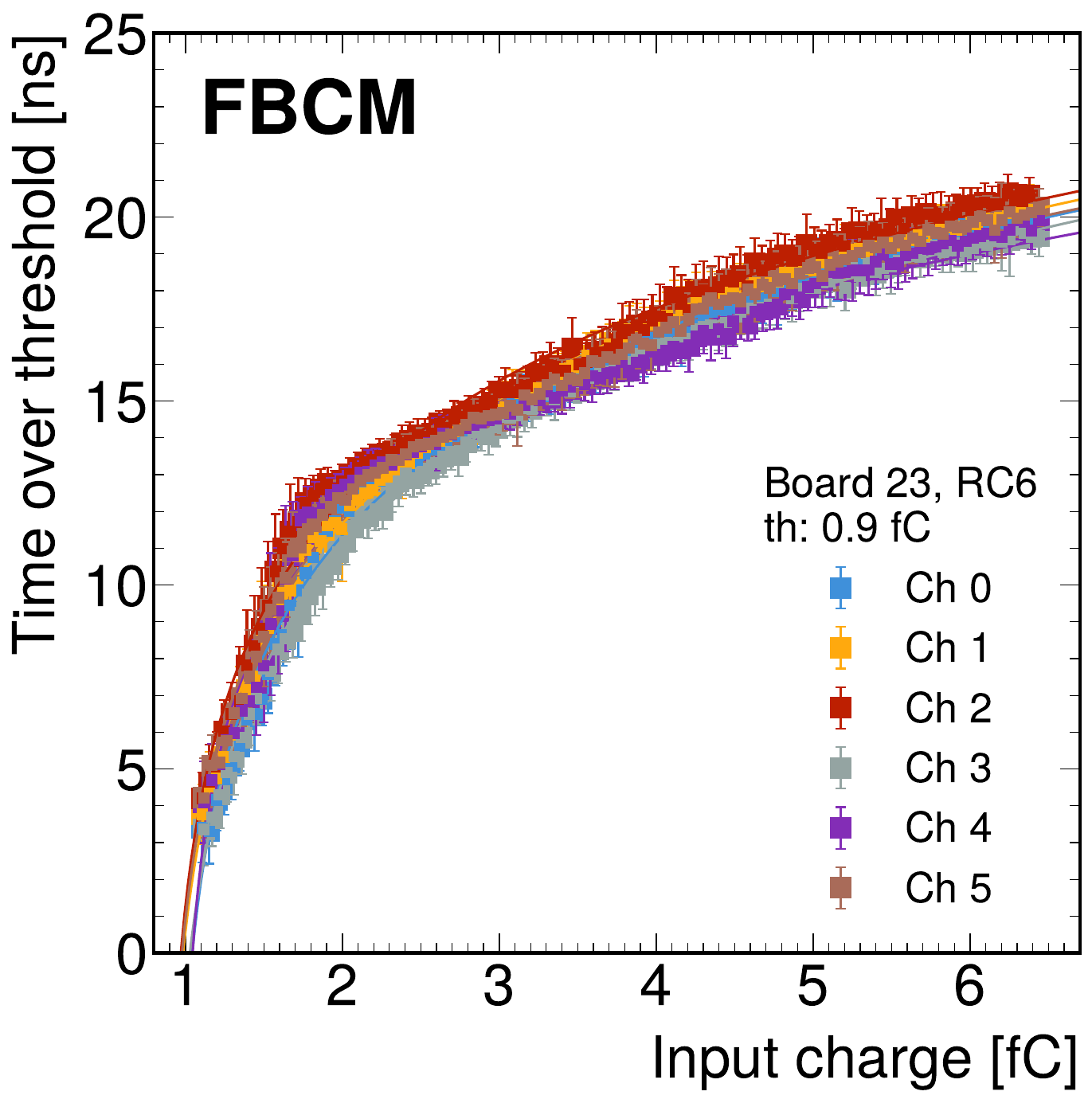}}
    \subfigure{%
        \includegraphics[width=0.45\linewidth]{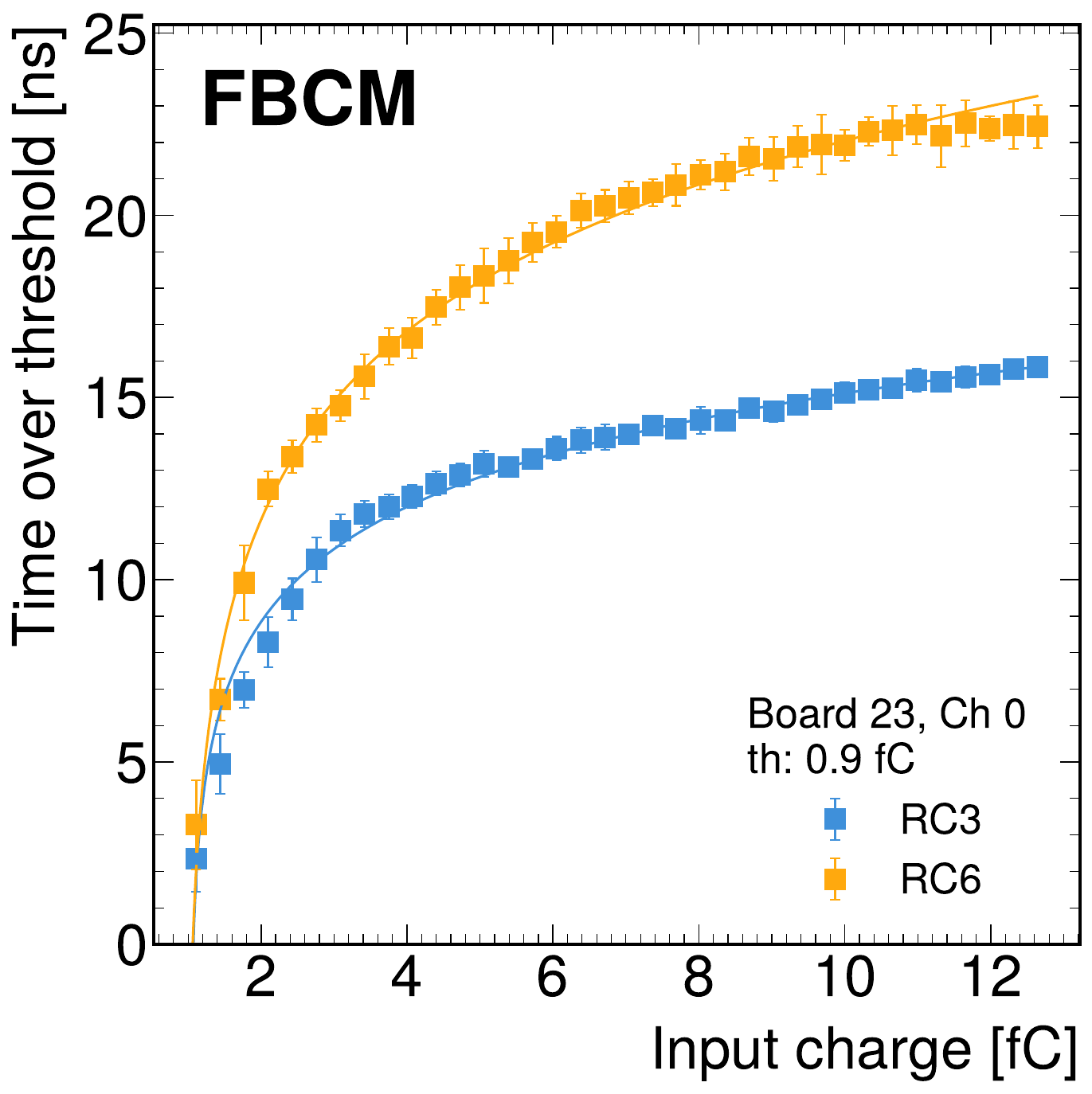}}
\caption{ Left: ToT calibration curves measured in the laboratory with a threshold of 0.9 fC and an RC6 setting for six channels of the same ASIC, obtained with a logarithmic fit to the data. Right: comparison of the calibrations curves for the same ASIC channel with RC3 and RC6 settings. }
\label{fig:ToT_calib}
\end{figure}

The calibration curve differs for every ASIC channel and depends on the ASIC RC setting as illustrated in Fig.~\ref{fig:ToT_calib} (left) and (right), respectively. It also depends on the ASIC threshold, therefore each channel must be calibrated separately for a particular threshold.
Dense sampling of the calibration curve, especially below 2.5 fC, allows for a more accurate approximation of the curve and reduces the bias on the extracted MPV.  However, the main uncertainty of the calibration curve is still caused by the difference between the shape of the calibration pulse and of the Si-sensor signal.    
It is to be noted that this check was done to quantify known difference between calibration curve approximation and real signal. During the detector operation only ToT measured in ns will be used for monitoring purposes and no conversion to the charge deposition is required.

\subsection{Effects of irradiation on time over threshold}
\label{subsec:All_boards_ToT_comparison}

\begin{figure}[htbp]
\centering
\includegraphics[width=1.0\textwidth]{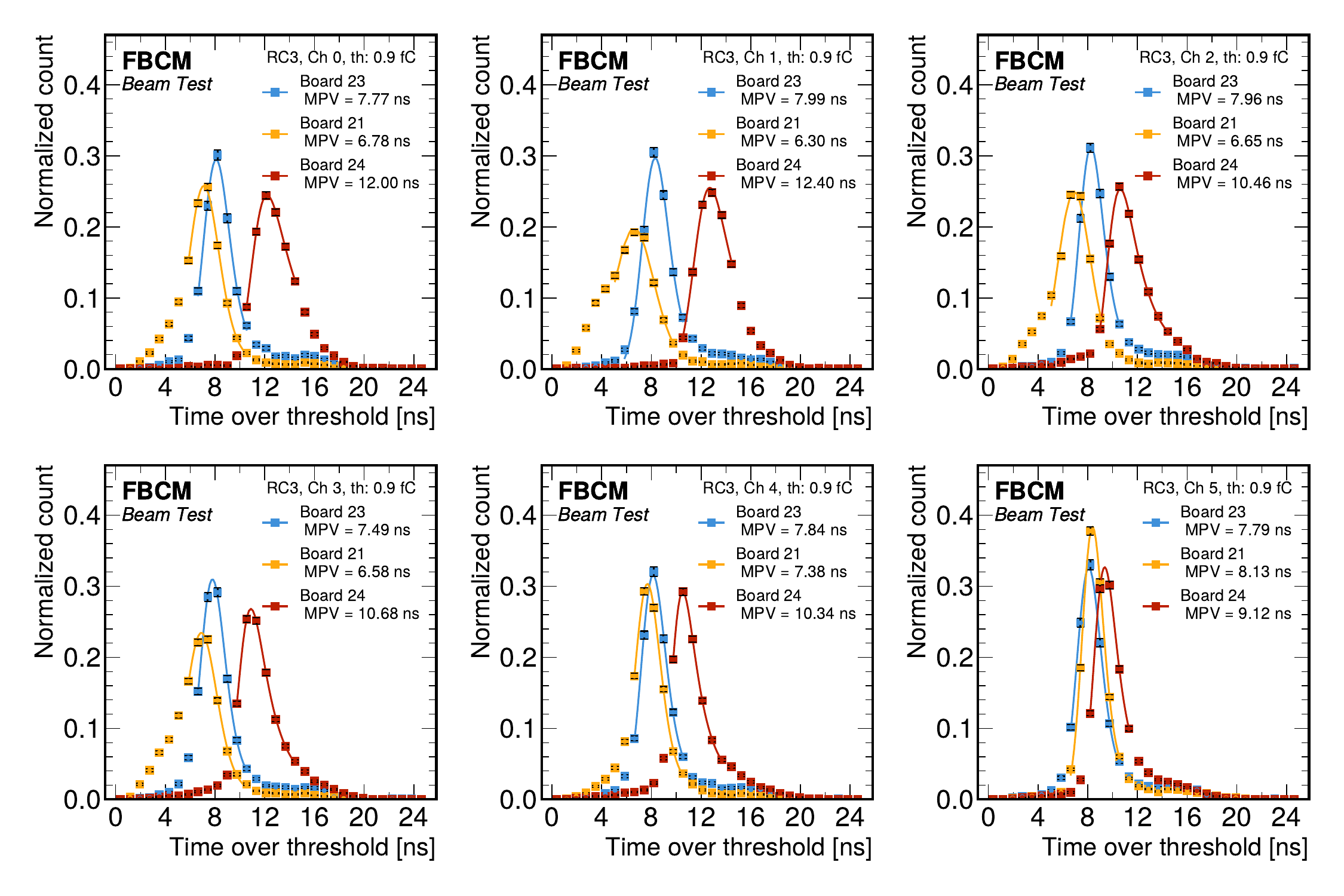}
\caption{
Time over threshold distributions for test boards with fresh and irradiated sensors. Board 23 (blue) contains unirradiated six-pad sensors, board 21 (yellow) contains irradiated six-pad sensors  (lower dose for channels 4-5), and board 24 (red) contains fresh (channels 1-2) and irradiated two-pad sensors (highest dose for channels 4-5). The most probable value of fits to convolution of a Landau and a Gaussian function is given in the legend. Data is taken at different bias voltages: board 23 at -100~V, board 21 at -300~V, board 24 at -800~V.
}
\label{fig:ToT_bd21_db23_db24_rc3}
\end{figure}

Under irradiation, defects are created in the bulk of the silicon sensors, which cause a reduction in charge collection efficiency and affect the shape of the analog pulse arriving to the ASIC input. Consequently, it is expected to measure lower ToT for irradiated sensors compared to fresh sensors of the same type. Between different types of sensor, it is expected to measure higher ToT for unirradiated two-pad sensors versus the six-pad sensors (290 vs. 150 $\mu$m), as the energy deposition in the bulk of the thicker sensor is larger. However, a thicker sensor also increases the likelihood of trapped charges with accumulated irradiation, which gradually offsets the larger energy deposition. Quantifying the effects of irradiation on the two types of sensors experimentally was one of the main objectives of the beam test. An example of such a comparison is shown in Fig.~\ref{fig:ToT_bd21_db23_db24_rc3} for three boards with direct bonding between the sensor pads and the ASIC. Fits to the ToT distributions using a convolution of a Landau and a Gaussian function are applied to extract the MPVs, which are listed in the legends. It is important to note that the pads on the sensors are irradiated heterogeneously and data is taken at different bias voltages: board 23 at -100~V, board 21 at -300~V, board 24 at -800~V. The observations are aligned with expectations and the conclusions of this comparison are listed below: 

\begin{itemize}
  \item Unirradiated vs. irradiated six-pad sensors: For board 21 (yellow), with irradiated six-pad sensors, the ToT peaks at lower values compared to board 23 (blue), with unirradiated six-pad sensors. However, channels 4 and 5 in board 21 were irradiated to a smaller dose, which explains why this trend is less pronounced.
  \item Unirradiated vs. irradiated two-pad sensors: For board 24 (red), the ToT distribution peaks at lower values in channels 2 to 5, with irradiated two-pad sensors, compared to the unirradiated channels 0 and 1 on the same board. 
  \item Unirradiated six-pad vs. two-pad sensors: For board 23 (blue), with unirradiated six-pad sensors, ToT is lower than for board 24 (red) channels 0 and 1, with unirradiated two-pad sensors. 
  \item Irradiated six-pad vs. two-pad sensors: For board 21 (yellow), with irradiated six-pad sensors, ToT is lower than for board 24 (red) channels 2 to 5, with irradiated two-pad sensors. But the difference is less than for unirradiated sensors. 
\end{itemize}

For some channels corresponding to the irradiated six-pad sensors (board 21), there is a significant difference between the expected shape described by a convolution of a Landau and a Gaussian function and the measured shape of the ToT distribution. This is most evident for channels 0 to 3 of board 21, where a ``shoulder'' is noticeable at lower ToT values. More pronounced ``tails'' than expected are also visible for all channels in all boards at higher ToT values. This deviation from the ideal shape varies depending on the threshold and is studied in more detail using the track pointing information described in Sec.~\ref{subsec:hit_position_study} and with the transient current technique described in Sec.~\ref{subsec:TCT}.

\subsection{ASIC signal threshold scans}
\label{subsec:TH_scans}

\begin{figure}[htbp]
\centering
\includegraphics[width=1.0\textwidth]{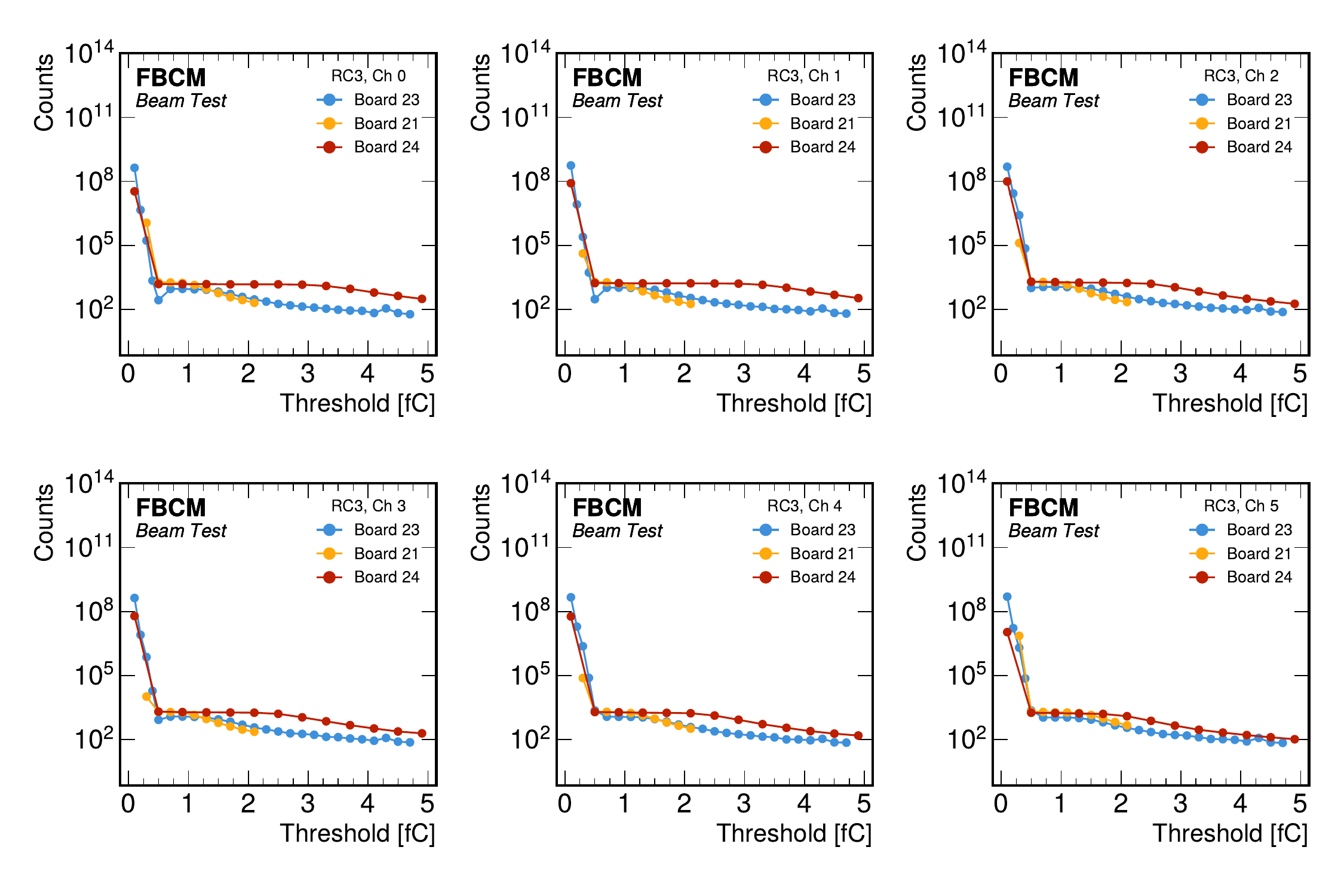}
\caption{Threshold scans for six channels of three test boards with fresh and irradiated sensors. The break at 0.5~fC identifies the minimum threshold to eliminate the noise.}
\label{fig:Th_scan_bd21_db23_db24_rc3}
\end{figure}

To define the minimum threshold above the noise, threshold scans were performed for boards with direct bonding of sensors to the ASIC. The thresholds were scanned from 0.1~fC up to 5~fC for test boards 23 and 24, and up to 2~fC for test board 21 due to limited  beam time. The integral of the ToT histogram, aggregated for a fixed time interval (17 spills) for each scanned threshold value, is shown in Fig.~\ref{fig:Th_scan_bd21_db23_db24_rc3}. Examples of individual ToT histograms were  shown in Fig.~\ref{fig:ToT_vs_TH_and_RC}. A minimum threshold of 0.5~fC was identified for boards with direct bonding to the ASIC to cut the noise independent of the board type and the delivered irradiation. 

\begin{figure}[htbp]
\centering
\includegraphics[width=1.0\textwidth]{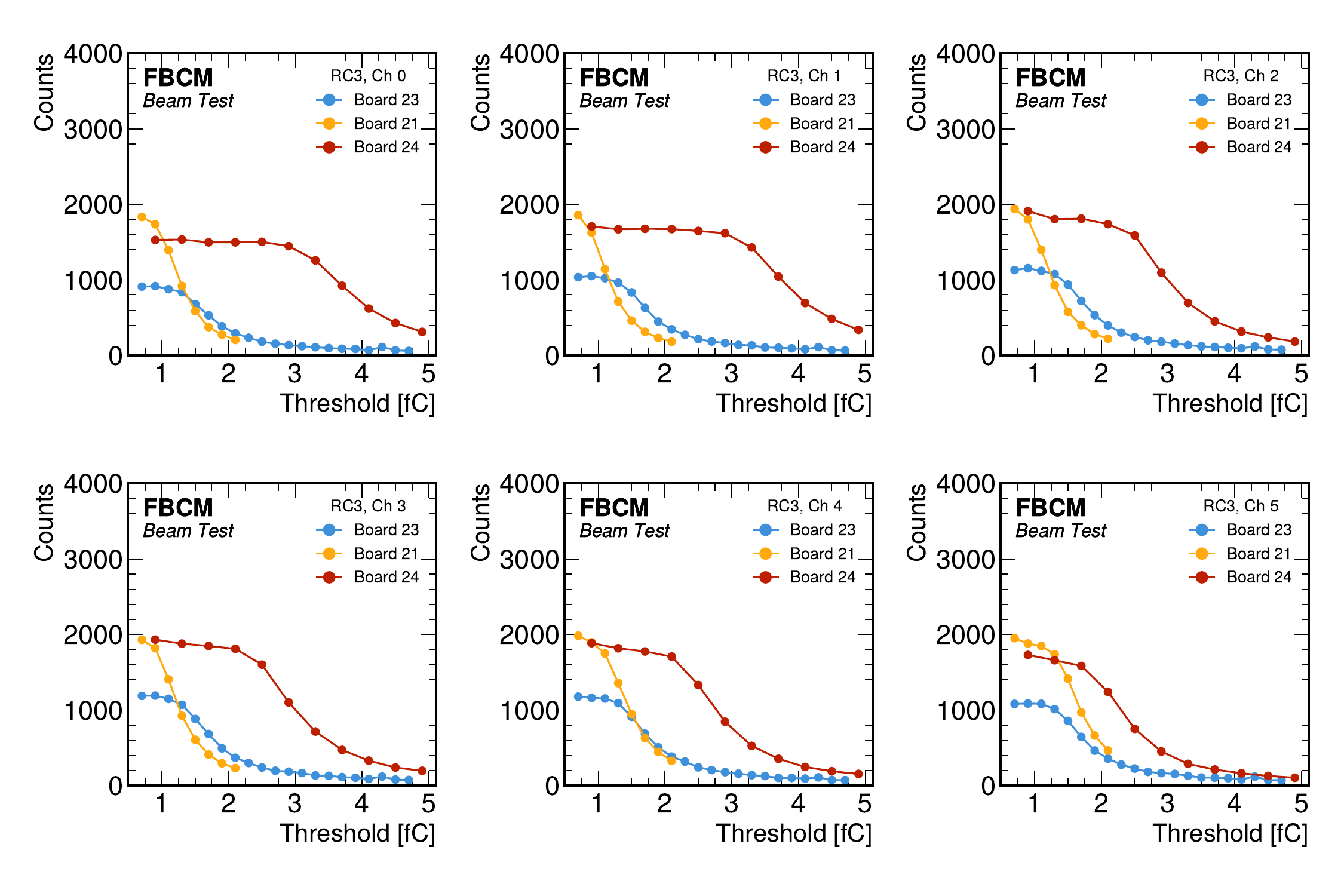}
\caption{Threshold scan above the noise threshold of 0.5 fC for test boards with fresh and irradiated sensors. Generally, irradiated channels exhibit a shaper drop in efficiency at lower threshold values.}
\label{fig:Th_scan_bd21_db23_db24_rc3_zoom}
\end{figure}

Fig.~\ref{fig:Th_scan_bd21_db23_db24_rc3_zoom} zooms on the results of the scans above the minimum threshold. 
For board 23 (blue), equipped with unirradiated six-pad sensors a drop in efficiency starts around 1.5~fC
. For board 21 (yellow), with irradiated six-pad sensors, a drop appears earlier at about 1~fC and is also steeper. For board 24 (red), with two-pad sensors, the drop in  efficiency is visible above around 3~fC for unirradiated channels 0 and 1, and above around 2~fC for all other irradiated channels.

\subsection{Sensor efficiency}
\label{subsec:sensor_efficiency}

\begin{figure}[htbp]
\centering
    \subfigure{%
        \includegraphics[width=0.45\linewidth]{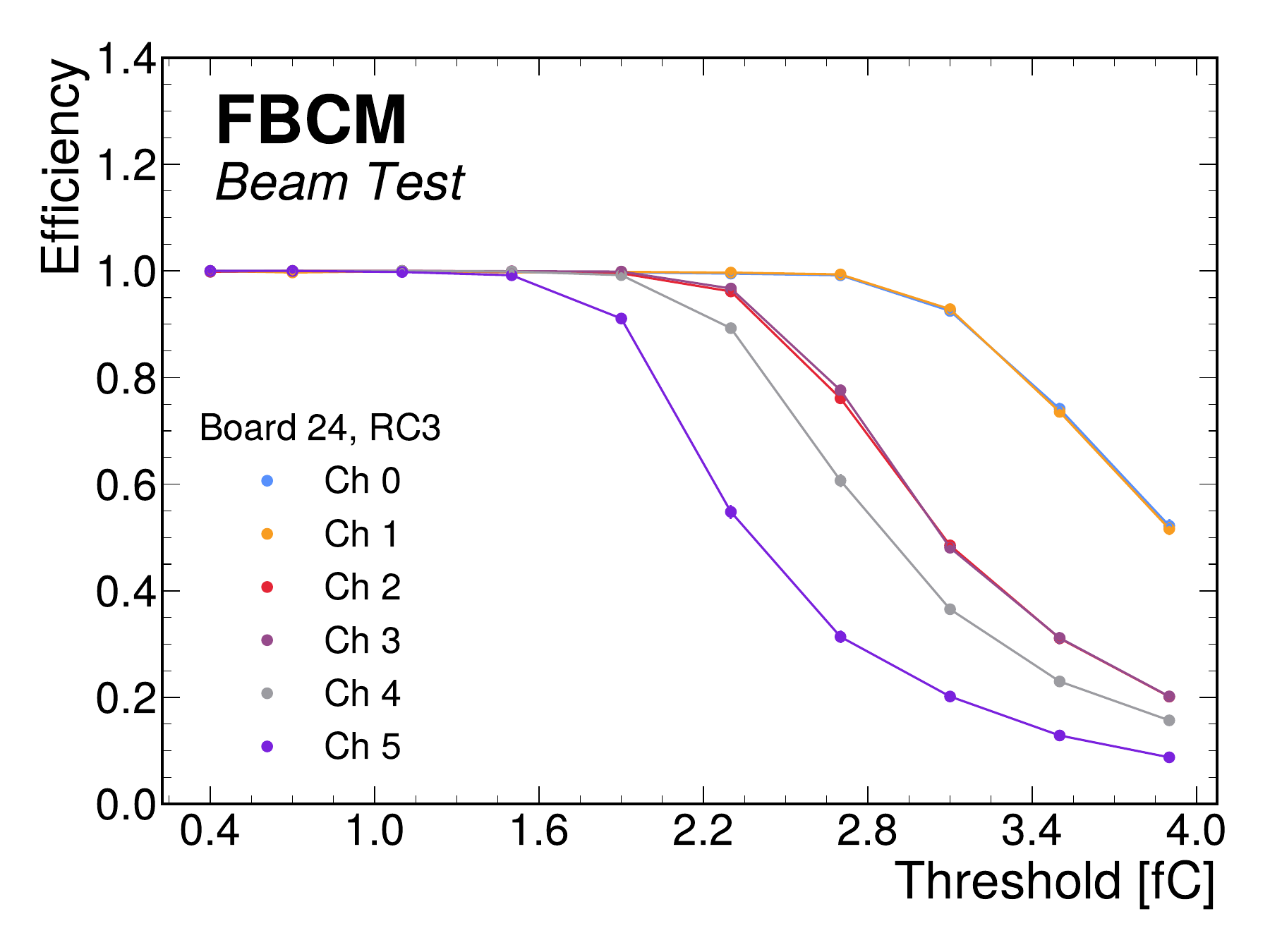}}
    \subfigure{%
        \includegraphics[width=0.45\linewidth]{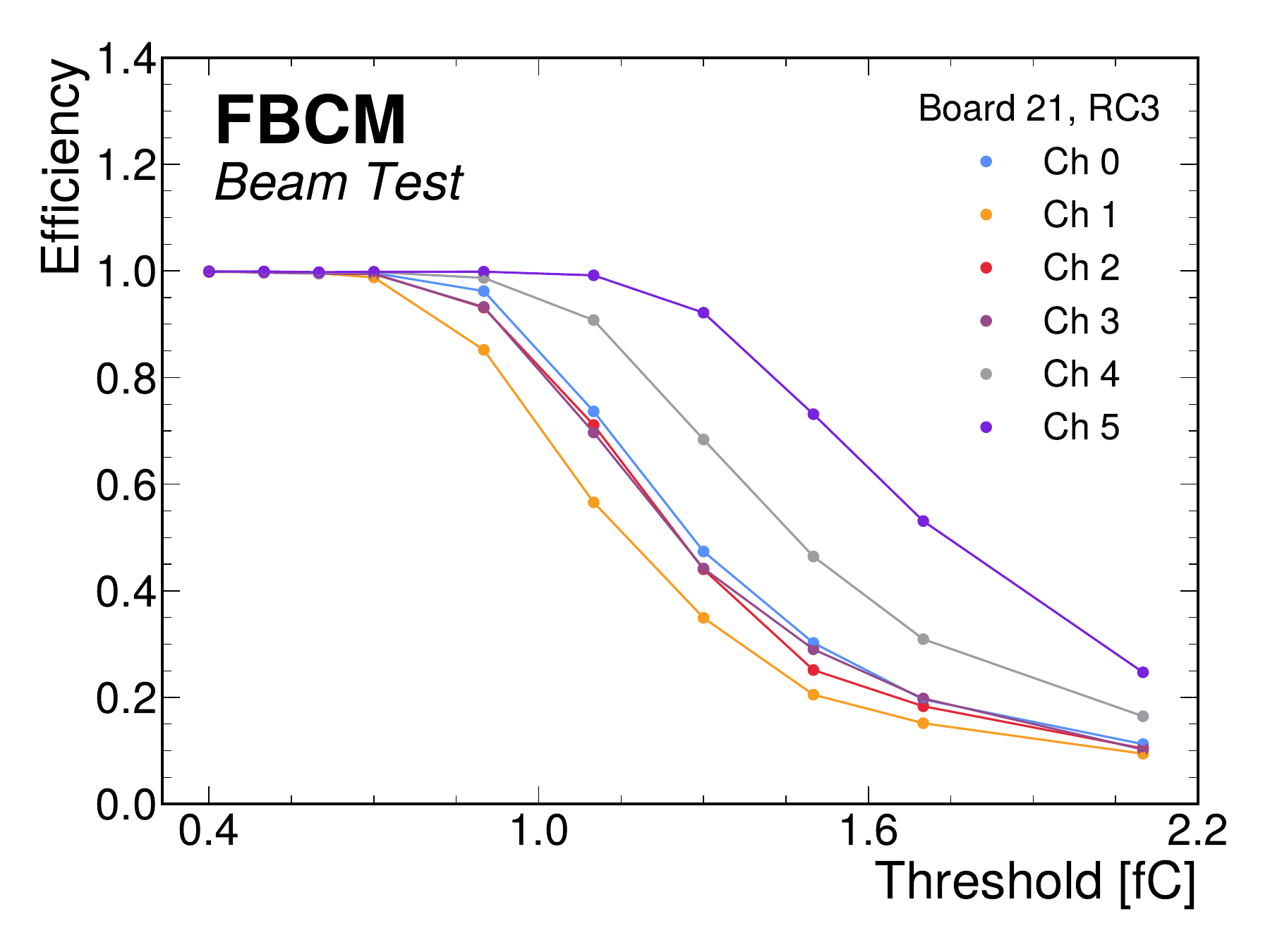}}
\caption{ Sensor efficiency as a function of the ASIC signal threshold for a two-pad sensor (left) and a six-pad sensor (right). }
\label{fig:Eff}
\end{figure}

Tracks reconstructed with the beam telescope were used to study the sensor efficiency which was defined as the ratio of tracks associated with a hit in an FBCM channel to the total number of reconstructed tracks pointing to the area of the corresponding FBCM sensor pad. Fig.~\ref{fig:Eff} shows the measured efficiency per channel for a two-pad sensor (left) and a six-pad sensor (right). As expected for higher energy deposition in thicker sensors, the operational range of full efficiency is broader,  with a plateau up to 1.6~fC compared to a plateau up to 0.8~fC for thinner sensors (especially for the most irradiated channels). These results are consistent with the conclusions from the ASIC signal threshold scans discussed in Sec.~\ref{subsec:TH_scans}.

\begin{figure}[htbp]
\centering
    \subfigure{%
        \includegraphics[width=0.48\linewidth]
        {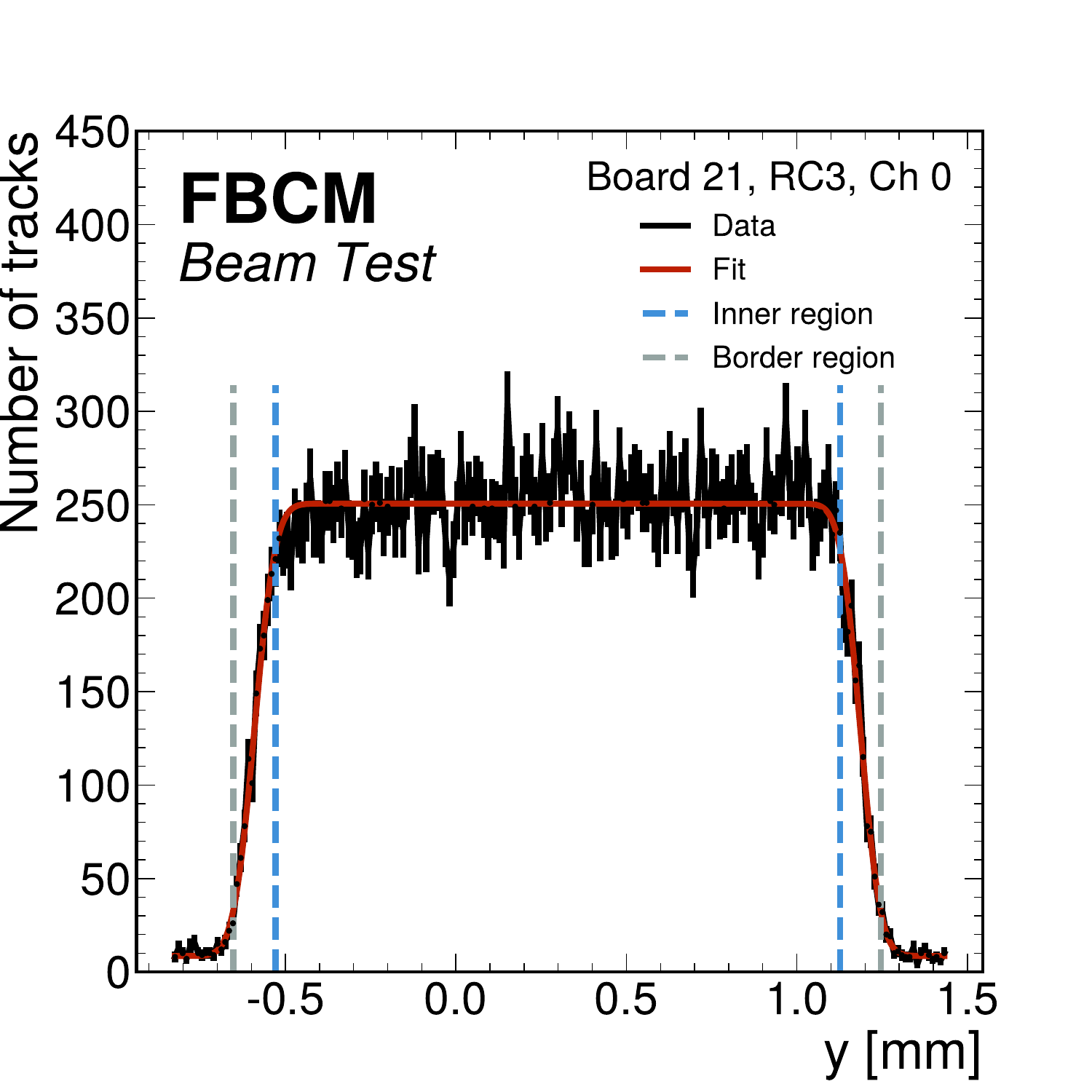}}
    \subfigure{%
        \includegraphics[width=0.44\linewidth]        {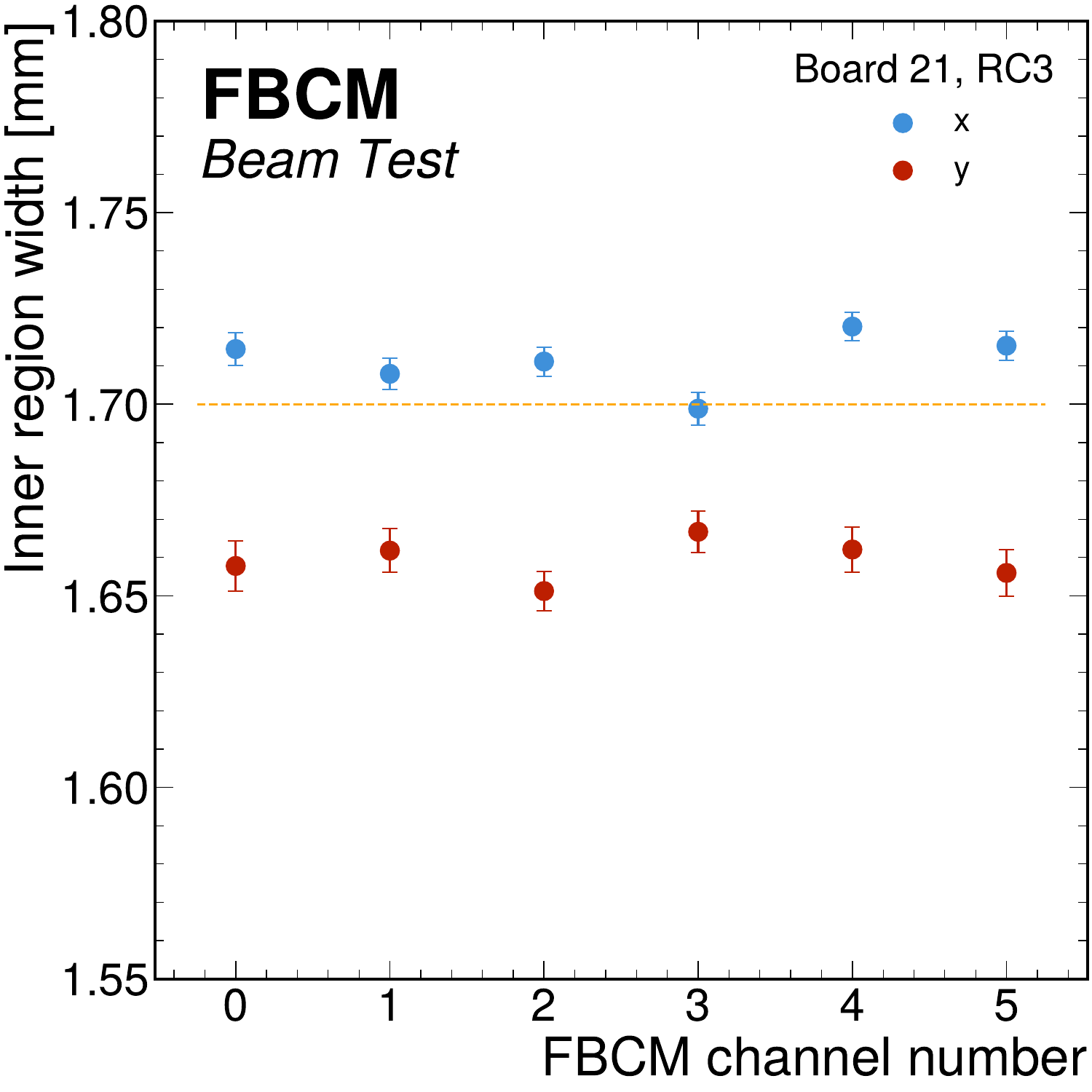}}
\caption{ Left: number of tracks associated with a hit in FBCM channel 0 as a function of the Y coordinate of the reconstructed track at the sensor plane. The inner region is defined between the blue dashed lines and the sensor metallization border region is defined between blue and gray lines on each side. Right: extracted width of the inner region for each FBCM channel in X and Y, where the orange line corresponds to the nominal size of the metallization.}
\label{fig:fed_region}
\end{figure}

After selecting tracks associated with a signal in each FBCM sensor pad, the active area of each pad can be determined. Fig.~\ref{fig:fed_region} (left) shows the number of tracks associated with a signal in one FBCM channel, as a function of the Y coordinate of the reconstructed track interpolated to the sensor plane. The width of the plateau of this distribution corresponds to the width of the inner (fiducial) region of the pad, while the falling edges correspond to the less efficient metallization border region.
After fitting a step function to this distribution with independent sets of parameters for the rising and falling edges, the fiducial region is defined at 90\% of the rectangle height (indicated by blue lines) and border region at 10\% of the rectangle height (indicated by gray lines). 
The width of the inner region is measured as the distance between the blue lines for each of the pads in X and in Y, and the summary plot is shown in Fig.~\ref{fig:fed_region} (right). The orange horizontal line shows the nominal  size of 1.7 mm of the metallization. The tracking resolution is about 3~$\mu$m as shown in Fig.~\ref{fig:telescope_residuals} and, thus, is expected to have a negligible contribution to the observed spread of the widths.

\subsection{Triggered and untriggered readout}
\label{subsec:Triggered_untriggered_readout}

\begin{figure}[htbp]
\centering
    \subfigure{%
        \includegraphics[width=0.45\linewidth]{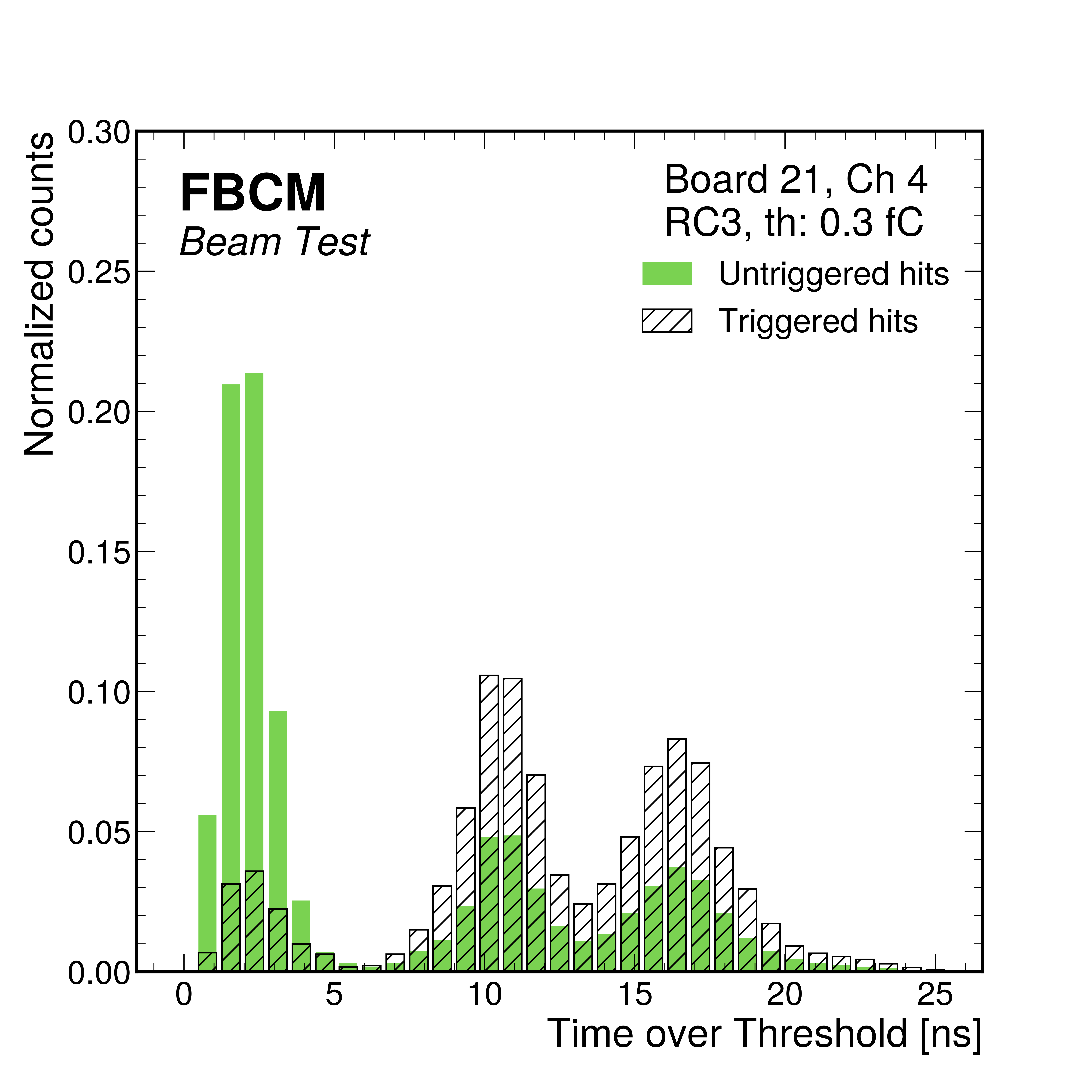}}
    \subfigure{%
        \includegraphics[width=0.45\linewidth]{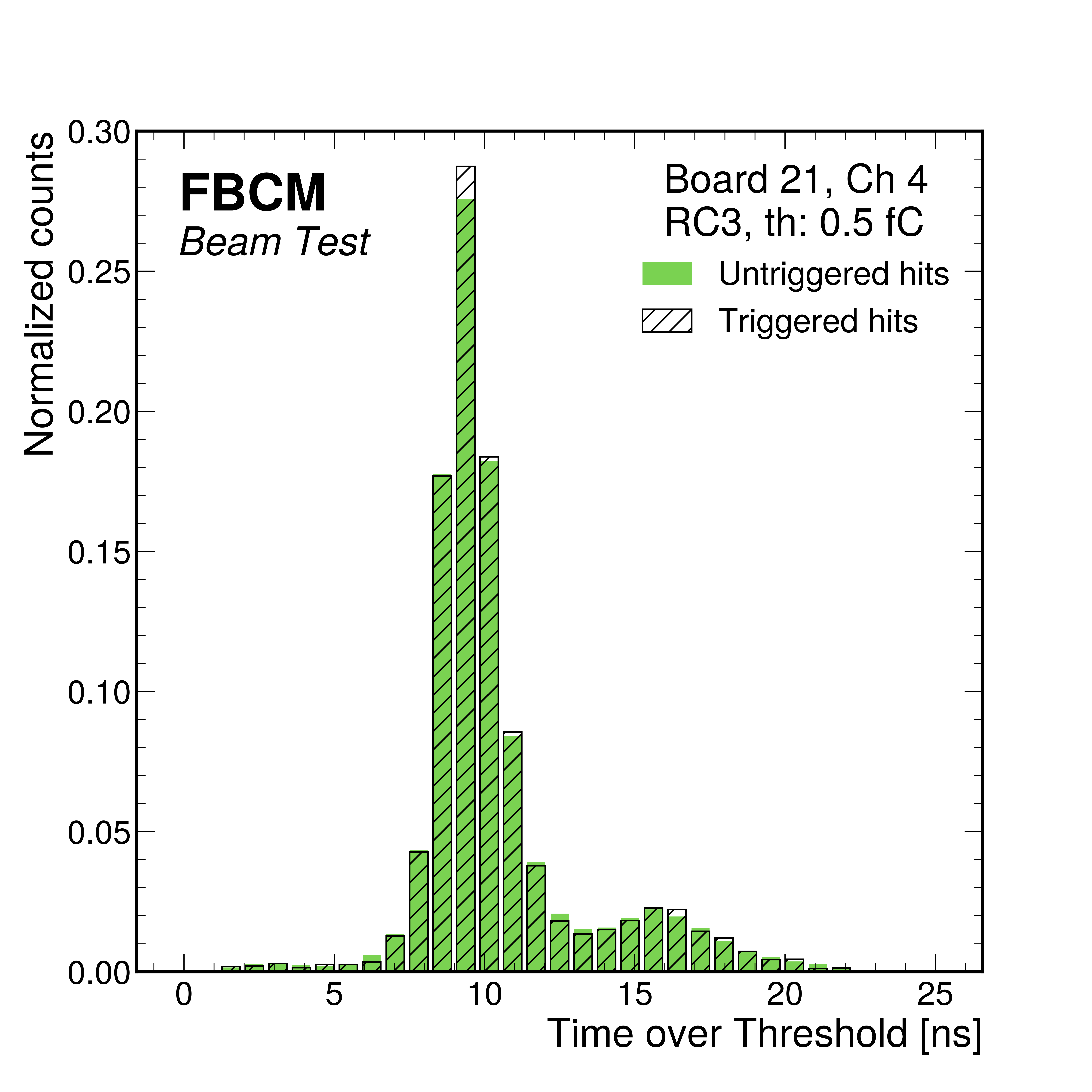}}
\caption{ Normalized distributions obtained with triggered and untriggered readout with a low threshold of 0.3~fC (left) and an increased threshold of 0.5~fC (right).}
\label{fig:Triggered_untriggered_readout}
\end{figure}

FBCM is designed for untriggered operation, however, as a cross--check during the beam test,  a  triggered readout was implemented. Details of the triggered readout are given in Sec.~\ref{subsec:data_acquisition}. At a low threshold of 0.3~fC, where significant contribution from noise (ToT values below 5~ns) is expected, the triggered ToT distribution differs from the untriggered one, as shown in Fig.~\ref{fig:Triggered_untriggered_readout} on the left. This efficient noise suppression is achieved requiring signal from the auxiliary detector (active mask of the SCC, see Sec.~\ref{subsec:data_acquisition}). With the threshold increased to 0.5~fC, the noise is already suppressed by the threshold, no additional suppression is added by triggered readout, and the two distributions are almost indistinguishable, as shown in Fig.~\ref{fig:Triggered_untriggered_readout} on the right. 

The presence of two signal peaks in the measured ToT distribution is unexpected. From the expected energy deposition only the peak around ToT of 10 ns was expected. However, double peak structure was reproduced with very low intensity electron and muon beams to verify that the effect is not dependent on the particle type or beam intensity. It was also investigated later in the laboratory with controlled laser pulse injections (see Sec.~\ref{subsec:TCT}).

\subsection{Sensor response vs. hit position}
\label{subsec:hit_position_study}

The unexpected features in the ToT distributions at low ASIC signal threshold choices are studied in this section, using geometric information of the hit positions in the FBCM sensor pads from tracks reconstructed in the beam telescope.

\begin{figure}[htbp]
\centering
    \subfigure{%
        \includegraphics[width=0.45\linewidth]{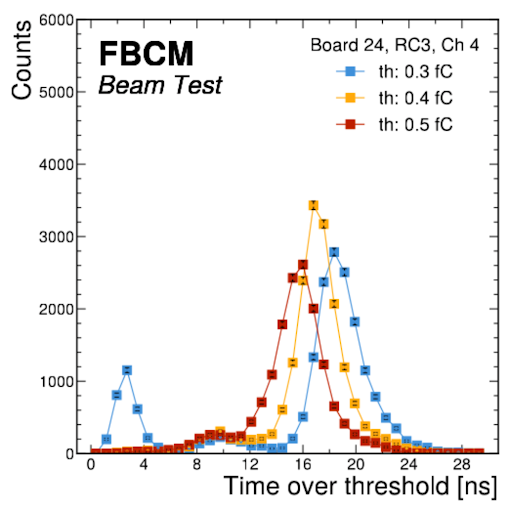}}
    \subfigure{%
        \includegraphics[width=0.45\linewidth]{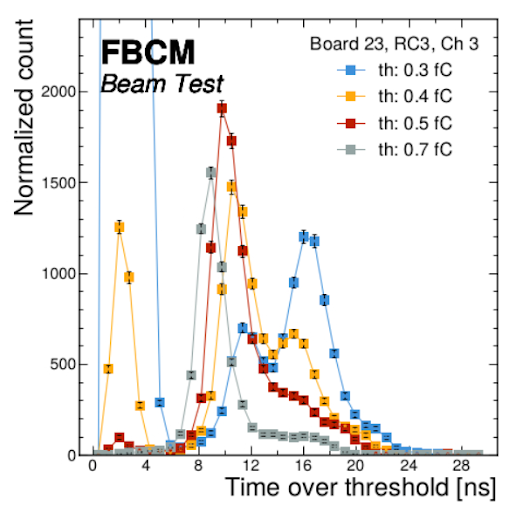}}
\caption{ Comparison of the ToT distributions at different ASIC signal thresholds for an irradiated two-pad sensor (left) and for an unirradiated six-pad sensor (right). ToT below 5~ns corresponds to noise (see Sec.\ref{subsec:Triggered_untriggered_readout}). }
\label{fig:ToT_double_peak}
\end{figure}

\begin{figure}[htbp]
\centering
    \subfigure{%
        \includegraphics[width=0.32\linewidth]{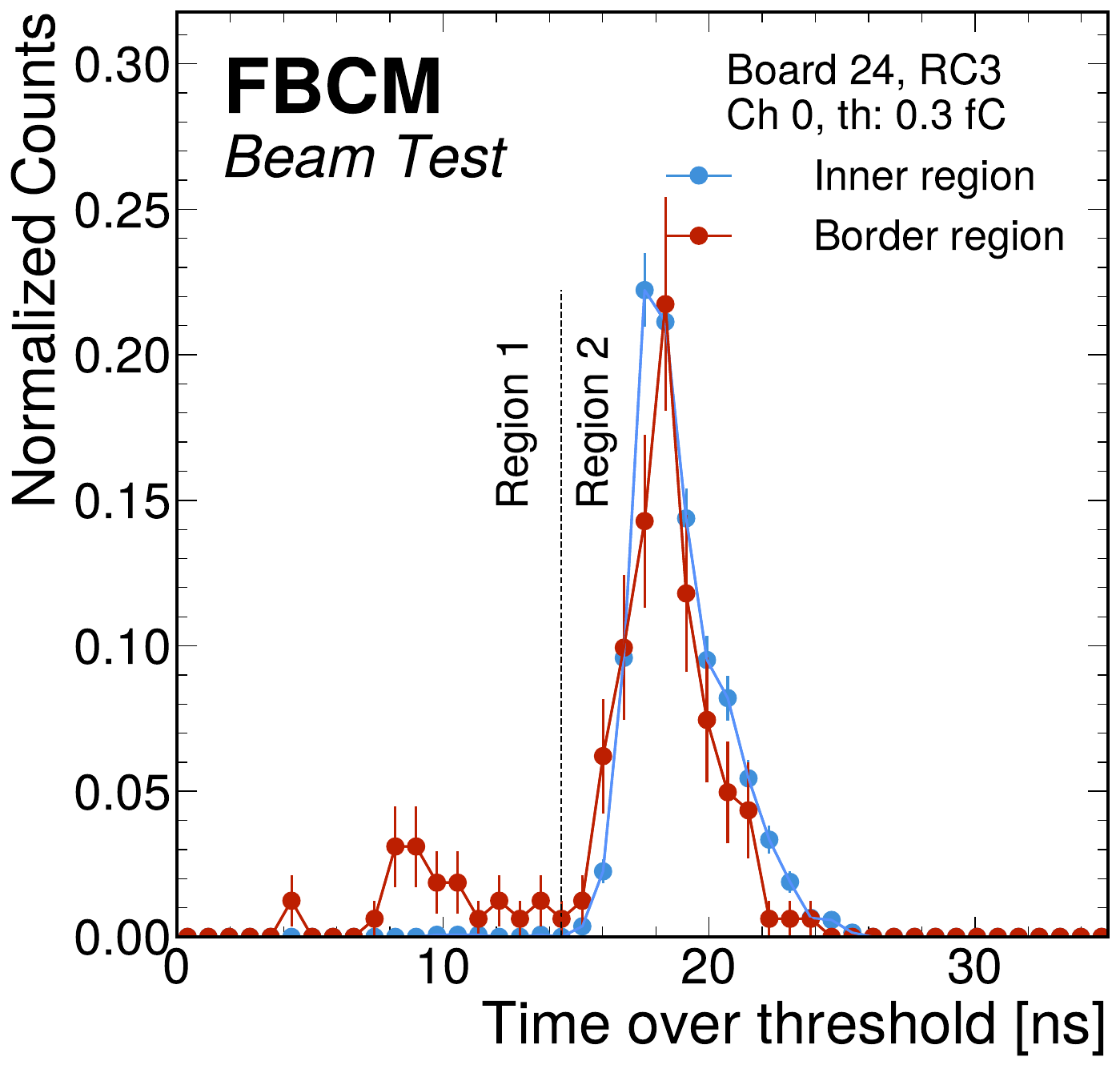}}
    \subfigure{%
        \includegraphics[width=0.3\linewidth]{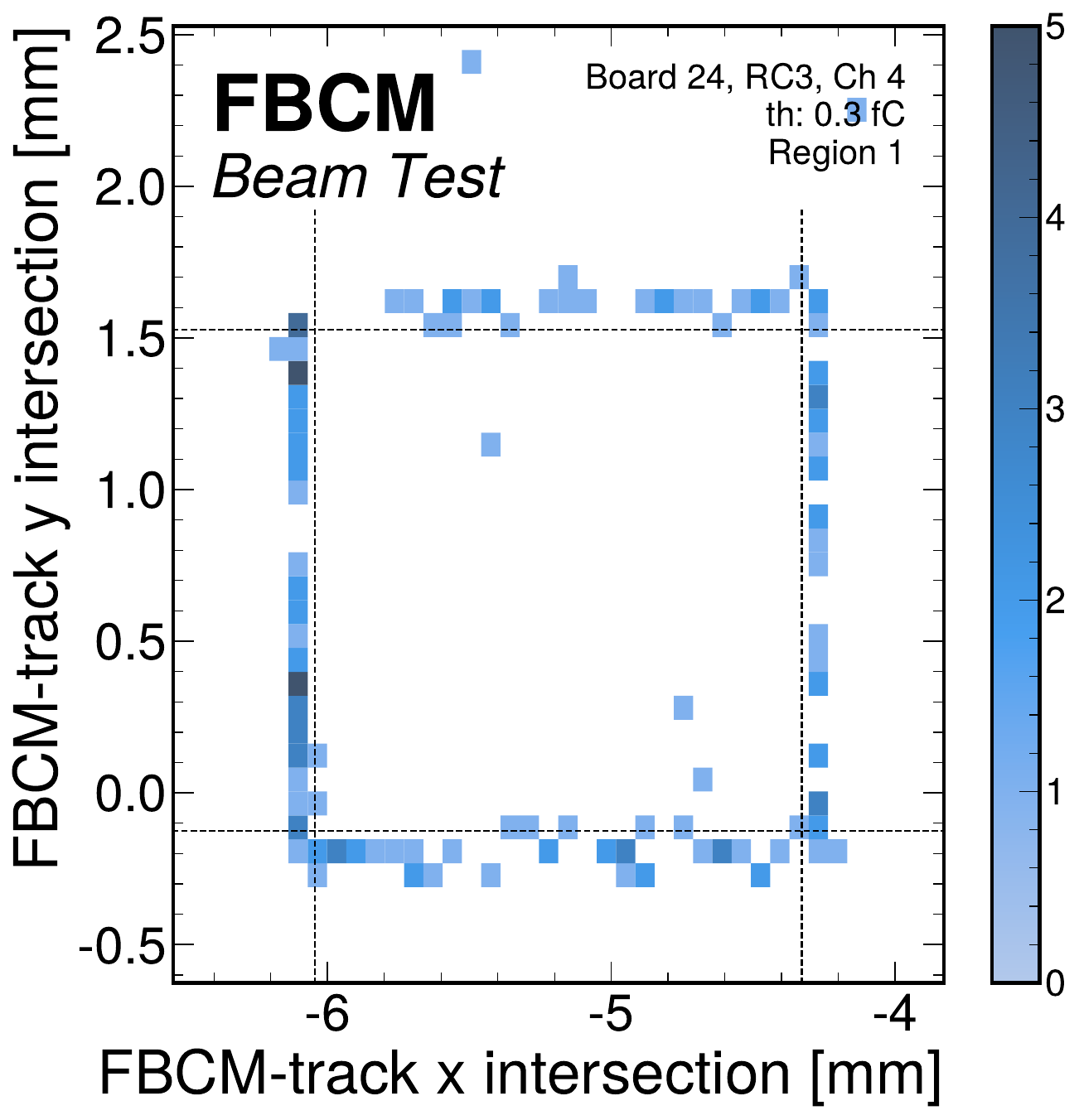}}
    \subfigure{%
        \includegraphics[width=0.3\linewidth]{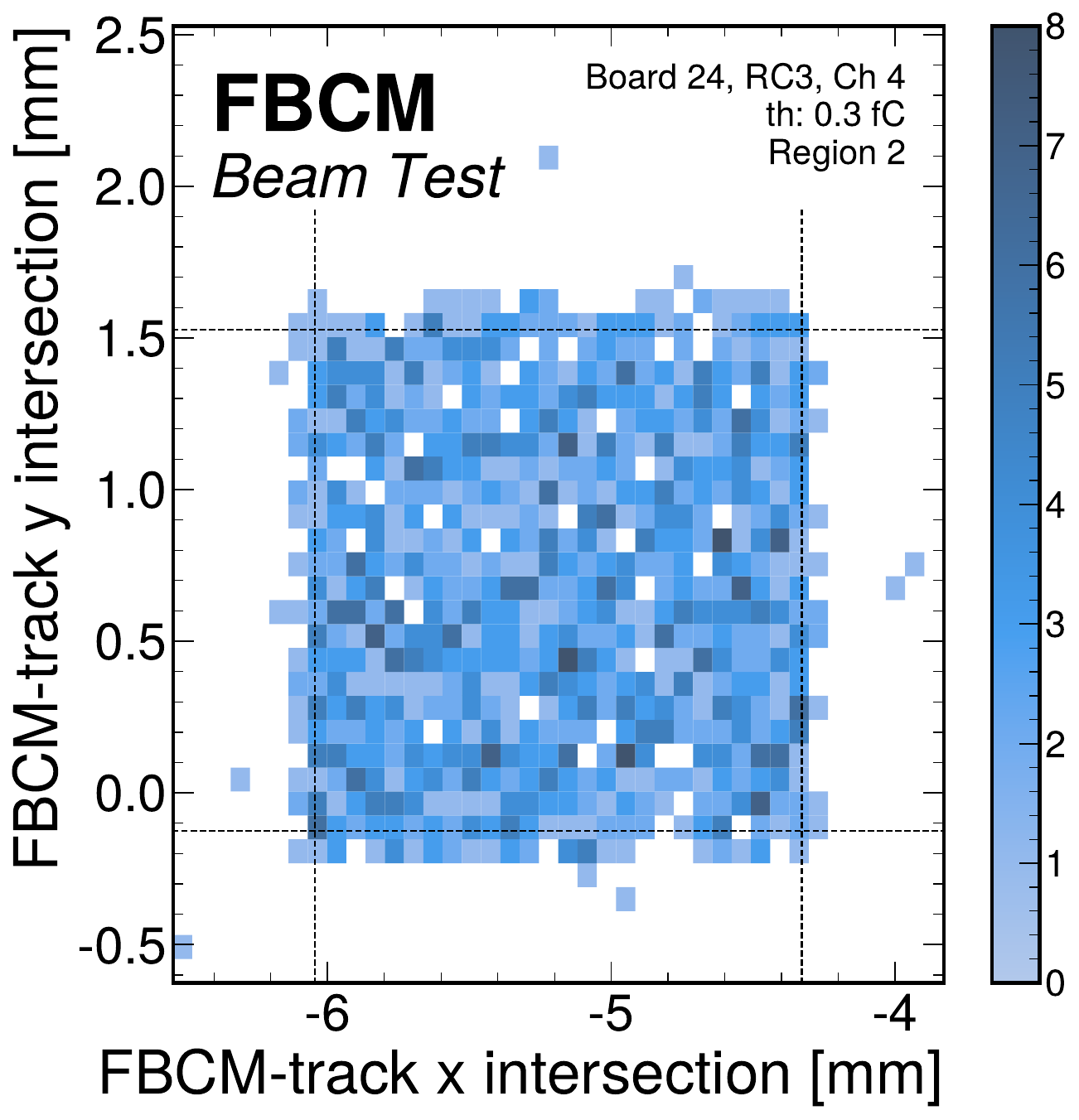}}
\caption{Left: ToT for inner and border regions for a two-pad sensor. Middle: hit map of the tracks associated with ToT below 13~ns. Right: hit map of the tracks associated with ToT above 13~ns. The dashed gray lines indicate the edge of the inner region.}
\label{fig:HitMap_2pad}
\end{figure}

Fig.~\ref{fig:ToT_double_peak} (left) shows the ToT distribution for an irradiated two-pad sensor. A small peak with a mean of around 9 ns is noticeable. This peak was much less pronounced for unirradiated two-pad sensors, presumably due to less defects and more homogeneous electric field at the edges, and is not visible for ASIC thresholds above 0.9~fC (see red line in Fig.~\ref{fig:ToT_bd21_db23_db24_rc3}). To study this effect, the ToT distribution for tracks associated with the inner region of the sensor were plotted separately from those associated with the border region. The definitions of the border and inner regions of the sensor are discussed in Sec.~\ref{subsec:sensor_efficiency}. From Fig.~\ref{fig:HitMap_2pad} (left), it is clear that tracks associated with the border region of the sensor (red) have a dominant contribution to the peak at low ToT values (below 13 ns). 

Moreover, tracks associated with low and high ToT values can be considered separately to investigate any correlation with the hit position. 
Fig.~\ref{fig:HitMap_2pad} (middle) shows that the reconstructed coordinates of tracks associated with ToT values below 13~ns primarily intersect the metallization at the border of the sensor.  Fig.~\ref{fig:HitMap_2pad} (right) shows the positional information of tracks associated with ToT values above 13 ns, which are homogeneously distributed around the inner region of the pad sensor. 
From this study, it was concluded that edge effects contributing to secondary a peak at smaller ToT values are more pronounced for irradiated sensors, and their effect on the ToT distribution can be minimized by increasing the ASIC signal threshold above 0.9 fC. It was shown in Sec.~\ref{subsec:sensor_efficiency} that irradiated two-pad sensors are fully efficient in this configuration.

\begin{figure}[htbp]
\centering
    \subfigure{%
        \includegraphics[width=0.46\linewidth]{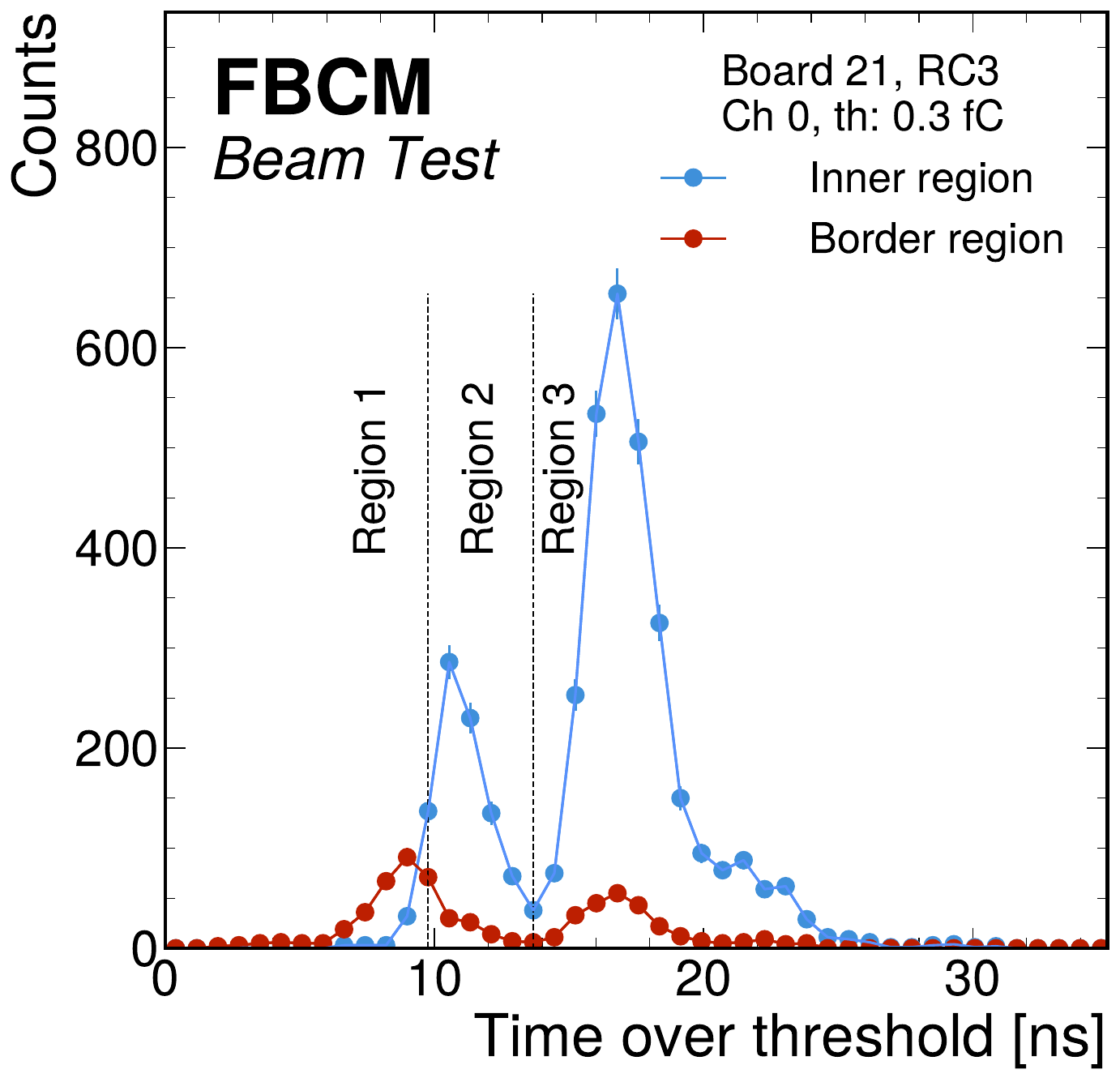}}
    \subfigure{%
        \includegraphics[width=0.45\linewidth]{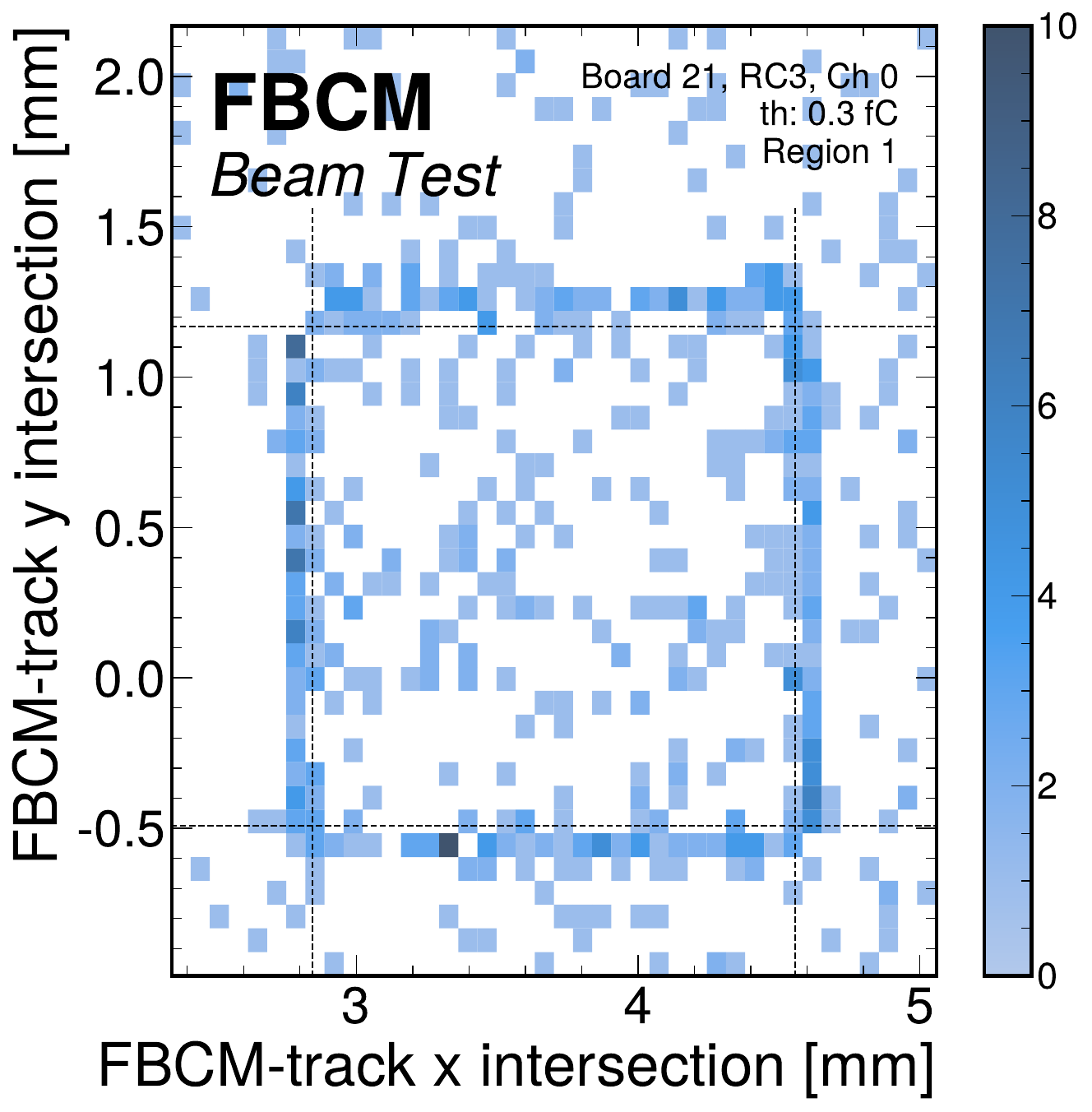}}
    \subfigure{%
        \includegraphics[width=0.46\linewidth]{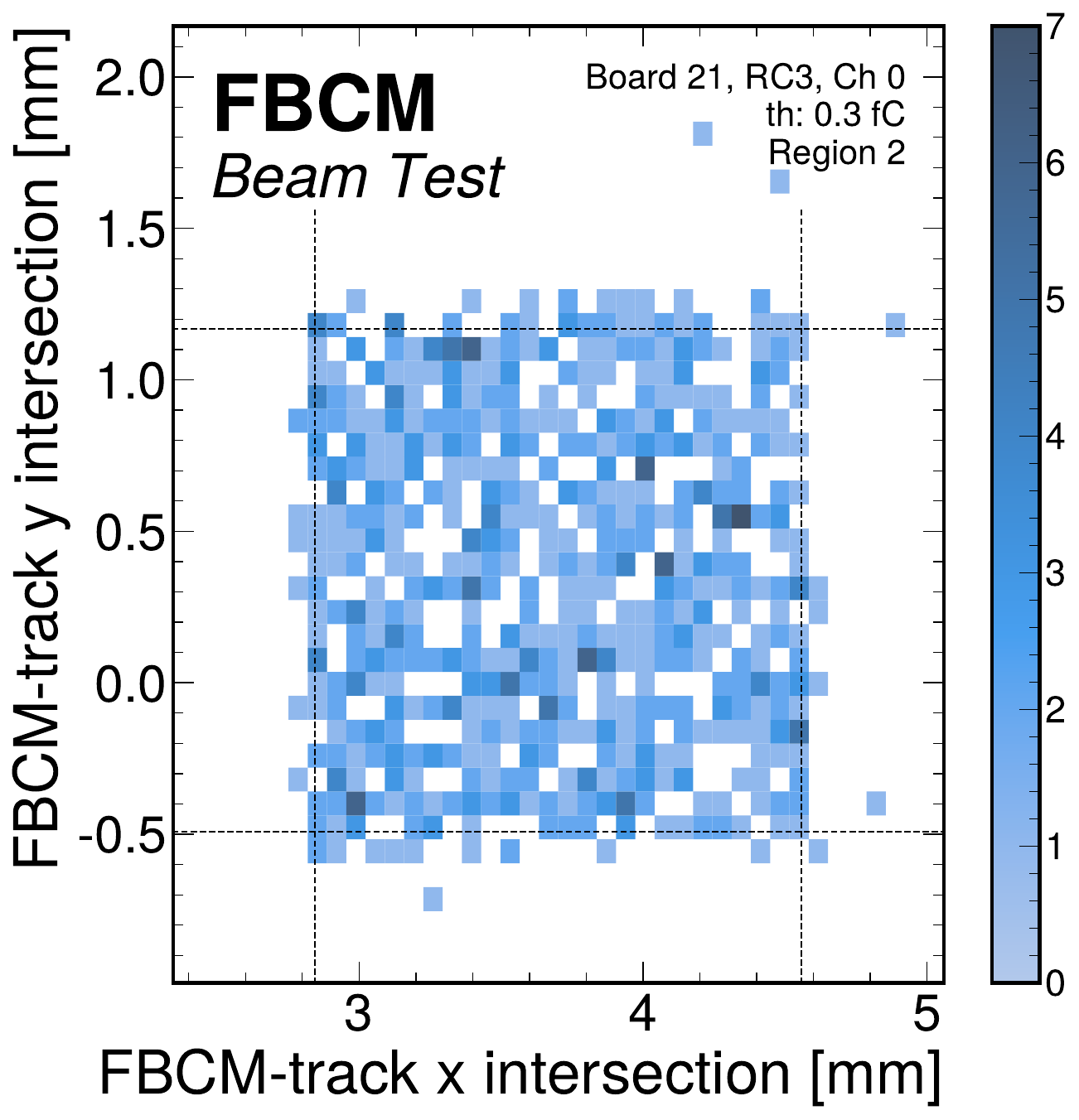}}
    \subfigure{%
        \includegraphics[width=0.46\linewidth]{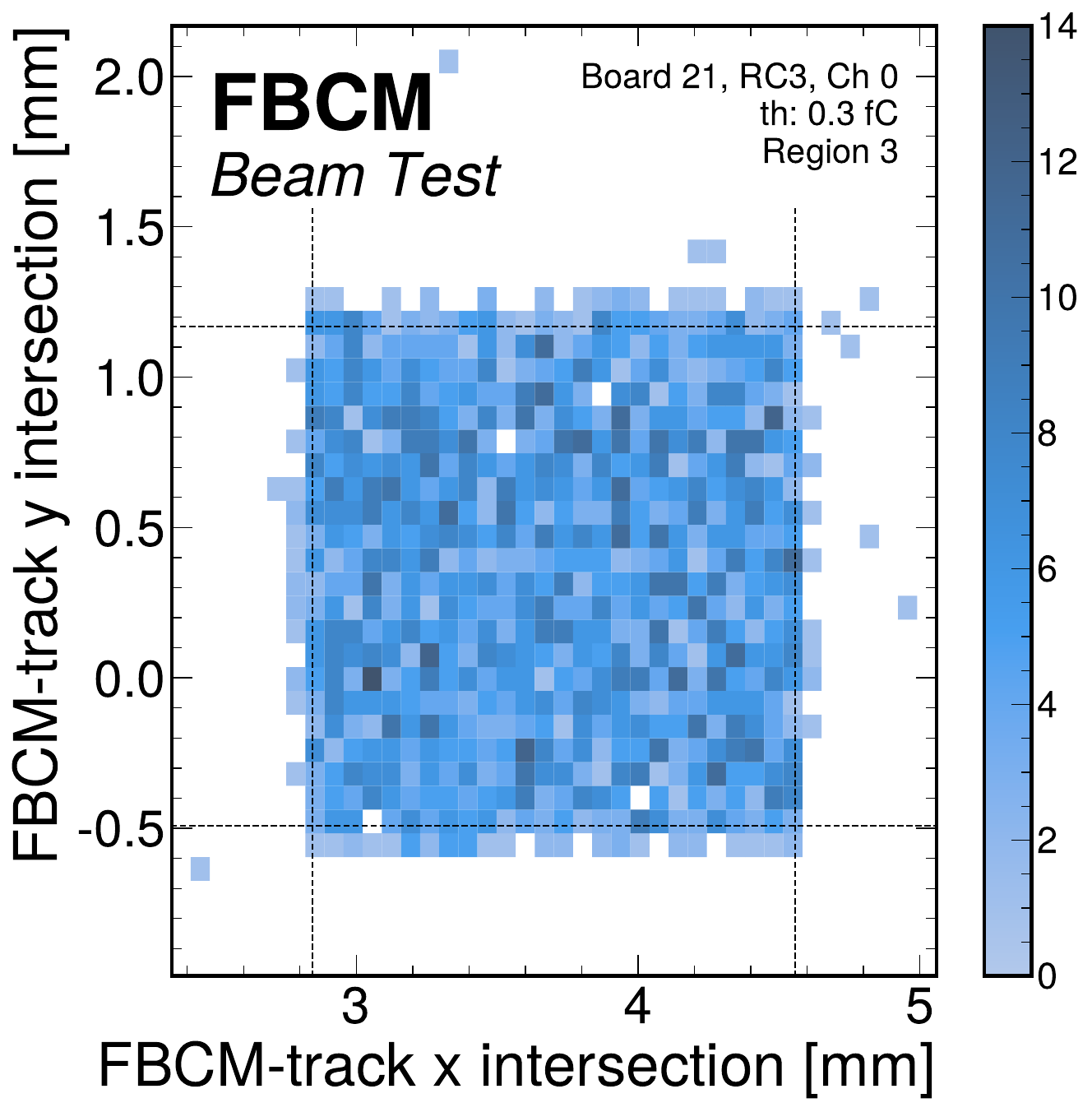}}
\caption{Top left: ToT for inner and border regions for a six-pad sensor. Top right: hit map of the tracks associated with ToT below 10~ns. Bottom left: hit map of the tracks associated with the first signal peak with ToT between 10 and 14~ns. Bottom right: hit map of the tracks associated with the second signal peak with ToT above 14~ns.}
\label{fig:HitMap_6pad}
\end{figure}
 
Fig.~\ref{fig:ToT_double_peak} (right) shows the ToT distribution for a six-pad sensor with different ASIC signal thresholds. A double-peak structure above 5~ns is noticeable, whose shape varies with the ASIC threshold. Note that ToT values below 5~ns correspond to noise (see Sec.\ref{subsec:Triggered_untriggered_readout}) and can be ignored in this study. The ToT distribution corresponding to a threshold of 0.3~fC (blue) shows a first signal peak at 12~ns and a second one at 16 ns. Both peaks shift to lower ToT values with increasing threshold, as expected from signal events, and the lower ToT peak becomes dominant at thresholds above 0.4~fC. It was already shown in Sec.~\ref{subsec:Triggered_untriggered_readout} that both peaks are signal-related. 

An analysis of the ToT distribution versus hit coordinate was carried out to investigate whether the effect can be attributed to tracks associated to a particular area of the sensor. The ToT distribution for tracks associated with the inner region of the sensor (blue) are shown separately from those associated with the border region (red) in Fig.~\ref{fig:HitMap_6pad} (top left). Both distributions show a double peak structure, but tracks corresponding to the border region have a larger contribution to lower ToT values. Three ranges of ToT values were considered separately to investigate the hit positions for tracks associated with each range: below 10~ns, between 10 and 14~ns, and above 14~ns. Region 1, low ToT are mostly associated with tracks intersecting the sensor border, as shown in Fig.~\ref{fig:HitMap_6pad} (top right). Tracks associated with ToT values larger than 10~ns are uniformly distributed around the inner region of the sensor pad, as shown in Fig.~\ref{fig:HitMap_6pad} (bottom left for Region 2 and bottom right for Region 3). 

The edge effects observed for two-pad sensors also affect the six-pad ones, but a further phenomena appears.
The transient current technique was explored to further investigate the double signal peak at larger ToT values (see Sec.~\ref{subsec:TCT}).

\subsection{Transient current technique study of six-pad sensors }
\label{subsec:TCT}

\begin{figure}[htbp]

\centering
    \subfigure{%
        \includegraphics[width=0.95\linewidth]{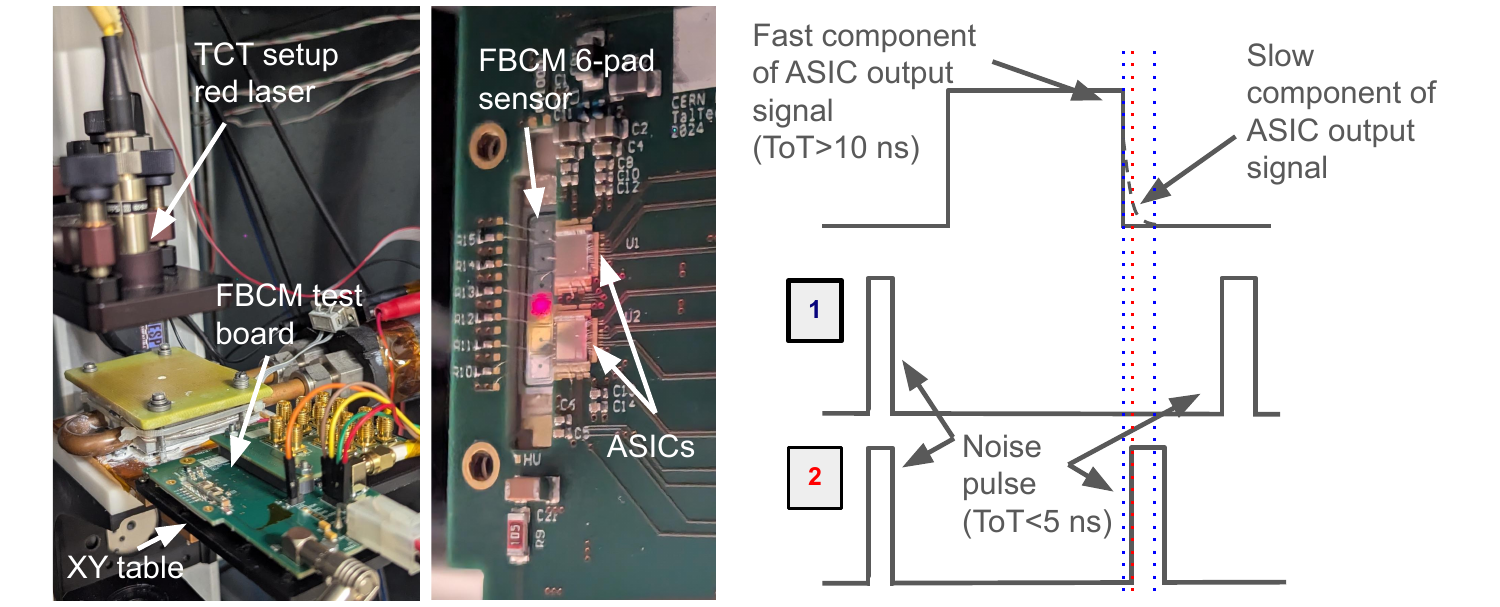}}
\caption{ Left: TCT setup at CERN. Middle: Detail of the six-pad sensor in FBCM test board with red laser pointing to channel 2. Right: sketch of the ASIC output signal and two scenarios of the noise pulses marked 1 and 2. The time window sensitive to after-pulsing is between the blue dotted lines. 
For the scenario number 1, noise arrives after the signal level returned to the baseline, so does not affect measured signal ToT. 
For the scenario number 2, the noise pulse arrival time (red dotted line) coincides with the slow falling edge of the signal, and as signal level is still above the baseline, noise pulse adds to the signal pulse length, causing a longer ToT. }
\label{fig:TCT_setup_photo}
\end{figure}

Transient Current Technique (TCT) exploits the signal induced in electrodes by the motion of
non-equilibrium free charge carriers in a semiconductor. In the TCT setup~\cite{TCT} available at CERN and operated by CERN EP-DT SSD laboratory, two lasers can be used: a red laser with
660~nm wavelength (1.9~eV) and an infrared with 1030 nm wavelength (1.2~eV). For the FBCM sensor test, the red laser was mostly used. The TCT setup and an enlargement of the six-pad sensor in the FBCM test board, with the red laser pointing into channel 2, are shown in Fig.~\ref{fig:TCT_setup_photo} (left and middle). 

FBCM is equipped with n-in-p type silicon sensors, with a negative bias voltage  applied to the back side of the sensor. The readout AC pad on the top surface of the sensor is wire-bonded to the ASIC input. With every red laser pulse,  electron-hole pairs are produced a few $\mu$m from the surface of the readout electrode, which allows the  study of the drift of one kind of charge carriers. Electrons reach the nearest electrode very quickly, while holes take a longer amount of time to travel through the full depth of the sensor. This results in the fast and slow component of ASIC output signal. The latter affects the falling edge of the signal as shown with smooth gray dashed line. Fig.~\ref{fig:TCT_setup_photo} (right) illustrates the expected ASIC response to the laser injection (upper pulse, ToT > 10~ns) and noise (middle and lower pulse, ToT<5 ns) corresponding to operation at low ASIC threshold.  
During the TCT measurements, the waveforms were recorded with an oscilloscope. The ToT of each recorded waveform was aggregated into histograms. 

\begin{figure}[htbp]

\centering
    \subfigure{%
        \includegraphics[width=0.95\linewidth]{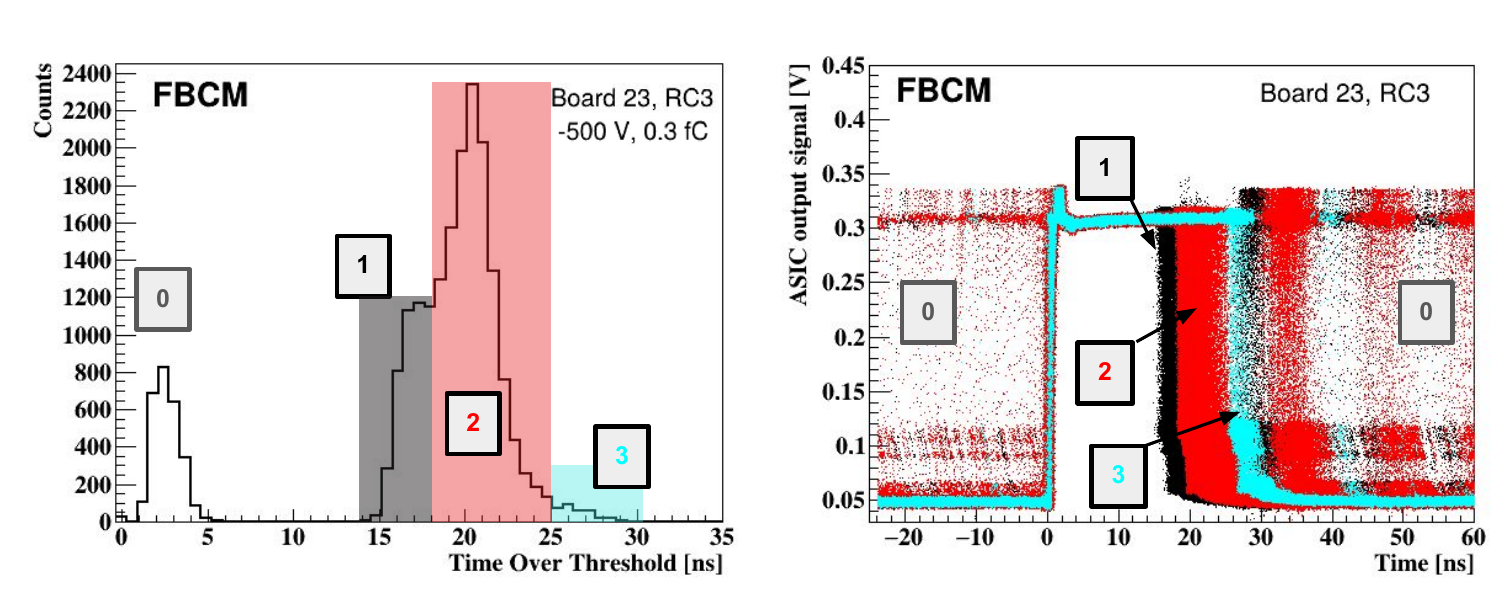}}
\caption{ Left: ToT histogram obtained using the TCT setup with red laser injections. The ASIC threshold is set to 0.3 fC with RC3 setting, and a bias of -500 V is applied to the sensor. Right: ASIC output generated by red laser pulse injections and recorded by the oscilloscope. Four groups of pulses are highlighted in different colors to show the correspondence between the ASIC output pulse (on the right) and the measured ToT (on the left). }
\label{fig:TCT_amp_andASICout_TH03fC}
\end{figure}

An example of the ToT histogram collected for a low ASIC threshold of 0.3~fC is shown in Fig.~\ref{fig:TCT_amp_andASICout_TH03fC} on the left and individual waveforms are shown on the right. Short noise pulses accumulate in the first peak of the ToT histogram, below 5~ns. The second peak is associated with laser injection, but it has a substructure. 
It results from the degraded timing response caused by the slow traveling charges, which affect the ToT and temporarily lower the discriminator's effective threshold following the prompt signal. This temporary threshold reduction increases the front-end’s susceptibility to after-pulsing, which may be induced by either preamplifier electronic noise or electromagnetic interference (EMI), i.e. electronic pickup. Electronic pickup noise is setup-dependent and not related to the ASIC or the sensor. The time window affected by the after-pulsing is captured between the blue dotted lines in Fig.~\ref{fig:TCT_setup_photo} (right). Noise pulses with arrival time in this time window (red dotted line) cause longer ToT, shown in region 2 of Fig.~\ref{fig:TCT_amp_andASICout_TH03fC} (left). The same mechanism and coincidence with a longer noise pulse correspond to region 3 of Fig.~\ref{fig:TCT_amp_andASICout_TH03fC} (left). If no noise pulse arrives within the window, the "nominal" ToT is observed in region 1.

\begin{figure}[h]

\centering
    \subfigure{%
        \includegraphics[width=0.95\linewidth]{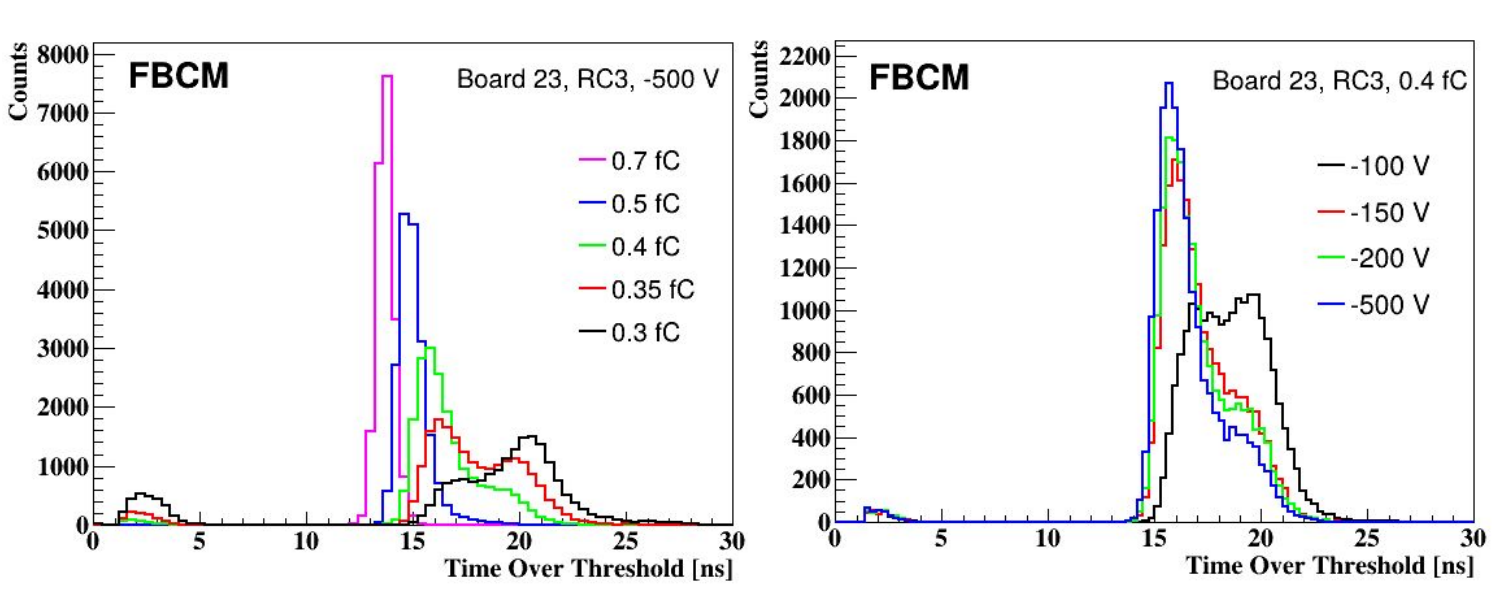}}
\caption{ Left: ToT histogram obtained in the TCT setup with red laser injections. A range of ASIC signal  thresholds is scanned and a bias of -500 V is applied to the sensor. Right:  ToT histogram versus a range of sensor bias voltages and a fixed 0.4~fC ASIC signal threshold. }
\label{fig:TCT_amp_andASICout}
\end{figure}

The ToT distributions obtained for a range of  ASIC signal thresholds at a fixed -500 V bias is shown in Fig.~\ref{fig:TCT_amp_andASICout} (left).
A double-peak structure, similar to the one observed during the beam test, is visible for low thresholds: 0.3~fC (black), 0.35~fC (red), and 0.4~fC (green). 
It was also shown that the applied bias voltage plays an important role and affects the ToT shape, as shown in Fig.~\ref{fig:TCT_amp_andASICout} (right). At -100~V the double-peak is very pronounced, while above -200~V it is suppressed, pointing to reduced charge collection efficiency  (CCE) at low bias voltage. 

\begin{figure}[htbp]
\centering
    \subfigure{%
        \includegraphics[width=0.49\linewidth]{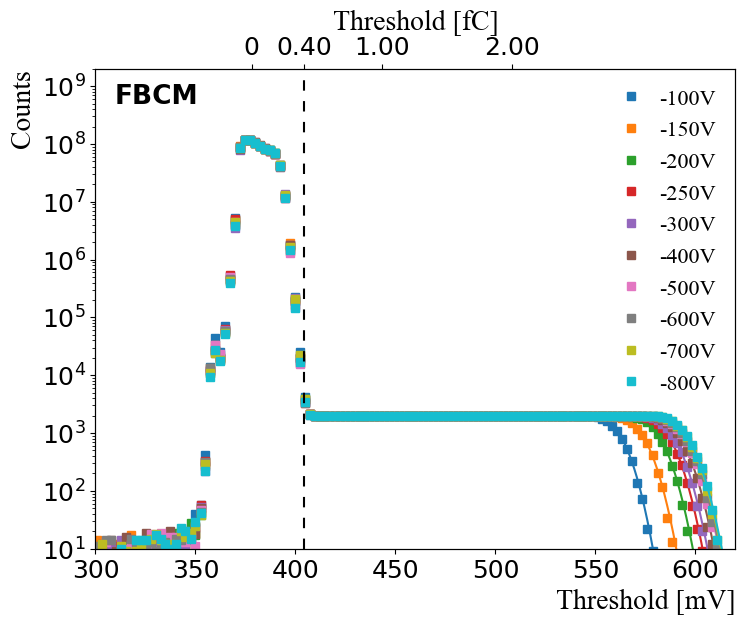}
        \includegraphics[width=0.5\linewidth]{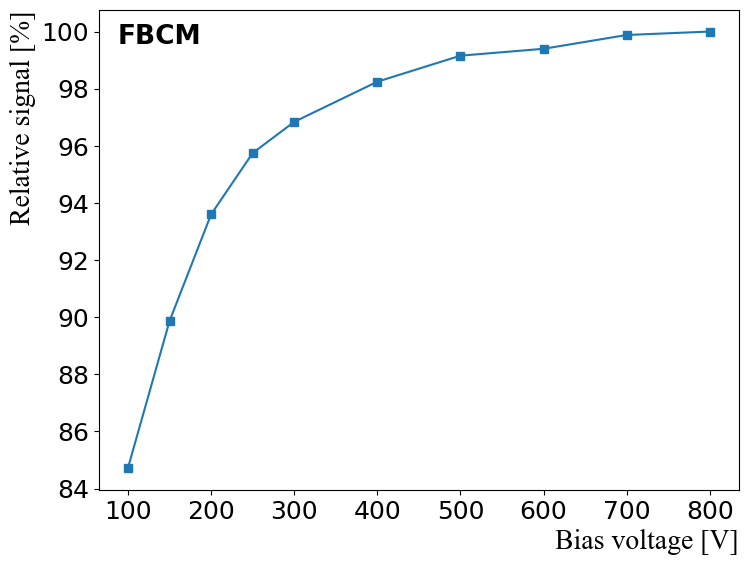}}
\caption{ Left: ASIC signal threshold scan obtained using the TCT setup at various sensor bias voltages, where the corresponding threshold in [fC] is shown on the top x-axis. Right: relative signal with respect to signal at -800~V as a function of the bias voltage. }
\label{fig:THscan_TCT}
\end{figure}

ASIC signal threshold scans at different sensor bias voltages were performed for fixed laser power and injection frequency and are shown in Fig.~\ref{fig:THscan_TCT} (left). Noise with a non-Gaussian shape is noticeable below a threshold of 0.4~fC, which corresponds to pickup noise. The falling edge of the threshold scan distribution shows significant efficiency improvement between -100~V and -200~V, and smaller improvements for voltage steps up to -800~V. Taking integrated counts at -800~V as a reference for full CCE, it was estimated that about 85\% of charge is collected at -100~V, as shown in Fig.~\ref{fig:THscan_TCT} (right). This scenario probes the worst-case CCE, since only the drift of the holes is observed in this kind of TCT experiment with red laser injections next to the electrode. 
The longer charge collection time of the holes in a sensor biased at -100~V with a ballistic deficit of about 15\%, as shown in Fig.~\ref{fig:THscan_TCT}, leads to timing degradation of the front-end response, as the latter represents a convolution of the input signal with the AC characteristics of the full readout chain. The degraded timing response impacts the ToT, which is therefore observed as a change in the ToT versus bias voltage.

Based on these measurements, important conclusions for the operation of the FBCM have been drawn. Pickup noise causes a double peak in the ToT distribution at low signal thresholds. This effect is strongly suppressed by increasing the bias voltage to a value above -200~V and the ASIC threshold above 0.4~fC. The latter condition is already required by the threshold scans presented in Sec.~\ref{subsec:TH_scans} which identified the minimum signal threshold at 0.5 fC. 

\section{Prospects and future work }
\label{sec:prospects}

Simulations of the FBCM ASIC response were conducted to evaluate the performance of the proposed system and optimize some of its parameters.
For example, the area and the position of the sensors were optimized.
The optimal location for 1.7 mm $\times$ 1.7 mm sensors for FBCM was found to be at $r\approx14.5~\text{cm}$~\cite{BRIL-TDR}. These simulation results were compared for two sensor thicknesses: 150~$\mu$m and 290~$\mu$m.
The linear response for both types of sensor was verified, finding deviations from linearity to be below $\pm$0.5\% up to a pileup of 200~\cite{Vahid_sim:2024}. 

The beam test results provide new prospects for tuning of the FBCM ASIC simulation based on real data for different ASIC parameters. 
In simulation, the input pulse shape can be varied to achieve the best agreement with the beam test results for fresh and irradiated sensors of different thicknesses. Then, this tuned simulation can be used to populate in the ASIC parameter space where direct beam test measurements are missing. For example, the simulation of the ToT spectra for intermediate irradiation and all possible RC settings should allow for an optimal choice of RC setting for different eras of detector operation, depending on the aggregated dose. 

As final FBCM deliverable used for luminosity is ToA measurement, more studies of the ToA have to be done. In particular, a decision must be taken on the number of samples per colliding bunch. 
Simulation can be also used to study possible oversampling options for the ToA distribution. In particular, to show if there is a significant ToA improvement in using 8 or 10 samples per bunch crossing, compared to 4 and 6 samples used in Run 3 for the BCM1F detector. Results of this study will be used to motivate the choice of the ToA sampling in the FBCM firmware. 
 
Work is ongoing to port the first version of the firmware tested at the beam test on the prototype FC7 board to the final Apollo board with a  Xilinx Virtex Ultrascale+ FPGA. Apollo FPGA resources allow the implementation of ToT and ToA histograming and processing of the data of three front-end modules served by a single service board, utilizing all available channels of the portcard. The portcard v2, with all the required lines for the ASIC configuration directly from the firmware, is already available.
Prototyping of the front-end module and service board is in progress. The assembly of the first functional FBCM unit for the next beam test is planned by August 2025. It will be the final test before the pre-production of the detector in 2026.

\section{Summary}

The development and prototyping of the Fast Beam Condition Monitor for the CMS Phase-2 upgrade for the HL-LHC is ongoing.
A variety of FBCM test boards with unirradiated and irradiated sensors and FBCM23 ASIC were successfully tested for the first time at a beam test in 2024. Two types of silicon-pad sensors were  considered. Using tracks reconstructed with a beam telescope, the sensor's response as a function of the hit position was studied in detail, and edge effects were found to be more pronounced for irradiated sensors. The silicon sensor efficiency was also measured as a function of the ASIC signal threshold. 
   
The design of the test board (which also served as the very first module prototype) has been optimized by replacing the pitch adapter with direct bonding between the sensor pads and the FBCM23 ASIC inputs to reduce pickup noise. The effect of the pickup noise on the time over threshold distribution was studied with the transient current technique, and an optimal bias voltage and ASIC threshold for FBCM operation were proposed. The current design of the test board with a 150~$\mu$m thick six-pad sensor with double guard rings and two FBCM23 ASICs was approved after the beam test and is being used as the starting point for the front-end module design. Stability of the new six-pad sensor was tested with X-ray, and despite shorter efficiency vs. threshold range (due to smaller sensors thickness) was prioritized above two-pad sensor. New front-end is expected to be ready and prototyped by summer 2025, in time for the next beam test with (close to) final front- and back-end components.

\acknowledgments
We acknowledge the support by the following institutes and funding agencies: CERN;   
the national research projects RVTT3 ``CERN Science Consortium of Estonia" and PUT PRG1467 ``CRASHLESS" (Estonia);
Helmholtz-Gemeinschaft Deutscher Forschungszentren (HGF) (Germany); National Research, Development and Innovation Office (NKFIH), including contract numbers K 143460 and TKP2021-NKTA-64 (Hungary);
the US CMS operations program, the US National Science Foundation (NSF), and the US Department of Energy (DOE) (USA).
We acknowledge support of the CERN EP-DT wire binding laboratory, CERN EP-DT SSD laboratory for TCT studies, CERN EP-ESE X-ray laboratory and IRRAD proton facility for ASIC and sensor irradiation. We are grateful to Andre Rummler for the beam telescope assembly and operation support, Michael Moll and Stefano Mersi for fruitful discussions and advice, and Giuseppe Pezzullo and Federico Ravotti for assistance with sensors preparation for irradiation.  

\bibliographystyle{JHEP}
\bibliography{references}

\end{document}